%% file: paper.tex
\documentclass[sigconf]{acmart}
\acmConference[ICSE 2024]{46th International Conference on Software Engineering}{April 2024}{Lisbon, Portugal}

\AtBeginDocument{%
  \providecommand\BibTeX{{%
    \normalfont B\kern-0.5em{\scshape i\kern-0.25em b}\kern-0.8em\TeX}}}
\usepackage{cleveref}

\setcopyright{acmcopyright}
\copyrightyear{2024}
\acmYear{2024}
\acmDOI{XXXXXXX.XXXXXXX}

\copyrightyear{2024}
\acmYear{2024}
\setcopyright{acmlicensed}\acmConference[ASE '24]{39th IEEE/ACM
International Conference on Automated Software Engineering }{October
27-November 1, 2024}{Sacramento, CA, USA}
\acmBooktitle{39th IEEE/ACM International Conference on Automated Software
Engineering (ASE '24), October 27-November 1, 2024, Sacramento, CA, USA}
\acmDOI{10.1145/3691620.3695075}
\acmISBN{979-8-4007-1248-7/24/10}

\usepackage{xcolor}
\usepackage{xspace}
\usepackage{pgf}

\usepackage{graphicx}
\usepackage{hyperref}
\usepackage{subcaption}
\usepackage{listings}
\usepackage{color}
\usepackage{tikz}

\definecolor{dkgreen}{rgb}{0,0.6,0}
\definecolor{gray}{rgb}{0.5,0.5,0.5}
\definecolor{mauve}{rgb}{0.58,0,0.82}

\usepackage{dcolumn}
\newcolumntype{d}[1]{D{.}{.}{#1}} % argument is number of decimal places
\newcolumntype{.}{d{0}} % "0" means decimal point is right-justified, not so different from right-justifying the column

\usepackage{datetime2}

\providecommand{\mathdefault}[1][]{}

\acmYear{2024}
\setcopyright{rightsretained}
\acmConference[ASE '24]{39th IEEE/ACM International Conference on Automated Software Engineering }{October 27-November 1, 2024}{Sacramento, CA, USA}
\acmBooktitle{39th IEEE/ACM International Conference on Automated Software Engineering (ASE '24), October 27-November 1, 2024, Sacramento, CA, USA}
\acmISBN{979-8-4007-1248-7/24/10}

\begin{document}

\input{macros.tex}

\title{Evaluation of Version Control Merge Tools}


% Using "\author" leads to ugly formatting.

% \author{Benedikt Schesch}
% \email{scheschb@ethz.ch}
% \affiliation{%
%   \institution{ETH Z\"urich}
%   \country{Switzerland}}
% 
% \author{Ryan Featherman}
% \email{ryan.featherman3@gmail.com}
% \affiliation{%
%   \institution{Microsoft}

% \author{Kenneth J. Yang}
% \email{kjy5@cs.washington.edu}

% \author{Ben R. Roberts}
% \email{brober3@cs.washington.edu}

% \author{Michael D. Ernst}
% \email{mernst@cs.washington.edu}

% \author{\mbox{Kenneth J. Yang \quad Ben R. Roberts \quad Michael D. Ernst}}
% \email{{kjy5,brober3,mernst}@cs.washington.edu}

% % Shared affiliation for last 3 authors.
% \affiliation{%
%   \institution{University of Washington}

\begin{teaserfigure}
\vspace*{-2em}
\begin{minipage}{.2\columnwidth}
\begin{centering}\LARGE
Benedikt Schesch \\ \large
b.schesch@googlemail.com \\
ETH Z\"urich \\
\end{centering}
\end{minipage}%
\hfill%
\begin{minipage}{.2\columnwidth}
\begin{centering}\LARGE
Ryan Featherman \\ \large
ryan.featherman3@gmail.com\\
Microsoft \\
\end{centering}
\end{minipage}%
\hfill%
\begin{minipage}{.5\columnwidth}
\begin{centering}\LARGE
Kenneth J. Yang \hfill Ben R. Roberts \hfill Michael D. Ernst \\  \large
\{kjy5,brober3,mernst\}@cs.washington.edu \\
University of Washington \\
\end{centering}
\end{minipage}%
\vspace*{1em}
\end{teaserfigure}

\begin{abstract}

A version control system, such as Git, requires a way to integrate changes
from different developers or branches. Given a merge scenario, a merge tool
either outputs a clean integration of the changes, or it outputs a conflict for
manual resolution.  A clean integration is correct if it preserves intended
program behavior, and is incorrect otherwise (e.g., if it causes a test
failure).  Manual resolution consumes valuable developer time, and
correcting a defect introduced by an incorrect merge is even more costly.

New merge tools have been proposed, but they have not yet been evaluated
against one another.
Prior evaluations
do not properly distinguish between correct and incorrect merges,
are not evaluated on a realistic set of merge scenarios,
and/or do not compare to state-of-the-art tools.
We have performed a more realistic evaluation.
The results differ significantly from previous claims, setting the record
straight and enabling better future research.
Our novel experimental methodology combines running test suites,
examining merges on deleted branches, and accounting for the cost of
incorrect merges.\looseness=-1

Based on these evaluations, we created a merge tool that
outperforms all previous tools under most assumptions.
It handles the most common merge
scenarios in practice.

\end{abstract}



\maketitle

\addtolength{\textfloatsep}{-.25\textfloatsep}
\addtolength{\dbltextfloatsep}{-.25\dbltextfloatsep}
\addtolength{\floatsep}{-.25\floatsep}
\addtolength{\dblfloatsep}{-.25\dblfloatsep}

\section{Introduction}

A merge occurs when developers using a version control system (VCS), such
as Git, integrate changes from different branches --- for example, after
two developers concurrently edit their
own copies of the code.
Merge tools, commonly utilized by VCSs, are designed to automatically
integrate changes.
The input to a merge tool is a pair of revisions or commits.

A
merge tool can produce three kinds of results.
In an \emph{unhandled} merge, the tool is
unable to integrate some changes and reports conflicts.  The programmer
must handle the conflicts.
A \emph{clean} merge has no unhandled changes (that is, no
conflicts); it can be further divided into the other two results, \emph{correct} and \emph{incorrect} merges.
In a correct
merge, the tool produces the desired output:  a runnable program version
that properly integrates both changes. In an incorrect merge,
the tool still produces a single merged program version without conflicts. However, that program version fails
to correctly integrate both changes. An incorrect merge could result in a
compilation error or in a runnable program that
fails its tests.

Merging is a well-known pain point for developers, so
many new merge tools have been proposed~\cite{Asklund1994,HuntT2002,ApelLBLK2011,TrindadeTavaresBCS2019,ApelLL2012,LessenichAL2014,CavalcantiBA2017,AsenovGMO2017,10.1145/2494266.2494277,LarsenFBM2023,ZHY2019,Hume2017,Lindholm2004,BakaoukasB2020,SkafMolliMRN2008,HorwitzPR89:TOPLAS,Binkley91,10.1109/CVSM.2009.5071721,SvyatkovskiyFGMDBJSL2022,Berzins1994}.
These papers present and evaluate many exciting and intriguing ideas,
such as structured merging.
Unfortunately, the state-of-the-art tools Hi-res
Merge~\cite{git-hires-merge}, IntelliMerge~\cite{ShenZZLJW2019}, and
Spork~\cite{LarsenFBM2023} have not been compared to one another.
{\ifanonymous{To the best of our knowledge, a}\else{A}\fi}ll previous
evaluations of merge tools suffer from at least one of these
three problems:  they do not evaluate \emph{merge correctness}, they are not
evaluated on \emph{representative merges}, or they do not compare with
\emph{state-of-the-art tools}.

We address these shortcomings and provide a more comprehensive
evaluation of merge tools.  To detect incorrect merges at scale, we
utilize automated testing.  To indicate how well a merge tool works under
a realistic mix of development scenarios, we collected merges from all
branches, including ones that have been deleted from the VCS\@.
% and also from pull requests' history.
We compared a broad set of merge tools (\cref{sec:evaluated-tools}),
including some that have never been evaluated before and new ones that
we created.

The contributions of this paper include:

\begin{itemize}
\item A novel experimental methodology for evaluating merge
  tools\ifanonymous\else{, which fixes flaws in previous evaluations}\fi.
  It validates merges via testing, includes non-main-branch merges, and
  quantitatively accounts for the cost of incorrect merges.
\item Open-source experimental infrastructure that implements our 
  methodology.
\item Comparison of both industrial and research merge tools.
\item New merge tools that outperform existing ones.
\end{itemize}

Here are some of the novel findings from our research:

\input{findings-list}

\section{Merge Algorithms}

\subsection{Terminology}
\label{sec:background-merge-algorithms}

The input to a VCS merge is a pair of commits or branch heads, called the
\emph{parent commits}.    The VCS stores the result in a merge commit, which is
a commit with two
%% This is true, but let's not introduce unneeded complexity here.
% (or more)
parents.  

A commit represents a single
state of the file system that the VCS manages.
To perform a merge, a VCS calls out to a three-way merge
tool~\cite{Mens2002}, passing the parent commits and their base commit.
The base commit is the nearest common ancestor in the VCS\@.
Use of a three-way merge tool simplifies the task of merging two file
system states to the task of integrating two sets of changes: the changes
from the merge base to parent 1, and the changes from the merge base to
parent 2.

To the best of our knowledge,
every three-way merge algorithm
has two phases: alignment and resolution.
The biggest differences are the program representation and
the change representation.

The \emph{alignment} or matching phase, identifies the unchanged sections in all
three versions, thus determining the relative position of changes.
The alignment phase is performed by a tool such as \<diff>.
%(typically between either 2 or 3 versions, though larger numbers are possible).
% (typically 2 or 3 versions).
A line-based diff consists of alternating common (unchanged) code sequences
and \emph{hunks}.
A hunk is a set of contiguous added, removed, and/or changed lines
between versions of a file.
For generality to non-line-based tools, this
paper uses the term ``change'' rather than ``hunk''.
The common code sequences are typically left implicit in
the diff representation.

The \emph{resolution} phase of three-way merging uses the following algorithm.
For each change C in a 3-way diff,
% (in a line-based tool, the change C is a hunk),
let C1 be the difference between the base and parent 1 and
% TODO: Could save a line here by saying "and likewise for C2".
let C2 be the difference between the base and parent 2.
C1 and C2 are at the same location in the source code.

% \vspace{-5pt}

\begin{itemize}
\item
  If C1 is the same as C2, use it;
  equivalently, if parent 1 is the same as parent 2, use it.
\item
  If C1 is empty, use C2;
  equivalently, if the base is the same as parent 1, use parent 2.
\item
  If C2 is empty, use C1;
  equivalently, if the base is the same as parent 2, use parent 1.
\item
  If C1 differs from C2, report a conflict;
  equivalently, if the base, parent 1, and parent 2 all differ, report a conflict.
\end{itemize}

% \vspace{-5pt}

Alternative merging schemes have been proposed that utilize a different
representation of a program than its lines.  For example, the Abstract
Syntax Tree (AST) is a parsed representation of a program that represents
program constructs with parent--child relationships.  The line-based
representation of two changes might be a conflict, but the tree-based
representation might not be a conflict because the two changes occur in
different places in the tree even though they appear on the same line in
the source code.
Tree-based~\cite{Asklund1994,HuntT2002,ApelLBLK2011,TrindadeTavaresBCS2019,LessenichAL2014,LessenichAL2014,CavalcantiBA2017,10.1145/1860559.1860600,10.1145/2494266.2494277,LarsenFBM2023,ZHY2019,Hume2017,Lindholm2004,BakaoukasB2020,SkafMolliMRN2008}
and
graph-based~\cite{HorwitzPR89:TOPLAS,Binkley91,10.1109/CVSM.2009.5071721}
merge algorithms can correctly merge edits that line-based tools consider a
conflict.

\subsection{Weaknesses of Merge Algorithms}

\begin{figure}
\begin{center}
    \begin{subfigure}[c]{0.13\textwidth}
        \begin{lstlisting}[basicstyle=\footnotesize\ttfamily,numbers=none]
def main():
    <@\textcolor{black}{n}@> = <@\textcolor{black}{128}@>
    print(<@\textcolor{black}{n}@>)
        \end{lstlisting}
        \vspace{-5pt}
        \centerline{\textbf{Merge base}}
    \end{subfigure}
\hfill
    \begin{subfigure}[c]{0.13\textwidth}
        \begin{lstlisting}[basicstyle=\footnotesize\ttfamily,numbers=none]
def main():
    <@\textcolor{orange}{n\_people}@> = <@\textcolor{black}{128}@>
    print(<@\textcolor{orange}{n\_people}@>)
        \end{lstlisting}
        \vspace{-5pt}
        \centerline{\textbf{Parent 1}}
%    \end{subfigure}
%    \hfill
%    \begin{subfigure}{0.13\textwidth}

\vspace{10pt}

        \begin{lstlisting}[basicstyle=\footnotesize\ttfamily,numbers=none]
def main():
    <@\textcolor{black}{n}@> = <@\textcolor{purple}{64}@>
    print(<@\textcolor{black}{n}@>)
        \end{lstlisting}
        \vspace{-5pt}
        \centerline{\textbf{Parent 2}}
    \end{subfigure}
\hfill
    \begin{subfigure}[c]{0.13\textwidth}
        \begin{lstlisting}[basicstyle=\footnotesize\ttfamily,numbers=none]
def main():
    <@\textcolor{orange}{n\_people}@> = <@\textcolor{purple}{64}@>
    print(<@\textcolor{orange}{n\_people}@>)
        \end{lstlisting}
        \vspace{-5pt}
        \centerline{\textbf{Merged}}
    \end{subfigure}
\end{center}
  \begin{tikzpicture}[overlay]
      \draw[->, thick] (-2.0, 2.3) -- (-1.4, 2.7); % Arrow from Merge base to Parent 1
      \draw[->, thick] (-2.0, 1.7) -- (-1.4, 1.3); % Arrow from Merge base to Parent 2
      \draw[->, thick] (1.2, 2.7) -- (1.8, 2.3); % Arrow from Parent 1 to Merged
      \draw[->, thick] (1.2, 1.3) -- (1.8, 1.7); % Arrow from Parent 2 to Merged
  \end{tikzpicture}
\halfprecaptionspace
\caption{Mergeable changes that line-based merge reports as a conflict.}
\label{fig:line-based-conflict}
\end{figure}
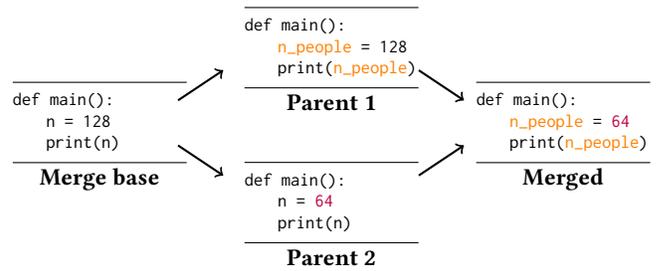

\begin{figure}
\begin{center}
    \begin{subfigure}[c]{0.13\textwidth}
        \begin{lstlisting}[basicstyle=\footnotesize\ttfamily,numbers=none]
def <@\textcolor{black}{mult}@>(a,b):
    return a*b
def main():
    a = <@\textcolor{black}{3*5}@>
    print(a)
        \end{lstlisting}
        \vspace{-5pt}
        \centerline{\textbf{Merge base}}
    \end{subfigure}
\hfill
    \begin{subfigure}[c]{0.13\textwidth}
        \begin{lstlisting}[basicstyle=\footnotesize\ttfamily,numbers=none]
def <@\textcolor{orange}{multiply}@>(a,b):
    return a*b
def main():
    a = <@\textcolor{black}{3*5}@>
    print(a)
        \end{lstlisting}
        \vspace{-5pt}
        \centerline{\textbf{Parent 1}}
%     \end{subfigure}
%     \hfill
%     \begin{subfigure}[c]{0.13\textwidth}

% \vspace{10pt}

        \begin{lstlisting}[basicstyle=\footnotesize\ttfamily,numbers=none]
def <@\textcolor{black}{mult}@>(a,b):
    return a*b
def main():
    a = <@\textcolor{purple}{mult(3,5)}@>
    print(a)
        \end{lstlisting}
        \vspace{-5pt}
        \centerline{\textbf{Parent 2}}
    \end{subfigure}
\hfill
    \begin{subfigure}[c]{0.13\textwidth}
        \begin{lstlisting}[basicstyle=\footnotesize\ttfamily,numbers=none]
def <@\textcolor{orange}{multiply}@>(a,b):
    return a*b
def main():
    a = <@\textcolor{purple}{mult(3,5)}@>
    print(a)
        \end{lstlisting}
        \vspace{-5pt}
        \centerline{\textbf{Merged (incorrectly)}}
    \end{subfigure}
    \begin{tikzpicture}[overlay]
      \draw[->, thick] (-6.2, 0.4) -- (-5.6, 0.8); % Arrow from Merge base to Parent 1
      \draw[->, thick] (-6.2, -0.2) -- (-5.6, -0.6); % Arrow from Merge base to Parent 2
      \draw[->, thick] (-3.2, 0.8) -- (-2.6, 0.4); % Arrow from Parent 1 to Merged
      \draw[->, thick] (-3.2, -0.6) -- (-2.6, -0.2); % Arrow from Parent 2 to Merged
  \end{tikzpicture}
\end{center}
\halfprecaptionspace
\caption{Conflicting changes that line-based merge cleanly, but
  incorrectly, merges.  Most previous evaluations count this as a
  successful merge.}
\label{fig:line-based-merge}
\end{figure}
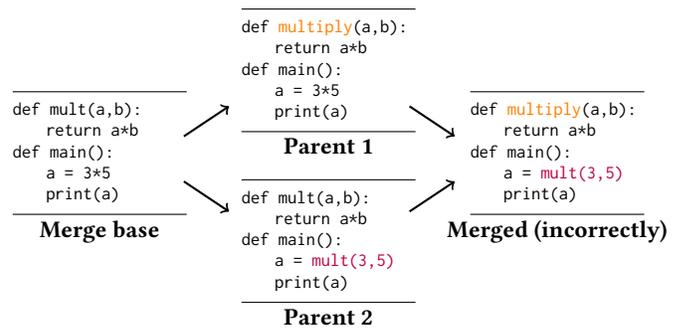

Every merge algorithm suffers from two complementary problems.
(1) It might fail to merge semantically independent changes to the same code
construct (that is, it might leave a conflict for the user to resolve),
% \todo{Can we pick a different term or better define code construct?}
(2) It might incorrectly merge changes in different
constructs that are semantically related.
\Cref{fig:line-based-conflict,fig:line-based-merge} give examples.

\Cref{fig:line-based-conflict} shows changes that line-based Git Merge reports
as a conflict, but a more sophisticated algorithm could  resolve.  Parent 1
renames variable \<n> to \<n\_people>.  Parent 2 changes the value of
\<n>. Since these changes occur on the same line, Git Merge reports a
merge conflict.  In fact, the two changes (rename a variable and change a
value) are semantically independent and
can be performed independently, in either order.  Another example of
an undesirable conflict is when Parent 1 changes the indentation of a line, while
Parent 2 makes a code change to that line.

\Cref{fig:line-based-merge} shows an example of an incorrect merge
that Git Merge would perform.  Parent 1 renames the function
\<mult> to \<multiply>, while Parent 2 adds an invocation of \<mult>. These
changes are on different lines, and so Git Merge integrates them cleanly.
In the merged version, the invocation to \<mult> on line 4 uses the old
name and fails because the program no longer contains a \<mult> routine.

\subsection{Evaluated Tools}
\label{sec:evaluated-tools}

\begin{figure}
\begin{tabular}{l|l}
    Name & Arguments \\
    \hline
    ort & (default) \\
    ort-ignorespace & \<-\relax -ignore-space-change> \\
    recursive-histogram & \<-s recursive -X diff-algorithm=histogram> \\
%    recursive-ignorespace & \<-s recursive -\relax -ignore-space-change> \\
    recursive-minimal & \<-s recursive -X diff-algorithm=minimal> \\
    recursive-myers & \<-s recursive> \\
    recursive-patience & \<-s recursive -X diff-algorithm=patience> \\
    resolve & \<-s resolve> \\
\end{tabular}
\precaptionspace
  \caption{Command-line arguments to \<git merge>.}
  \label{fig:git-merge-command-line-arguments}
\end{figure}

% Our experiments evaluated the following merge tools.

\subsubsection{Previous tools}

% TODO: Could save a line in the following paragraph.

\textbf{Git Merge} is \emph{line-based}: it treats a software artifact as a
sequence of lines and resolves merge conflicts based on textual equivalence of lines.
Line-based merging is the default in most version control systems, including Git, the
most widely used~\cite{stack-overflow-developer-survey:2022}.

Git Merge permits customization of both the alignment and resolution stages.
Git Merge uses the term ``strategy'' (the \<-s> command-line argument) for
the resolution algorithm, and it uses command-line arguments, which vary by
strategy, for the alignment algorithm.
We ran Git Merge in the configurations shown in
\cref{fig:git-merge-command-line-arguments}.

\input{git-merge-variants}

\textbf{Hires merge}~\cite{git-hires-merge}, for ``high-resolution'' merge,
ignores whitespace and automatically resolves non-overlapping changes at
the \emph{character} level, rather than by lines as Git Merge does.
It calls \<git merge-file>, and it does more work
only if that leads to a merge conflict.
The implementation puts every character on a separate
line, merges them using standard tools\todo{using what algorithm?}, and then removes the extra
newline characters.  (This
implementation is reminiscent of a technique for version 
control of trees by translating to a syntax that line-based version control
systems can merge~\cite{AsenovGMO2017}.)
If this
merge strategy fails, a conflict is reported.
% the changes must overlap at the character level, so
% human intervention is necessary. The tool is called git-hires-merge (it
% was previously named git-subline-merge),
% but despite ``git'' in its name, it is not part of the standard Git toolset.

\textbf{IntelliMerge} \cite{ShenZZLJW2019} is \emph{graph-based} and
exclusively targets Java files.  It is a refactoring-aware
merge algorithm.
Refactorings,
such as the rename operations in
\cref{fig:line-based-conflict,fig:line-based-merge}, can often lead to
conflicts or erroneous merges in other merge algorithms.
IntelliMerge converts the base and parent versions of a
source file into program element graphs (PEGs).  A PEG's nodes are
classes, methods, fields, etc.  The edges define relationships, such as a
class implements an interface or a class defines a class
member. IntelliMerge performs a matching or alignment step to detect
refactorings and code edits, merges the PEGs, and then serializes the
merged PEGs back into text files.

% ...In their (IntelliMerge) paper, the evaluation protocol compares the merges of the tool to the merge commit resolution which is pushed by the developer to fix merge conflicts. Since it compares all the conflicts which appear in each merge most of them are resolved by the git merge conflict resolution, which biases the results of the tool in favor of git and are not representative of the correctness of a merge. This is why they obtain a precision of $99.46\%$ for Git Merge.

\textbf{Spork} \cite{LarsenFBM2023} is \emph{tree-based} and exclusively targets Java files.  It
uses a tree-based merge algorithm called
3DM-Merge that operates on ASTs (abstract syntax trees).  It uses GumTree
\cite{FalleriMBMM2014} for its alignment phase. It uses domain
knowledge to resolve specific conflicts during the merge process, such as
the fact that import statements and method declarations can be ordered
arbitrarily.
% It is designed to be used alongside Git as a merge driver so that it can
% handle Java files, while Git handles everything else.
It is the most
starred AST-based merge tool on GitHub.

% ... The evaluation of the tool (Spork) mostly relies on the number of conflicts, which is a poor metric as we have seen before. Qualitative arguments for the merge operated by the tool are based upon a few simple and artificial merges. The issue of incorrect merges is not addressed during evaluation.

% \vspace{-5pt}

\subsubsection{Previously-proposed algorithms}

\textbf{Adjacent} is our implementation of a proposal by
Nguyen et al.~\cite{NguyenI2017}.  It resolves non-overlapping changes on
adjacent lines --- for example, when Parent 1 modifies line $n$ but not
$n+1$ and Parent 2 modifies line $n+1$ but not $n$.
In 51,007 conflicting
merges, programmers resolved
24--85\% of adjacent-line conflicts by applying both
changes, as opposed to taking just one of the changes, neither of the
changes, or making a different change~\cite{NguyenI2017}.  They conclude ``Git should merge
two adjacent-line [sic] instead of considering them as a conflict.''  
We implemented this recommendation in order to evaluate it.\looseness=-1

By contrast to Adjacent, Git Merge reports a conflict if adjacent lines are
changed.  This strategy conservatively avoids some incorrect merges, at the
cost of creating unhandled merges for the programmer to address.
We reviewed the literature that discusses the fact that
Git Merge gives a conflict for
adjacent-line edits.  Every discussion we found treats this as a design
decision.  However, this behavior falls out of Git Merge's
implementation of alignment, which requires a line to be in all three
versions (base, parent1, and parent2) --- no alignment occurs in Git when adjacent
lines are edited.\looseness=-1

\subsubsection{New tools}

\textbf{Imports} resolves conflicts among Java \<import> statements.  Both
IntelliMerge and Spork implement special-case handling of import statements.  Neither
paper~\cite{ShenZZLJW2019,LarsenFBM2023}
mentions this fact nor evaluates how important it is to their results.
% (FSTMerge~\cite{ApelLBLK2011} also special-cases order-independent
% constructs such as \<import>s, \<implements>, and declarations.) 
We
implemented the Imports merge tool in order to answer this important
question.  In other words, the Imports tool enables an \emph{ablation
study} that compares the importance of resolving import statements
to the importance of the non-import-statement parts of a merge tool.

To our surprise, Imports performed \emph{better} than
IntelliMerge and Spork.  This suggests that --- if their \<import> handling
is implemented well --- the other parts of those tools are a net
negative.  This also suggests that tool builders
and researchers should first focus on doing simple things well before
adding complexity that may be unneeded or undesirable.  Unfortunately,
simple but effective approaches are often rejected by the academic
community.

%% TODO: Can save a line by changing "illustrates" to "shows".

Our Imports tool first runs Git Merge.  Then, it re-merges any changes
% not "hunks", because it acts even if Git Merge merged import statements cleanly.
that
involve only import statements.
The Imports tool may re-introduce an import
statement that was removed by either parent, if the imported symbol is used
in the merged code.
We found that simply unioning the \<import> lines in each git conflict
performed much worse.
This illustrates that a merge tool should consider the context
(such as the rest of the file), not just the text of the conflict itself.\looseness=-1
% Refactoring-aware merge tools already consider context.

\textbf{Version-Numbers} first runs Git Merge, then resolves remaining
conflicts among version numbers.  When
version numbers differ in all three versions of a program, and those in the
two edited versions are both larger than that in the base version, it chooses
the largest one.  It requires version numbers to contain at least one
period (``\<.>'').  We implemented this tool based on our observations
(\cref{sec:qualitative}) of common merge
tool failures.  It is useful not just in Java files, but also in buildfiles
such as Maven \<pom.xml> files.

\textbf{\Plumelib} combines the Imports and Version Numbers
functionality.  It shows what is achievable without complex algorithms and
implementations. Some authors of this paper use \Plumelib for our daily work;
we find that it saves us work.

% Omit "https://" from the text to save a line.
The new tools are available at
\url{https://github.com/plume-lib/merging},
which gives instructions for using tham as git merge drivers, git
mergetools, or to clean up conflicts at an arbitrary time..\looseness=-1

\section{Research Questions}

% Our research questions are:

\begin{description}
\item[RQ1] Which Git Merge configurations save the most developer time?
  Are Git's defaults the best for merging?
\item[RQ2] Which merge tools save the most
  developer time?  Do these conclusions differ from those claimed in
  previous work?
\item[RQ3] Are there differences in merge tool performance on
  merges from different sources (main branch, other branches)?
  Is performance on the main branch indicative of
  overall performance?
\item[RQ4] What are the most common causes of merge tool failures
  (conflicts and incorrect merges)?
\end{description}

\section{Measurements}
%talk about assumptions and approach

\subsection{Correctness of Merge Resolutions}\label{sec:merge-outcomes}
A correct merge resolution integrates both parent changes
without introducing any defects that violate the program's
specification.
If a merge tool reports any conflicts, we label the (entire) merge resolution as
\emph{\textbf{unhandled}}.
Otherwise, our methodology uses the project's test suite as a proxy for
correctness~\cite{BrunHEN2011,SeibtHCBA2022}.  Our experimental
infrastructure uses
the tests in the
repository, including any files that have just been merged.
%
% \todo{Future work could evaluate their (partial) correctness, such as by
%   replacing remaining conflict markers with the human-committed merge
%   result.}
%
If the merge is clean and the test suite passes, the merge
resolution is \emph{\textbf{correct}}.
If the merge is clean and compilation or testing fails, the
merge resolution is \emph{\textbf{incorrect}}.  We also use the
\emph{\textbf{incorrect}} label for uncaught exceptions and timeouts.
% \Cref{sec:threats-to-validity} discusses limitations of this methodology.

We reclassified some merges from ``unhandled'' to ``incorrect'', when the
output both a conflict \emph{and} an incorrectly-merged hunk.
For every ``unhandled'' merge, we ran a fixup tool that never makes a
mistake in merging.  If that tool merges all the remaining hunks, but the
resulting code fails its tests, then the original merge must have contained
at least one incorrectly-merged hunk, and our experimental infrastructure
reclassifies it as ``incorrect''.
The fixup tool is \Plumelib.  We manually inspected hundreds of \Plumelib
executions.  In every case, the preceding merge tool was responsible for
the test failure and \Plumelib was not.

% \vspace{-5pt}

\subsection{Merge Tool Performance:  Developer Time}
\label{sec:effort-reduction}

The best merge tool is the one that saves developers the
most time.
Let the average cost to a developer of manually merging a conflict be
\|UnhandledCost|, and likewise for
\|CorrectCost| and \|IncorrectCost|.

Suppose that a developer
encounters 100 merge scenarios, and merge tool $T$ yields 85 correct merges,
10 conflicts, and 5 incorrect merges.  Then the total cost to the developer
is $85 \times \|CorrectCost| + 10 \times \|UnhandledCost| + 5 \times
\|IncorrectCost|$.

We expect that $\|IncorrectCost| \gg \|UnhandledCost| \gg
\|Correct|\-\|Cost| \approx 0$.
Incorrect merges may
require a developer to debug a failure, in order to locate a defect caused by the
incorrect merge, when the tests are next run.
Unhandled merges are easier because they come with
location and diff information.
The
best merge tool will prioritize reducing incorrect merges, while also minimizing the
number of unhandled merges as a second priority.
Like us, \cite{CavalcantiBA2017} speculates that \|IncorrectCost| (which they
call ``false negatives'') is high.
Future work should measure these costs.

Setting \|CorrectCost| to zero simplifies the equation to $\|Cost|(T) =
10 \times \|UnhandledCost| + 5 \times
\|IncorrectCost|$.  Let $k$, the \emph{cost factor} for incorrect merges, be
$\|IncorrectCost| / \|UnhandledCost|(T)$.  This further simplifies the
example to $\|Cost|(T)
% TODO: I can save a line here, by commenting this line of LaTeX.
= 10 \times \|UnhandledCost| + 5 \times k \times \|UnhandledCost|
= (10 + 5 \times k) \times \|UnhandledCost|$.

Generalizing beyond the 100 commits, the developer cost when using tool $T$ is 
$\|Cost|(T) = (\|numUnhandled|(T) + \|numIncor|\-\|rect|(T) \times k) \times \|UnhandledCost|$.
If the developer used no merge tool, the developer cost would be
$\|ManualCost| = \|numMerges| \times \|UnhandledCost|$.
A merge tool $T$ is valuable iff it reduces the developer's effort:
  $\|Cost|(T) < \|ManualCost|$.
The benefit of using tool $T$ is\looseness=-1

% \vspace{-7.5pt}

\begin{equation*}
  \begin{gathered}
  \|EffortReduction|(T)
   =  \frac{\|ManualCost| - \|Cost|(T)}{\|Manual\_Cost|} \\
   =  1 - \frac{\|Cost|(T)}{\|Manual\_Cost|} \\
   =  1 - \frac{\|numUnhandled|(T) + \|numIncorrect|(T) \times k}{\|NumMerges|}
  \end{gathered}
\end{equation*}

% \vspace{-2.5pt}

\noindent
Larger values of \|EffortReduction| are better.
A tool that produces only correct merges gets a score of $1$, while a tool
that only produces unhandled merges (equivalent to a purely manual
resolution strategy) gets a score of $0$. A tool that costs a developer
more than manual resolution gets a negative score.
The score depends on the value of $k$, which future
work should experimentally measure.

\section{Methodology}
\subsection{Data Set}\label{sec:dataset}
\input{results/combined/defs.tex}
\input{results/greatest_hits/defs.tex}

\begin{figure}
\setlength{\tabcolsep}{.9\tabcolsep}
\begin{tabular}{l|r|r|r|r|r|r|r}
    & & & \multicolumn{5}{|c}{Merge size} \\
    & & & $\cap$ & \multicolumn{3}{|c}{total} & \multicolumn{1}{|c}{~} \\
    Phase & Repos & Merges & \#f & \#f & hunks & lines & imp \\
    \hline
    Java repos  & \combinedReposInitial & - & - & - & - & - & - \\
    Head passes & \combinedReposValid & \combinedMergesInitial & 1 & 11 & 28 & 299 & 73\% \\
    Java diff & \combinedReposJavaDiff & \combinedMergesJavaDiff & 3 & 32 & 96 & 1190 & 94\% \\
    Parents pass & \combinedReposJavaDiffAndParentsPass & \combinedMergesJavaDiffAndParentsPass & 3 & 24 & 73 & 823 & 93\% \\
\end{tabular}
  \precaptionspace
  \caption{Repositories and merges at each phase of collection for our data
    sets.  ``$\cap$ \#f'' gives the median number of files changed by both parent 1 \emph{and}
    parent 2.  The ``total'' lines are medians for the changes in 
    parent 1 \emph{or} parent 2.  ``Imp'' is the percentage of merges that
    involve Java import statements.}
  \todo{Add a line just after ``Java repos'' for ``uses Gradle or Maven''.}
  \todo{Add more ``intersection'' data.}
  \label{fig:data-set-attrition}
\end{figure}

We used GitHub repositories from two datasets of high-quality code:
GitHub's Greatest Hits~\cite{GitHubGreatestHits} and
Reaper~\cite{MunaiahKCN2017}.
GitHub's Greatest Hits contains GitHub's 17,000 most popular and
depended-upon repositories, as ranked by a combination of popularity (star
count) and dependency extent.
Reaper is a subset of the GHTorrent~\cite{Gousios2013} dataset.
Reaper consists of only ``engineered software projects \ldots\ that
leverage sound software engineering practices'', as determined by
community, continuous integration, documentation, history, issues, license,
and unit testing.

We selected all Java projects from GitHub's Greatest Hits.
We selected all Java projects from Reaper that have a
\|unit\_test| score (the ratio of test lines of code to source lines of
code) of at least $0.25$ and have at least $10$ stars.
There are \combinedReposInitial unique Java repositories.

We retained repositories that utilize the Maven or Gradle build
automation tools and whose head (latest) commit on the main branch passes its
tests (under JDK 8, 11, or 17) within \combinedParentTestTimeout
minutes.\footnote{We used an Intel i9-13900KF @ 5.8GHz.}
%% \footnote{We used 4 computers.
%% The first computer has 4 Intel
%% Xeon E7-4850 v2 @ 2.30GHz CPUs with a total of 96 cores, the second and
%% third computers have 4 Intel Xeon CPUs E7-8850 v2 @ 2.30GHz with a total of
%% 96 cores, and the fourth machine has an Intel Core i9-10980XE CPU @
%% 3.00GHz with a total of 36 cores.}
This yielded a total of
\combinedReposValid projects containing at least one merge. We collected
\combinedMergesInitial merges from these projects.

We retained only merges that contain a diff within a
\<.java> file, and both of whose parents pass tests within a timeout of
\combinedParentTestTimeout minutes.
This left us with \combinedMergesJavaDiffAndParentsPass merges.

Requiring both parents to pass their tests makes it highly likely that a test failure in the
merged result can be attributed to the merge tool.
This was always the case, in our manual examination of hundreds of merges.

Our methodology depends on running test suites.  If some programs have poor
test coverage, then the fact that all tests passed might just mean that the
tests didn't cover the code that was merged.  To assess this threat
to validity, we used JaCoCo~\cite{JaCoCo} to compute the coverage of 100
randomly-chosen projects. The average code coverage was 53\% and the
25/50/75 percentile values were 28/54/77\%.

\Cref{fig:data-set-attrition} shows how the filtering affected
the set of merges.
%
% Measurements in the ``Java diff'' row are larger than those in the ``Head
% passes'' row.
%
Merge size in the ``parents pass'' row is smaller than
in the ``Java diff'' row.
However, merges where the parents pass tests are still larger than
merges overall (the ``head passes'' row).
The most important metric is the number of files that were edited by both
parents, ``$\cap$ \#f''.  Only these files can be involved in a conflict.
This is unaffected by requiring that the parents pass tests.

\subsection{Flaky tests}

Flaky tests --- those that sometimes pass and sometimes fail --- are
prevalent in software
projects~\cite{LuoHEM2014,RahmanR2018,LeongSPLTM2019,LamSOZEX2020,MudduluruWME2021},
so an experiment that runs
tests must account for them.  We wished to avoid nondeterministic,
unrepeatable results.  However, it is essential to utilize the entire test
suite, all of which provides information to developers.
Therefore, throughout our experiments
we ran every test 5 times, counting the test as a success if any run succeeded.

\def\combinedAverageTriesUntilPass{1.0149383127709237\xspace}
\def\combinedNumberofMergesWith1TriesUntilPass{44569\xspace}
\def\combinedNumberofMergesWith3TriesUntilPass{99\xspace}
\def\combinedNumberofMergesWith5TriesUntilPass{35\xspace}
\def\combinedNumberofMergesWith2TriesUntilPass{256\xspace}
\def\combinedNumberofMergesWith4TriesUntilPass{26\xspace}

There were 44985 merges for which tests passed.  On average, each of these
merges had to be tested 1.01 times before the tests passed.  256 of the
merges first passed on the 2nd test run, and 35 of the merges first
passed on the 5th test run.

\subsection{Merge Inputs}
%% TODO: I can maybe save a line.
We collected merge commits from two sources:
the main branch and other branches.
% , and pull request branches that have been deleted. We are able to detect merge commits since they have more than one parent commit.

The main, or mainline, branch is often named \<main>, \<master>, or \<trunk>.
Other branches include feature branches, release branches, etc.
Some other-branch merges are
also accessible from the main branch, so we removed any duplicates and only
labeled a merge as an other-branch merge if it does not appear as a main
branch merge.

% The second source is pull request branches that have been deleted after a
% pull request is closed.  Pull request branches may be in the repository
% itself (in which case they were feature branches before being deleted) or
% in a fork of the repository.  We collected pull request branches by
% querying the version control hosting service (in our case, GitHub) for the
% commits in every pull request. If a merge found in a pull request does not
% appear on an existing branch, then it must have happened on a deleted
% branch. We refer to these as \emph{pull request merges}.

Main branches are often meant to represent
clean snapshots of a project.  They might encounter different activity and
therefore different merges than other branches, which could contain
sloppier or finer-grained changes that nonetheless need to be merged.
In 2017, among the 19 most popular
Java projects on GitHub, only 13 had any merges in Java
files on the main branch~\cite{AsenovGMO2017}.

% \todo{still unsure if the other papers use mainline only - we might need to
%   look at artifacts to be very sure but the next part assumes they do}

Feature branches are deleted after being merged, so they are not available
from the git repository.  We obtained the information from GitHub, which
retains information about deleted branches.

\subsection{Evaluation of Merge Correctness}\label{sec:evaluation-of-merge-correctness}

% \todo{Add an architectural diagram of our experimental infrastructure.}

% We evaluated each Git Merge configuration described in \cref{sec:evaluated-tools}.
% We then picked the best configurations, and evaluated them against
% the other merge tools.

If a tool produces a clean merge, our experimental infrastructure checks
whether the merge passes its test suite (\cref{sec:merge-outcomes}),
using a timeout of \combinedMergeTestTimeout minutes per run.  This timeout is
larger than the 30 minutes of \cref{sec:dataset} in case both Parent 1 and Parent 2 added tests
to the base version, in which case the merged tests may take longer to run than the tests
on either Parent 1 or Parent 2.

%% TODO: This whole section is boring and could be cut.

\subsection{Tool Implementations}
We wrote wrapper scripts that invoke each merge tool on a set of merge inputs.
\subsubsection{Git Merge Wrapper}
For Git Merge, the wrapper checks out the first parent branch, then invokes
\<git merge> with the second parent branch as an input. Extra inputs are
used for specifying specific strategies.
\subsubsection{Hires-Merge Wrapper}
The wrapper configures Hires-Merge as a Git merge driver,
by writing to the \texttt{.gitattributes} file after cloning a repository.
Then, \<git merge> can be invoked as usual.
\subsubsection{Spork Wrapper}
The wrapper configures Spork as a Git merge driver for Java files,
by writing to the \texttt{.gitattributes} file after cloning a repository.
This makes Git use Spork to resolve conflicts between Java files, and use
its default merge strategy (ort) for all other files.
Then, \<git merge> can be invoked as usual.
\subsubsection{IntelliMerge Wrapper}
IntelliMerge only outputs merge results for Java files.  Our wrapper first runs
IntelliMerge to merge Java files, and stores them in a temporary
location. It then runs Git Merge to generate merge results for all files
in-place. Finally, it overwrites all the Java files with the IntelliMerge
versions. File copying has a negligible effect on run time.

One complication is conflict detection --- IntelliMerge's exit code only
indicates whether the tool completed without exception, rather than whether
the merge was clean as merge tools are expected to do.
% Furthermore, when
% Git Merge (as called by the wrapper) reports a conflict, the merge should
% be classified as Unhandled only if the conflict occurs in a non-Java file,
% but ignored if the conflict was in a Java file.
Therefore, our wrapper
ignores exit codes; it determines whether a merge is successful by searching for
conflict markers 
(e.g., ``\codeid{<\relax <\relax <\relax <\relax <\relax <\relax <}'') that
appear in conflicted files.

\section{Results}
\subsection{Git Merge Configurations (RQ1)}

\begin{figure}
\begin{smaller}
\setlength{\tabcolsep}{.9\tabcolsep}
\input{results/combined/tables/git/table_summary.tex}
\end{smaller}
\precaptionspace
\caption{Performance of different Git Merge configurations.
  \Cref{fig:costplot-git} visualizes this data.}
% (+ 2630 2905 120)
\label{fig:git-results}

\centering
\resizebox{\linewidth}{!}{\input{results/combined/plots/git/cost_without_manual.pgf}}
\precaptionspace
\precaptionspace
\caption{Effort reduction as a function of $k$, the
  relative cost of incorrect merges. This graph visualizes
  the data of \cref{fig:tool-results}.
  The best merge tool is Gitmerge-ort or Gitmerge-ort-ignorespace,
  depending on $k$.
}
% \todo{Benedikt: This will be a bit of a hassle but will make the graph much clearer:
%   merge lines that are identical.  That is, change the ``Gitmerge-ort''
%   label to ``Gitmerge-ort and Gitmerge-recursive-histogram'', and remove
%   the ``Gitmerge-recursive-histogram'' label and line.  Likewise, merge
%   ``Gitmerge-recursive-minimal'' and ``Gitmerge-recursive-myers''.}
\label{fig:costplot-git}
\end{figure}

\Cref{fig:git-results,fig:costplot-git} show the performance of Git Merge
configurations.  The differences are relatively small,
but these small differences are important.
For example, \cref{sec:race-condition-merge} gives an example of an
incorrect merge that compiles and passes some tests, but it contains a race
condition that was not in either parent.
A merge may create a defect that is not detected by the test suite.  The
resulting bug might be detected long afterward (making it more expensive to
fix) or might be deployed to production (which is even more expensive).
Thus, bad merge resolutions can have very significant implications.
\cite{PerrySV2001} statistically justifies special attention to
merge failures.

% \label{sec:auto-crash-analogy}
% By analogy, there are about 0.25 crashes per million vehicle km in the
% EU\@.
% Although these events are rare, they
% are still important to mitigate.
% , and likewise for merge conflicts.

\subsubsection{Resolution Phase}

Git Merge supports three resolution strategies.  Ort, the newest, is tied
for the best, with now fewer correct merges and no more incorrect merges
than any other strategy.  The ort and recursive strategies, behave the
same on this dataset, when both use the Myers alignment algorithm.  This
surprised us, given the hate that the recursive strategy received on
forums.

\subsubsection{Alignment Phase}

The recursive strategy permits selecting a diff algorithm to use for the
alignment phase.
Patience is a popular diff algorithm because it is considered to create the
most human-readable diffs~\cite{Cohen2010}.
Although it has the most correct \emph{and} incorrect merges, it is nearly
indistinguishable from the other alignment algorithms.
The ``minimal'' diff algorithm creates the smallest possible diffs, at the
cost of run time, but it performs identically (on our dataset) to myers, a
simple greedy algorithm.
Differences that are important to software developers when viewing diffs do
not seem to matter much to line-based merge tools.
However, we found that setting git to use the zdiff3 conflict style (rather
than diff3) hindered downstream tools.
% todo: explain
zdiff3 moves lines common to both parents out of a hunk and into the surrounding
common text.  This is considered better for human
inspection because the hunk is shorter, but it gives an incorrect view of the
base text.

\subsubsection{Ignoring Space Changes}
\label{sec:ignorespace}

In our dataset, ignoring
  whitespace decreases unhandled merges by
  % (/ 3962 3750.0)
  5\%, but increases incorrect merges by
  % (/ 3962 3750.0)
  10\% (\cref{sec:ignorespace}).
\Cref{fig:costplot-git,fig:costplot-tools} shows that Git Merge Ignorespace outperforms Git Merge if $k <
5$\todo{Double-check this number.}; Git Merge is better if $k > 5$.

This supports the folk wisdom that spacing conflicts can cause issues for
merge tools, and it supports the inclusion of the \<-\relax
-ignore-space-change> argument to Git.  

The absolute difference between Git Merge and Git Merge Ignorespace is small.
Ignoring whitespace changes could be catastrophic in other
languages, such as Python and YAML, where the amount of white space (indentation) is
semantically significant (see example in \cref{sec:yaml-bad-merge}).

\begin{figure}
    \centering
    \resizebox{\linewidth}{!}{\input{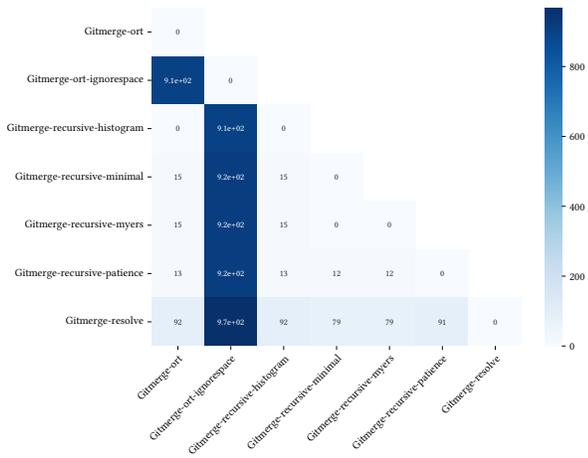}}
    \precaptionspace
    \precaptionspace
    \caption{Number of clean merges that are different
      % (i.e., not syntactically identical)
      between each pair
      of Git Merge configurations.
      Each Git Merge configuration produces around 7000 clean merges.}
    %% These todos still need to be done.
    %% \todo{In all the heatmaps, switch from absolute numbers to percentages
    %%   (with, say, 2 digits of precision).}
    %% \todo{In all the heatmaps, make the text in the cells larger; it is not
    %%   readable currently.}
    %% \todo{In all the heatmaps, don't use exponential notation; instead, use
    %%   ``0.00...'', which is easier to interpret.}
    \label{fig:git-heatmap}
\end{figure}

\subsubsection{Differences in Merge Output}

\Cref{fig:git-heatmap} shows, for each pair of configurations, the number
of times they produced clean merges with \emph{distinct} contents.
Apart from Gitmerge-ort-ignorespace, all the
configurations produce the same output in most situations.
With regard to syntax (\cref{fig:git-heatmap}), Gitmerge-ort-ignorespace is
quite different from other git configurations.
However, with regard to semantics (\cref{fig:git-results}),
Gitmerge-ort-ignorespace is relatively similar to other git configurations.
Perhaps most of the textual differences were formatting differences rather
than semantic changes. 

\medskip

We selected Git Merge and Git Merge Ignorespace to compare with other merge tools
(\cref{sec:results-all-tools}).  Henceforth, just ``Git Merge'' means ort,
which is Git's default configuration.

\begin{figure}
\begin{smaller}
\input{results/combined/tables/tools/table_summary.tex}
\end{smaller}
\precaptionspace
\caption{Performance of merge tools (visualized in \cref{fig:costplot-tools}).}
\label{fig:tool-results}

  \centering
  \resizebox{\linewidth}{!}{\input{results/combined/plots/tools/cost_with_manual.pgf}}
  \precaptionspace
  \precaptionspace
  \caption{Effort reduction as a function of $k$, the
    relative cost of incorrect merges. This graph visualizes
    the data of \cref{fig:tool-results}.}
  \label{fig:costplot-tools}
\end{figure}

\subsection{Comparison Among All Merge Tools (RQ2)}
\label{sec:results-all-tools}

\begin{figure}
    \centering
    \resizebox{\linewidth}{!}{\input{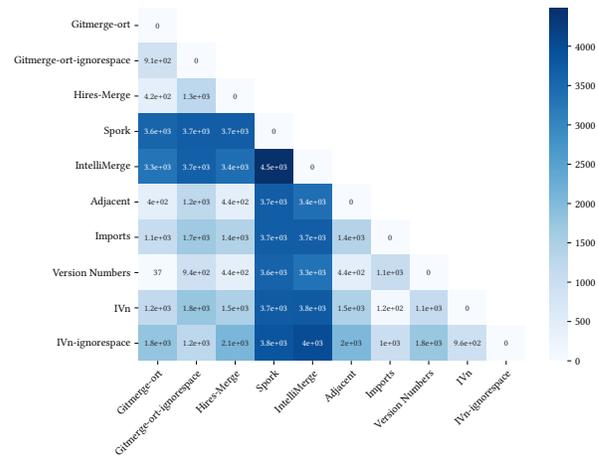}}
    \precaptionspace
    \precaptionspace
    \caption{Number of clean merges that are different (i.e., not syntactically
      identical) between each pair of tools.  Each merge tool
        produces around 7000 clean merges.}
    \label{fig:tool-heatmap}
\end{figure}

\Cref{fig:tool-results,fig:costplot-tools} show the experimental results
among the merge tools.
\Cref{fig:tool-heatmap} shows the number of times each pair of tools
produced a clean merge that differed.
The tree- (``structured'') and graph-based merge tools Spork and
IntelliMerge are the most different from all the other merge tools --- but
the single greatest difference was between Spork and IntelliMerge.

\Plumelib or \Plumelib-ignorespace is the best tool except when $k \sim 1$,
in which case Spork is best.

``Structured'' merge tools (such
as Spork and IntelliMerge) produce both more correct merges and more
incorrect merges~\cite{SeibtHCBA2022}, leading to a relatively steep slope in \cref{fig:costplot-tools}.
Adjacent and Hires Merge have the next lowest number of unhandled merges.

Whether Spork is better than Git Merge (and the other tools) depends on the
relative cost of incorrect merges (see \cref{fig:costplot-git,fig:costplot-tools}).  If an
incorrect merge is no worse than an unhandled merge, then Spork is the best
tool.  If an incorrect merge is at least {\ortSporkIntersection} times as bad as an unhandled
merge, then Spork is the \emph{worst} tool other than IntelliMerge.  If an incorrect merge is
6\todo{Check this number.} times as
bad as an unhandled merge, then using Spork is worse than using no merge
tool at all (that is, manually resolving every merge).

\subsubsection{Comparison to previous results}
The IntelliMerge~\cite{ShenZZLJW2019} paper found a significant difference between itself and Git Merge when measuring hunks and lines:
``Comparing with GitMerge, IntelliMerge reduces the number of conflict
blocks by 58.90\% and the lines of conflicting code by
90.98\%''~\cite{ShenZZLJW2019}. In contrast, our evaluation found
IntelliMerge significantly worse. Here are two possible reasons for this discrepancy.

First, IntelliMerge's data set only includes merges that Git Merge was
unable to merge cleanly \emph{and} where a refactoring change was part of the
conflict. In other words, it was only applied to exactly the merges it is
designed to resolve \emph{and} where its competition failed.
We applied IntelliMerge to 
a more representative set of merges.

Second, the success metric of the IntelliMerge paper is reduction
in the number of conflict lines and hunks.
Perhaps that metric does not correlate with successful merges that lead to
correct behavior.

The IntelliMerge paper acknowledges, as a limitation, that it doesn't evaluate
the tool's impact on Incorrect Merges, which it refers to as \textit{False
  Negative Conflicts}.

% The Spork paper~\cite{LarsenFBM2023} did not compare Spork to IntelliMerge
% (which predates Spork) nor to Git Merge. Therefore, no comparison with its
% results is possible; our research provides the first such comparison.

\subsubsection{Run Time}

\begin{figure}
    \input{results/combined/tables/tools/table_run_time.tex}
    \precaptionspace
    \caption{Merge tool run time.}
    \label{fig:run-time}
\end{figure}

\Cref{fig:run-time} shows the run times of each tool.  Each number is the
median of 3 runs.
Spork and IntelliMerge
most often cause noticeable pauses. The IntelliMerge~\cite{ShenZZLJW2019} paper
reports a median run time of 0.54 seconds, which is very close to our
measurement of 0.50 seconds.  The IntelliMerge paper did not report the
mean, but our data show that it is twice as high.
Adjacent's maximum run time is high because of its use of a 3-way dynamic
programming algorithm.  Perhaps that algorithm could be cut
off earlier without materially affecting the output.

% do we want to talk about run time at all
% talk about agreement numbers
% what is the best tool
% how does this differ from the papers

\subsection{Differences Between Merge Sources (RQ3)}
\label{sec:merge-sources}

\begin{figure}
\input{results/combined/tables/tools/table_feature_main_summary.tex}
\precaptionspace
\caption{Results broken down by merge source: the main branch
  or other branches. Each percentage
  indicates the fraction of merges from that source that yielded that outcome.}
\label{fig:sources}
\end{figure}

% [RQ2 Q1] Are there notable differences in tool performance on merges from different sources (main branch, feature branch, pull request)?

% [RQ2] Are there notable differences in tool performance on merges from different sources (main branch, feature branch, pull request)? Is performance on main branch merges a good indicator of overall performance?
Our dataset contains \combinedMainBranchMerges
(\combinedMainBranchMergesPercent{}\%) main branch merges and
\combinedOtherBranchMerges (\combinedOtherBranchMergesPercent{}\%) other
branch merges.

\Cref{fig:sources} shows merge results by merge source (main branch merges
vs.\ other branch merges).
Previous evaluations only use main branch merges.

Merge tools perform better on main branch merges than on other
branch merges.  An evaluation on only main branch merges is misleading with
respect to absolute performance.  In real-world usage, merge tools will
perform worse than in previous studies.

However, the \emph{relative} performance of tools is similar between main
and other branches.  The differences do cause different rankings depending
on $k$, but primarily for tools whose Effort Reduction values were already similar.
An evaluation on only main branch
merges is therefore an acceptable way to measure \emph{relative} performance, with
a few caveats.

% \section{Discussion}

% could add future work but doesn't belong in conclusion
% Future work on the evaluation procedure can aim to increase the amount of labelled merges and make sure to reduce flaky tests. Another direction of work can aim to analyze the cause of the disappointing performance of Spork and IntelliMerge with respect to Git Merge.

\section{Qualitative assessment (RQ4)}
\label{sec:qualitative}

\input{paper-qualitative.tex}

\section{Threats to Validity}
\label{sec:threats-to-validity}

\textbf{Construct validity.}
Testing is an imperfect proxy for correctness.  If tests fail, then most
likely the merge is incorrect, but if tests pass, the merge might still be
incorrect.  For instance, the merge might be wrong in files that are not
executed by the test suite.
Therefore, our measured
number of incorrect merges understates the problem of incorrect merges.

Though testing is an imperfect proxy, we believe it is
better than the alternatives.  One alternative would be an automated proof
that the merged program is equivalent to what is in the version control
repository or is equivalent to a merge of the
branches~\cite{HorwitzPR89:TOPLAS}, though verification is too expensive
and unscaleable;
furthermore, what is committed to the repository may be wrong, as discussed
in \cref{sec:resolution-in-the-commit-history}.
The testing proxy is also used by the program repair community, who discovered
very serious errors in papers that did not test repairs~\cite{QiLAR2015}.
That community calls a patch that passes a test suite ``plausible'', and
reserves ``correct'' for one that matches the programmer's intent (which
is, in general, unknowable).

The values \|UnhandledCost| and \|IncorrectCost| are averages.  A
particular merge tool might produce better- or worse-than-average conflicts
and incorrect merges.
It seems likely that all tools perform better on merges that a human would
find easy and perform worse on merges that a human would
find hard.

\textbf{Internal validity.}
The difference in performance between merge tools is small.
It is possible that a larger experiment would change the results.
Even small differences are
important because of the high cost of manually reverse-engineering
incorrect merges.

Flaky tests can lead to spurious test failures.  We mitigated this by
running each test suite multiple times.

Our experimental infrastructure ignores merges in which either parent fails
its tests.  It is conceivable that merge tools have different success rates
on such merges.

\textbf{External validity.}
Our experiments are on Java programs, because state-of-the-art merge tools
are designed for Java.  Merge tools (including state-of-the-practice ones)
may behave differently in other contexts, such as for Python and YAML where
whitespace is semantically significant.

Our experiments use 42092 Java projects from two well-known datasets of
high-quality repositories.  It is possible that other repositories might
have different characteristics.

Our paper presents a metric for determining quality of merge tools.  The
metric is for use by researchers and merge tool authors, not by end users
of merge tools.

Our ranking of tools depends on $k$, the \emph{average} relative cost of incorrect
merges.  Our work makes no prediction about any specific merge.
The average for $k$ does not capture differences in developer time within these
groups.  For example, larger diffs may be harder for both tools
and developers to manage, though small
code changes do not get merged faster than large ones~\cite{KudrjavetsNR2022}.
It is possible that merge tools disproportionately succeed on ``easier''
merges; nonetheless, automating these cases does reduce developer effort.
It is possible that a tool ranked higher by our metrics correctly handles
more merges, but simpler ones, than a tool ranked lower by our metrics.
Future work should develop heuristics to reward tools for producing
unhandled merges that are easy to manually resolve, or producing incorrect
merges that are easy to debug.

% Our evaluation relies on automated testing infrastructure, which is subject
% to flaky tests. Flaky tests are tests that might fail arbitrarily and
% usually pass after multiple retries. This is compounded by our use of
% parallelism in our implementation, which runs merge tools and tests on
% multiple merges in parallel. This might cause unwanted interference between
% the tests, causing undesired test failures. Our initial evaluation relied
% on a single run of the test suite, but we observed instances where
% rerunning the test yielded a different outcome. Switching to three test
% runs has improved the reliability of our outputs, but it remains possible
% for a tool's correct resolution to be labeled as incorrect due to these
% factors.

% Our automated testing methodology cannot evaluate individual changes or
% hunks within an unhandled merge.  A non-clean merge that is one conflict
% away from a correct merge is more useful to a developer than one that is 50
% conflicts away.  We believe that our methodology does capture when one
% merge tool is better than another.

\section{Related Work}

\ifanonymous\else
\subsection{Terminology}

A line-based merging technique treats a program as a sequence of atomic lines.
A tree-based merging technique treats a program as an AST (abstract syntax tree).
Some literature on tree merging~\cite{ApelLBLK2011,LessenichAL2014,CavalcantiBA2017} calls tree merging ``structured merging''
and calls line-based merging ``unstructured merging''.
However, line-based merging does
account for some structure (lines), and tree merging fails to account for
some structure (def--use edges, etc.).
``Semistructured merging''~\cite{ApelLBLK2011} treats a program as an AST down to, say,
procedures, and uses text-based merging on the procedure bodies.
\fi

\subsection{Other merge tools}
\label{sec:other-merge-tools}

This section discusses tools that we wished to include in our study, but were
unable to.

JDime~\cite{LessenichAL2014} first runs a faster, less capable
algorithm (line-based merging).  Only when that algorithm fails does it use
a slower, more capable algorithm (tree-based merging).  The paper calls
this staged approach ``auto-tuning''.  JDime
is unsuitable for practical use because it
discards comments (as do some
other tools~\cite{AsenovGMO2017}), discards file headers (as do some other
tools~\cite{ZhuH2018}), and
arbitrarily reorders methods and fields (as do some other tools
\cite{HorwitzPR89:TOPLAS,LessenichAL2014,TavaresBCS2020,LarsenFBM2023}),
and runs very slowly.
Performing a merge with JDime takes ``about 15 min''~\cite{ApelLL2012},
% \footnote{Single-thread
% performance has increased by over 3$\times$ since then.}
which the paper claims ``can be safely neglected.''
A previous evaluation~\cite{LarsenFBM2023}
found that Spork outperforms JDime.
We fixed some bugs in JDime, and
the JDime authors helpfully addressed
a dozen bugs that we reported, but
remaining unfixed bugs prevented us from using it.
In addition, JDime does not handle the full syntax of Java 8 
(released in 2014){\ifanonymous\else{ and
its authors marked our issue about this
as ``won't fix''\todo{cite}}\fi}.\looseness=-1

\todo{Cut or shorten?}
The Spork paper~\cite{LarsenFBM2023} did evaluate against JDime.
Maybe JDime's poor performance in those
experiments is due its bugs.
Or maybe Spork's experiments (containing only 1740 merges)
exercised less of JDime.  For example, Spork's experiments might have
focused on programs written
in Java 8 that do not use all of Java 8's functionality. By contrast, our
experiments used Java code as recent as Java 17, which was released in
2021.

AutoMerge~\cite{ZhuH2018,ZHY2019} (sometimes incorrectly called
``Auto\-Merge\-PTM''~\cite{LarsenFBM2023}) represents the set of all possible
merges via version space algebra.  It ranks all the merge possibilities;
the developer must choose among them.  The AutoMerge tool is not
publicly available.

DeepMerge~\cite{DinellaMSBNL2023} and
MergeBERT~\cite{SvyatkovskiyFGMDBJSL2022} are neural (that is, deep
learning) approaches to merging.  Neither tool is publicly available.

Many merge tools exist that depend on GUI interaction\todo{cite some},
rendering them unsuitable for our experiments, which are fully automated
with no human interaction.  We experimented with these tools and found that
they give the same results as Git Merge; we speculate that they use it
internally.  In other words, their main differentiation from competitors is
their GUI, not their merge algorithm.
RefMerge~\cite{EllisND2023} uses a different merge algorithm (that of
MolhadoRef~\cite{DigNMJ2006}), but it is implemented as an IntelliJ plugin,
so we could not include it in our experiments.
FSTMerge~\cite{ApelLBLK2011} depends on the visual kdiff3 tool.

\subsection{Comparisons of merge tools}
\label{sec:comparisons-of-merge-tools}

\todo{Explicitly relate to the limitations at the beginning of the paper.}

% How much Myers and Histogram diff output was *not* identical.
% (average 4.2 4.1 8.2 4.1 7.4 1.7 5.0 6.0 2.4 5.2 5.6 3.3 3.9 7.0)

Nugroho et al.~\cite{NugrohoHM2020} compared the alignment algorithms
that are built into Git.  (Git calls them ``diff algorithms.'')
Their focus is
on use in academic research.  In 52 papers, all used the default diff
algorithm (at the time, Myers).  They investigated whether use of a
different alignment algorithm might have changed the experimental results.
The Myers and Histogram diff output was identical for over 95\% of
commits.  Our experiments show that the 5\% of differences do not affect merging.

Ellis et al.~\cite{EllisND2023} compared two operation-based
refactoring-aware merge tools (RefMerge and IntelliMerge) against
Git Merge.  They considered only merge scenarios that contain
refactoring-related conflicts.  For 2001 such merge scenarios from 20
open-source Java projects on GitHub, IntelliMerge produced a clean merge
3\% of the time, and RefMerge produced a clean merge 6\% of the time.
They did not run tests to verify the merges.

% Our project started in October 2022.
% Their paper has "date of publication 27 October 2021".

Seibt et al.~\cite{SeibtHCBA2022} compared three merge strategies: unstructured,
structured, and semi-structured.  They used one merge tool:  the JDime
structured merge tool~\cite{LessenichAL2014}, extended to perform
unstructured and semi-structured merge.  They evaluated all 7 (=
$2^3 - 1$) combinations of unstructured, semi-structured, and structured
merging.  They considered 10 repositories compared to our \combinedReposJavaDiffAndParentsPass.
Notably, they used testing as a proxy just as we did.  They reported
\todo{the number of?}merge failures
and test failures; we additionally offer quantitative
guidance regarding which tool programmers should use.  Our work
complements theirs.

\subsection{Weaknesses of previous comparisons}

All previous
evaluations of merge tools suffer from at least one of these
three problems:  they do not evaluate \emph{merge correctness}, they are not
evaluated on \emph{representative merges}, or they do not compare with
\emph{state-of-the-art tools}.

\subsubsection{Merge correctness}

Most evaluations of merge tools assume that every \emph{clean} merge is a
\emph{correct} merge.  They falsely assume that no merge tool ever makes a mistake.
Thus, they ignore the cost of incorrect merges~\cite{CavalcantiBA2017}.
In an industrial case study~\cite{PerrySV2001}, 1\% of clean merges
were incorrect.
% Most of these incorrect merges resulted in compilation errors.

\label{sec:resolution-in-the-commit-history}

Other evaluations assess correctness by comparing the
output of a tool with
the resolution in the commit history~\cite{ShenZZLJW2019,DinellaMSBNL2023,SvyatkovskiyFGMDBJSL2022}.
This is the wrong metric if a developer
uses Git Merge and Git Merge cleanly but incorrectly merges the two
branches.
Such a situation is not uncommon:  in one
study~\cite{BrunHEN2013}, over
% From Figure 4:
% (/ (+ 2 53 7 51 15 5) (+ (+ 2 53 7 51 15 5) 1080 113 102.0))
9\%
of textually clean merges produced by Git Merge
failed to build or failed to pass tests.
In other words, a comparison with the
VCS history will reward Git Merge while penalizing a correct tool.  In one
evaluation~\cite{ShenZZLJW2019}, the authors state that this bias explains
why they computed $99.5\%$ precision for Git Merge.

A third approach is manual labeling\todo{citations}. For example,
\cite{EllisND2023} evaluated 50 merges where IntelliMerge and a
proposed tool produced different resolutions.
Especially when performed by the authors of a merge tool,
this approach can lead to bias or
labeling errors.  It is difficult for a maintainer to evaluate correctness,
much less an outsider who is unfamiliar
with the codebase. A manual approach also scales
poorly.  A small sample suffers the risk that the results may not
generalize to larger samples.

\label{sec:tests-as-oracle}
A few previous studies have used test suites as a proxy for merge
correctness~\cite{BrunHEN2011,SeibtHCBA2022}.  However, this approach has
not caught on; \cite{SvyatkovskiyFGMDBJSL2022} notes it as a rare and
unusual approach.  \cite{CavalcantiBA2017} claims to ``explore build or
test failures'', but their experimental methodology does not run tests.
The program repair community discovered
serious errors in papers that did not test repairs~\cite{QiLAR2015}.

\subsubsection{Representativeness of merges}

Some evaluations use synthetic
merges~\cite{HuntT2002,ApelLBLK2011,LessenichAL2014}, which are merges
between arbitrary commits in the VCS\@.
Synthetic merges may differ from real-world merges.
% \todo{Cite the mutation literature.}

Some evaluations~\cite{ApelLBLK2011,ShenZZLJW2019}
consider only merges that are demonstrative of a
certain scenario. For example, \cite{ShenZZLJW2019} uses only
refactoring-related merge commits when evaluating IntelliMerge, which
performs refactoring detection.  They did not evaluate how well the tool
performs on merges without refactoring, such as whether it mistakenly
detects refactorings in them.

Previous evaluations only collect merge commits that lie on a project's
main branch.  At least, they do not mention considering other branches.
\cite{GZ2014} explicitly excluded non-main-branch merges from consideration.
\cite{TavaresBCS2020} states that there is no reason to
believe that merges not on the main branch differ from those they analyzed.
Our \cref{sec:merge-sources} shows that they do differ.
It is common for projects to maintain a
linear VCS history on their main branch, for example by using
squash-and-merge for pull requests.
The many merges into other branches, such as
pulling the main branch into a feature branch, are qualitatively
different from the few merges that do appear on the main branch.
We also consider
merges into long-lived branches, such as release or development branches.

\subsubsection{Comparison to state-of-the-art tools}

The state-of-the-practice merge tool is Git Merge.
The behavior of Git Merge can be customized by command-line arguments, and
these arguments affect the quality of its merges.  All previous comparisons
against Git Merge have only considered its default configuration, without
considering these command-line arguments.

To the best of our knowledge, the only publicly-available functional
command-line merge
tools for Java are Hires-Merge~\cite{git-hires-merge},
IntelliMerge~\cite{ShenZZLJW2019}, Spork \cite{LarsenFBM2023}, and our new tools~\cite{PlumeLibMerging}.
(\Cref{sec:other-merge-tools} discusses
other
tools such as JDime, AutoMerge, RefMerge, etc.)  Hires-Merge predates
IntelliMerge and Spork, but has not yet been experimentally evaluated.
The Spork
paper adopts IntelliMerge's experimental protocol, but 
excludes IntelliMerge and Git Merge from its evaluation.\footnote{The Spork
paper only evaluates  ``structured merge algorithms'' --- that is, only tree merge
algorithms, not line-based nor graph-based algorithms.  By contrast, they
classify IntelliMerge as ``semi-structured''.  Inconsistently, their
evaluation includes JDime even though JDime contains both an unstructured and
a structured pass.}  Our paper answers the important
research question of how these merge tools perform in a head-to-head
comparison.

\subsection{Other research on merging}

\cite{MenezesTPMPMC2020} found that ``attributes such as the number of
committers, the number of commits, and the number of changed files seem to
have the biggest influence in the occurrence of merge conflicts'', in terms
of statistical correlation.

\cite{GhiottoMBvdH2020} found that programmers resolved most hunks
by choosing
one of the two parent versions.
% \todo{Do we know whether these were trivial
%   merges or not?  That is, might the resolution have been made by a tool
%   rather than the programmer?}
% These correspond to the trivial cases of
% \cref{sec:background-merge-algorithms}, which would be automatically
% resolved by a line-based tool.\looseness=-1

\section{Conclusion}

\todo{Rewrite.  Is this section needed, given \cref{sec:findings-list}?}
Merging allows simultaneous work on multiple tasks during collaborative
software development. Multiple merge tools have been proposed with
promising evaluations. We showed that these evaluations can be misleading
if they fail to include all merges or do not assess merge correctness. To address such
shortcomings, we propose a novel evaluation protocol that includes merges
from deleted branches, uses
a resolution's test suite to identify whether a merge is correct or
incorrect, and quantitatively accounts for incorrect merges.
We created new merge tools that outperform existing ones.
We found that representative merge sources are essential for evaluating the
absolute performance of tools, but are less important in comparing them.
In several cases, our experimental results answer the questions and
update the claims of previous
papers, leading to clearer understanding of the strengths and weaknesses of
merge tools.
We
hope our work contributes to a more principled and experimentally driven approach to the development of new merge tools.\looseness=-1

\paragraph*{Data availability}

Our experimental data are publicly available, along with the programs that
produced and analyzed them. The code to compute all the results can be
found at \url{https://github.com/benedikt-schesch/AST-Merging-Evaluation}
and the computed data can be found at \url{https://zenodo.org/records/13366866}.
The new merging tools are available at
\url{https://github.com/plume-lib/merging}.

\bibliographystyle{plain} % We choose the ACM-Reference-Format reference style, defined in ACM-Reference-Format.bst
\bibliography{refs,plume-bib/bibstring-unabbrev,plume-bib/ernst,plume-bib/soft-eng,plume-bib/testing,plume-bib/version-control,plume-bib/crossrefs}

\end{document}

\title{Evaluation of Version Control Merge Tools (Appendix)}


% \author{Benedikt Schesch}
% \email{scheschb@ethz.ch}
% \affiliation{%
%   \institution{ETH Z\"urich}
%   \country{Switzerland}}
% 
% \author{Ryan Featherman}
% \email{feathr@cs.washington.edu}
% 
% \author{Ben R. Roberts}
% \email{brober3@cs.washington.edu}
% 
% \author{Kenneth J. Yang}
% \email{kjy5@cs.washington.edu}
% 
% \author{Michael D. Ernst}
% \email{mernst@cs.washington.edu}
% 
% % Shared affiliation for last 3 authors.
% \affiliation{%
%   \institution{University of Washington}

\begin{teaserfigure}
{\vspace*{-2em}
\begin{minipage}{.2\columnwidth}
\begin{centering}\LARGE
Benedikt Schesch \\ \large
b.schesch@googlemail.com \\
ETH Z\"urich \\
\end{centering}
\end{minipage}%
\hfill%
\begin{minipage}{.2\columnwidth}
\begin{centering}\LARGE
Ryan Featherman \\ \large
ryan.featherman3@gmail.com\\
Microsoft \\
\end{centering}
\end{minipage}%
\hfill%
\begin{minipage}{.5\columnwidth}
\begin{centering}\LARGE
Kenneth J. Yang \hfill Ben R. Roberts \hfill Michael D. Ernst \\  \large
\{kjy5,brober3,mernst\}@cs.washington.edu \\
University of Washington \\
\end{centering}
\end{minipage}%
\vspace*{1em}} \\
\begin{minipage}{\textwidth}
\strut\hfill\Large September 25, 2024\hfill\strut
\end{minipage}
\vspace*{.5em}
\end{teaserfigure}

\maketitle

This appendix accompanies the ASE 2024 paper ``Evaluation of Version
Control Merge tools''.

% \section{Variants of Git Merge}
% 
% \input{git-merge-variants}

\section{Git terminology}

Git uses the terms ``merge strategy'', ``merge driver'', and ``merge tool''
to mean different parts of the merging process.  These different parts are
implementation details that are not relevant to this paper.  We
use ``merge tool'' more generically; the closest Git equivalent to this
paper's ``merge tool'' is a Git ``merge driver''.

\section{Qualitative comparison of merging algorithms}

Sections~\ref{sec:first-qualitative}--\ref{sec:last-qualitative}
analyze the strengths and weaknesses of various merging algorithms.
Indexes such as ``222-42'' refer to specific merges in our dataset. 

Git Merge was considered the baseline algorithm to which we compared other algorithms. To find strengths we filtered for merge instances where an algorithm passed tests while Git Merge did not. Specifically, we were filtering for when Git Merge did \emph{not} pass tests this would also capture scenarios where Git Merge failed to merge and tests timing out in addition to outright failing tests. Weaknesses of algorithms were found by filtering for instances where Git Merge passed tests and the algorithms did not.

We also considered the "programmer" merge as a ground truth for what the "correct" merge should have been when comparing the behaviors of both Git Merge and the various algorithms to it.
\section{Spork}
\label{sec:first-qualitative}

\subsection{Strengths}

There were 765 instances from our dataset where Spork passed tests post-merge while Git Merge did not.

\subsubsection{Overlapping Unique Additions (35091-165)}

\strut\\ Spork's strategy of parsing code into an AST tree and identifying differences at comparable nodes was quite successful in examples where different branches both added differing methods. These overlapping additions were the bread and butter of Spork's successes.

\begin{lstlisting}[basicstyle=\footnotesize\ttfamily,numbers=none]
<@\leftmarker@>
public void testAsStringColumn() {
  column1.appendCell("1923-10-20T10:15:30");
  column1.appendMissing();
  StringColumn sc = column1.asStringColumn();
  assertEquals("Game date strings", sc.name());
  assertEquals(2, sc.size());
  assertEquals("1923-10-20T10:15:30.000", sc.get(0));
  assertEquals(StringColumnType.missingValueIndicator(), sc.get(1));
<@\basemarker@>
<@\baserightsepmarker@>
public void testFormatter() {
  column1.setPrintFormatter(DateTimeFormatter.ISO_LOCAL_DATE_TIME, "NaT");
  column1.append(LocalDateTime.of(2000, 1, 1, 0, 0));
  column1.appendMissing();
  assertEquals("2000-01-01T00:00:00", column1.getString(0));
  assertEquals("NaT", column1.getString(1));
<@\rightmarker@>
\end{lstlisting}

Git-Merge failed this conflict since the lines that the methods were added overlapped. Spork, however, successfully merged both methods alongside the programmer. The biggest weakness of line-based merging is that it ignores the context of the conflicting file. A programmer's vein of thinking alongside Spork, however, uses this context to realize that both branches were adding non-semantically overlapping changes. The Spork algorithm thrives on clear-cut examples where both branches added non-semantically overlapping changes.

\subsubsection{Resolving Refactorings (38-142)}

\strut\\ Spork does a good job handling variable and function renames. For example, in \<DateTimeColumnType.java>, Git Merge could not resolve the following merge:
\begin{lstlisting}[basicstyle=\footnotesize\ttfamily,numbers=none]
<@\leftmarker@>
<@\texttt{\textcolor{red}{long packedDateTime}}@> = row.getPackedDateTime(newColumn.name());
newDateTime.appendInternal(<@\texttt{\textcolor{red}{packedDateTime}}@>);
<@\basemarker@>
PackedDateTime dateTime = row.getPackedDateTime(newColumn.name());
newDateTime.appendInternal(dateTime.getPackedValue());
<@\baserightsepmarker@>
PackedDateTime dateTime = row.getPackedDateTime(newColumn.name());
<@\texttt{\textcolor{red}{dateTimes}}@>.appendInternal(dateTime.getPackedValue());
<@\rightmarker@>
\end{lstlisting}
In this scenario, \<dateTime>'s type changed and is now used in place of
\<newDateTime>. This change in type, usage, and variable name, is all on
adjacent lines confused Git Merge's line-based approach. However, Spork was able to reason through these changes. Spork merged this like so:
\begin{lstlisting}[basicstyle=\footnotesize\ttfamily,numbers=none]
<@\texttt{\textcolor{red}{long packedDateTime}}@> = row.getPackedDateTime(newColumn.name());
<@\texttt{\textcolor{red}{dateTimes}}@>.appendInternal(<@\texttt{\textcolor{red}{packedDateTime}}@>);
\end{lstlisting}

\subsubsection{Successful Refactoring (158-459)}

\strut\\ Spork often succeeded at implementing small or large refactorings when variables were named differently across the file.

\begin{lstlisting}[basicstyle=\footnotesize\ttfamily,numbers=none]
<@\leftmarker@>
Object [] getExecutionParameters();
Object [] getCurrentContext();
<@\basemarker@>
Object [] getExecutionParameters();
<@\baserightsepmarker@>
Map<String, Pair<Serializable, Boolean>> getExecutionParameters();
<@\rightmarker@>
\end{lstlisting}
Here, getExecutionParamters has its type changed to \codeid{Map<}\-\codeid{String, Pair<Serializable, Boolean>{}>} in only the right branch. Git Merge fails to merge since another line next to it also changes a different variable. Git Merge often fails preemptively when adjacent unrelated lines are involved.
Spork correctly chooses to factor \<getExecutionParameters> to a map alongside the programmer. Overall, Spork was effective at refactoring due to its AST strategy of creating nodes for each part of the code. This way, it could logically identify the same variable node as it was renamed or had its type change and change accordingly. This example was not unique.

\subsection{Weaknesses}

There were 266 instances from our dataset where Spork did not pass tests when Git Merge passed.

Most of the issues found with Spork are due to bugs with its implementation rather than issues with the underlying algorithm.

\mysection{Non-Compilable}

\subsubsection{Spork Sometimes Places Parentheses Incorrectly (345-203)}

\strut\\ For instance, in \<TokenTester.java> the expected merge should have been
\begin{lstlisting}[basicstyle=\footnotesize\ttfamily,numbers=none]
Assertions.assertThat(ex.getMessage().contains(expectedError)<@\texttt{\textcolor{red}{)}}@>;
\end{lstlisting}
however, Spork produced
\begin{lstlisting}[basicstyle=\footnotesize\ttfamily,numbers=none]
Assertions.assertThat(ex.getMessage()<@\texttt{\textcolor{red}{)}}@>.contains(expectedError);
\end{lstlisting}
The Spork-merged code does not compile.
Such errors might be due to bugs in Spork's underlying parser, Spoon.

\subsubsection{Classes With the Same Name (1322-24)}

\strut\\ Spork can confuse classes from different packages with the same names with each other (1322-24). This is potentially a fault with the underlying GumTree matcher where two classes (with the same names) are matched when they should have been unique tokens. For instance, in \<SQSMessageConsumerPrefetch.java>, Git Merge
correctly generated
\begin{lstlisting}[basicstyle=\footnotesize\ttfamily,numbers=none]
protected javax.jms.Message convertToJMSMessage ...
\end{lstlisting}
while Spork returned
\begin{lstlisting}[basicstyle=\footnotesize\ttfamily,numbers=none]
protected <@\texttt{\textcolor{red}{Message}}@> convertToJMSMessage ...
\end{lstlisting}
In Spork's output, \<Message> resolves to \<com.amazonaws.services.sqs\-.model.Message> instead of \<javax.jms.Message>.
This issue also can be observed in 3498-8.

\subsubsection{Omission of Types (1741-4)}

Another common problem was the omission of types, especially in
\<catch> clauses. For instance, in \<BaseProcessor.java> Spork produced
\begin{lstlisting}[basicstyle=\footnotesize\ttfamily,numbers=none]
} catch (<@\texttt{\textcolor{red}{\textvisiblespace}}@>e) {
\end{lstlisting}
which omitted the \<Exception> type. This is non-syntactical Java
code that does not compile.

\subsubsection{Incorrect Placement of Generics (2955-13)}

\strut\\ In \<UnsignedVariableBitLengthType.java>, Git Merge returned
\begin{lstlisting}[basicstyle=\footnotesize\ttfamily,numbers=none]
new UnsignedVariableBitLengthType( img, nBits ) );
\end{lstlisting}
while Spork returned
\begin{lstlisting}[basicstyle=\footnotesize\ttfamily,numbers=none]
new <@\texttt{\textcolor{red}{<nBits>}}@>UnsignedVariableBitLengthType(img));
\end{lstlisting}
The addition of \<<nBits>> was invalid Java code. This behavior is also seen in 26275-370 and 1215-1573.

\subsubsection{Dropping Escape Characters (4595-12)}

\strut\\ Spork occasionally drops escape characters in strings which would result in invalid Java code (and strings). In \<JDBC3Database\-Meta\-Data.java>, Git Merge merged
\begin{lstlisting}[basicstyle=\footnotesize\ttfamily,numbers=none]
return "\"";
\end{lstlisting}
while Spork gave
\begin{lstlisting}[basicstyle=\footnotesize\ttfamily,numbers=none]
return "<@\texttt{\textcolor{red}{"}}@>";
\end{lstlisting}
This is invalid Java code. This pattern also appears in 1215-245.

\subsubsection{Merge Cascaded Conditionals (1885-445)}

\strut\\ For \<StartMojo.java>, Git Merge returned
\begin{lstlisting}[basicstyle=\footnotesize\ttfamily,numbers=none]
} else {
    if (!daemon) {
        log.info("K3PO started (CTRL+C to stop)");
    }
    else {
        log.info("K3PO started");
    }
}
\end{lstlisting}
while Spork returned
\begin{lstlisting}[basicstyle=\footnotesize\ttfamily,numbers=none]
} else if (!daemon) {
    log.info("K3PO started (CTRL+C to stop)");
} else {
    log.info("K3PO started");
}
\end{lstlisting}
While this change happened to be acceptable for this repository, it is a change to the source code. This is not enough to conclude that this is an algorithmic error when handling nested conditionals as opposed to a shortcut taken when converting the AST back to source code, but it is a risky behavior and deviates from the intent of merging code. For reference, the programmer chose the pattern presented by Git Merge.

\mysection{Compilable}
%% Unfortunately, this request is impossible because of "stretchable glue"
%% in TeX.  (Well, not impossible, but I'm not fixing it now.)
% \todo{delete extra line between this header and the next section. I'm not sure how}

\subsubsection{Spork Fails to Include Method Body (2162-99).}

\strut\\ Spork's strategy of parsing code into an AST tree should also theoretically succeed in cases with mixes of changes. In this example, one programmer removes throwing the exception while another changes code inside the method.

\begin{lstlisting}[basicstyle=\footnotesize\ttfamily,numbers=none]
<@\leftmarker@>
public void onPlayerChat(String message) {
    Utill.broadcastMessage("<" + this.getPlayerName() + "> " + message);
<@\basemarker@>
public void onPlayerChat(String message) <@\texttt{\textcolor{red}{throws Exception}}@> {
    Utill.broadcastMessage("<" + this.getPlayerName() + "> " + message);
<@\baserightsepmarker@>
public void onPlayerChat(String message) <@\texttt{\textcolor{red}{throws Exception}}@> {
    if (message.startsWith("/")) {
        final String fullCommand = message.substring(1);
        final String[] split = fullCommand.split(" ");
        final String[] args;
        if (split.length != 1) {
            args = new String[split.length - 1];
            System.arraycopy(split, 1, args, 0, args.length);
        }
<@\rightmarker@>
\end{lstlisting}
The left branch removes the exception and the right changes the inside of
the method. Git Merge fails due to the adjacency of the differences.

\noindent
Spork:
\begin{lstlisting}[basicstyle=\footnotesize\ttfamily,numbers=none]
public void onPlayerChat(String message) {
    Utill.broadcastMessage("<" + this.getPlayerName() + "> " + message);
\end{lstlisting}

\noindent
Programmer:
\begin{lstlisting}[basicstyle=\footnotesize\ttfamily,numbers=none]
public void onPlayerChat(String message) {
    if (message.startsWith("/")) {
        final String fullCommand = message.substring(1);
        final String[] split = fullCommand.split(" ");
        final String[] args;
        if (split.length != 1) {
            args = new String[split.length - 1];
            System.arraycopy(split, 1, args, 0, args.length);
        }
\end{lstlisting}
Spork correctly includes the exception, however it fails to pick the newer code inside the method. This highlights a weakness of the AST strategy: it's much harder to execute in complicated examples.

In general, changing one section of code can create/remove the need to change other parts like the exception. Without compiling the code, any merge algorithm choosing one branch over the other would run the risk of incorrect merges. A thorough programmer looking to avoid future errors might choose to deal with a merge conflict from Git Merge rather than risk an incorrect merge.

\subsubsection{Spork Adds Parentheses (2995-13)}

\strut\\ For example, in \<Unsigned128BitType.java>, Git Merge returns
\begin{lstlisting}[basicstyle=\footnotesize\ttfamily,numbers=none]
this( ( NativeImg< ?, ? extends LongAccess > ) null );
\end{lstlisting}
while Spork produces
\begin{lstlisting}[basicstyle=\footnotesize\ttfamily,numbers=none]
<@\texttt{this(\textcolor{red}{(}(NativeImg<?, ? extends LongAccess>) \textcolor{red}{(}null\textcolor{red}{)}\textcolor{red}{)});}@>
\end{lstlisting}
Spork added parentheses around \<null> and then again around the whole expression. While these changes do not affect the behavior, and such merges are marked as correct by our methodology since they pass code tests, they do make an unnecessary change that a programmer must manually revert.

\subsubsection{Spork Added Extra Semicolons (1741-4)}

\strut\\ In \<BaseProcessor.java>:
\begin{lstlisting}[basicstyle=\footnotesize\ttfamily,numbers=none]
int to = (int) docTrees.getSourcePositions().getEndPosition(pkgTree.getCompilationUnit(), doc, node);<@\texttt{\textcolor{red}{;}}@>
\end{lstlisting}
While this is legal Java code, it is an artifact that should not have been there. This issue is also found in 26642-300 and 12269-213 (including many more). This pattern can also occur with three semicolons like in 12870-226.
\begin{lstlisting}[basicstyle=\footnotesize\ttfamily,numbers=none]
SearchStrategyModule stratModule = new SearchStrategyModule() {
    [...]
};<@\texttt{\textcolor{red}{;;}}@>
\end{lstlisting}

\subsubsection{Spork Swaps Qualifier Positions (4959-12)}

\strut\\ Swaps to \<static> and \<final> occur and are harmless, but non-idiomatic.
In \<JDBC3\-Data\-base\-Meta\-Data.java>, Git Merge returned
\begin{lstlisting}[basicstyle=\footnotesize\ttfamily,numbers=none]
private final static Map<String, Integer> RULE_MAP = new HashMap<String, Integer>();
\end{lstlisting}
while Spork returned
\begin{lstlisting}[basicstyle=\footnotesize\ttfamily,numbers=none]
private <@\texttt{\textcolor{red}{static final}}@> Map<String, Integer> RULE_MAP = new HashMap<String, Integer>();
\end{lstlisting}

\section{IntelliMerge}

\subsection{Strengths}

There were 87 instances from our dataset where IntelliMerge passed tests post-merge while Git Merge did not.

\subsubsection{IntelliMerge Resolves Dense Multi-line Conflicts (8323-1022)}

\strut\\ For example, in \<EntityFactory.java>, Git Merge was unable to
resolve the following merge:

\begin{lstlisting}[basicstyle=\footnotesize\ttfamily,numbers=none]
<@\leftmarker@>
import de.hochschuletrier.gdw.ss14.ecs.components.AnimationComponent;
import de.hochschuletrier.gdw.ss14.ecs.components.BehaviourComponent;
import de.hochschuletrier.gdw.ss14.ecs.components.CameraComponent;
import de.hochschuletrier.gdw.ss14.ecs.components.CatBoxPhysicsComponent;
import de.hochschuletrier.gdw.ss14.ecs.components.CatPhysicsComponent;
import de.hochschuletrier.gdw.ss14.ecs.components.CatPropertyComponent;
import de.hochschuletrier.gdw.ss14.ecs.components.DogPropertyComponent;
import de.hochschuletrier.gdw.ss14.ecs.components.EnemyComponent;
import de.hochschuletrier.gdw.ss14.ecs.components.ExitMapPhysicsComponent;
import de.hochschuletrier.gdw.ss14.ecs.components.ExitMapPropertyComponent;
import de.hochschuletrier.gdw.ss14.ecs.components.InputComponent;
import de.hochschuletrier.gdw.ss14.ecs.components.JumpDataComponent;
import de.hochschuletrier.gdw.ss14.ecs.components.JumpablePhysicsComponent;
import de.hochschuletrier.gdw.ss14.ecs.components.JumpablePropertyComponent;
import de.hochschuletrier.gdw.ss14.ecs.components.LaserPointerComponent;
import de.hochschuletrier.gdw.ss14.ecs.components.LightComponent;
import de.hochschuletrier.gdw.ss14.ecs.components.MovementComponent;
import de.hochschuletrier.gdw.ss14.ecs.components.ParticleEmitterComponent;
import de.hochschuletrier.gdw.ss14.ecs.components.PlayerComponent;
import de.hochschuletrier.gdw.ss14.ecs.components.RenderComponent;
import de.hochschuletrier.gdw.ss14.ecs.components.ShadowComponent;
import de.hochschuletrier.gdw.ss14.ecs.components.WeldJointPhysicsComponent;
import de.hochschuletrier.gdw.ss14.ecs.components.WoolPhysicsComponent;
import de.hochschuletrier.gdw.ss14.ecs.components.WoolPropertyComponent;
<@\basemarker@>
import de.hochschuletrier.gdw.ss14.ecs.components.AnimationComponent;
import de.hochschuletrier.gdw.ss14.ecs.components.BehaviourComponent;
import de.hochschuletrier.gdw.ss14.ecs.components.CameraComponent;
import de.hochschuletrier.gdw.ss14.ecs.components.CatBoxPhysicsComponent;
import de.hochschuletrier.gdw.ss14.ecs.components.CatPhysicsComponent;
import de.hochschuletrier.gdw.ss14.ecs.components.CatPropertyComponent;
import de.hochschuletrier.gdw.ss14.ecs.components.DogPropertyComponent;
import de.hochschuletrier.gdw.ss14.ecs.components.EnemyComponent;
import de.hochschuletrier.gdw.ss14.ecs.components.InputComponent;
import de.hochschuletrier.gdw.ss14.ecs.components.JumpDataComponent;
import de.hochschuletrier.gdw.ss14.ecs.components.JumpablePhysicsComponent;
import de.hochschuletrier.gdw.ss14.ecs.components.JumpablePropertyComponent;
import de.hochschuletrier.gdw.ss14.ecs.components.LaserPointerComponent;
import de.hochschuletrier.gdw.ss14.ecs.components.LightComponent;
import de.hochschuletrier.gdw.ss14.ecs.components.MovementComponent;
import de.hochschuletrier.gdw.ss14.ecs.components.ParticleEmitterComponent;
import de.hochschuletrier.gdw.ss14.ecs.components.PlayerComponent;
import de.hochschuletrier.gdw.ss14.ecs.components.RenderComponent;
import de.hochschuletrier.gdw.ss14.ecs.components.ShadowComponent;
import de.hochschuletrier.gdw.ss14.ecs.components.WeldJointPhysicsComponent;
import de.hochschuletrier.gdw.ss14.ecs.components.WoolPhysicsComponent;
import de.hochschuletrier.gdw.ss14.ecs.components.WoolPropertyComponent;
<@\baserightsepmarker@>
import de.hochschuletrier.gdw.ss14.ecs.components.*;
<@\rightmarker@>
\end{lstlisting}
There were lines from the left and right branches that needed to be merged and lines from the base branch needed to be deleted. IntelliMerge produced:
\begin{lstlisting}[basicstyle=\footnotesize\ttfamily,numbers=none]
import de.hochschuletrier.gdw.ss14.ecs.components.ExitMapPhysicsComponent;
import de.hochschuletrier.gdw.ss14.ecs.components.ExitMapPropertyComponent;
<@\texttt{\textcolor{red}{import de.hochschuletrier.gdw.ss14.ecs.components.*;}}@>
\end{lstlisting}
Here, the right branch replaced all of the imports from the base branch with a \<*> import. The left branch had most of the same imports as well except for \<ExitMapPhysicsComponent> and \<ExitMapPropertyComponent> which persisted into the output. These two were interleaved among the rest of the imports that were removed from the base branch and IntelliMerge was able to handle this change that spanned multiple lines and sections.

\subsection{Weaknesses}

There were 1285 instances from our dataset where IntelliMerge failed to pass tests when Git Merge passed.

\mysection{Non-Compilable}

\subsubsection{IntelliMerge Duplicates Fragments (2967-105)}

\strut\\ Most of IntelliMerge’s erroneous outputs come from duplicated fragments tokens or phrases that are repeated twice. For example, in \<BGP4Update.java>, methods are declared
\begin{lstlisting}[basicstyle=\footnotesize\ttfamily,numbers=none]
public <@\texttt{\textcolor{red}{public}}@> BGP4 ...
\end{lstlisting}
instead of 
\begin{lstlisting}[basicstyle=\footnotesize\ttfamily,numbers=none]
public BGP4 ...
\end{lstlisting}
which results in incorrect Java code.

Exception declarations are also duplicated in \<InterASTEv2LSA.java> of
2967-105.
Git Merge returned
\begin{lstlisting}[basicstyle=\footnotesize\ttfamily,numbers=none]
private void decode() throws MalformedOSPFLSAException{
\end{lstlisting}
IntelliMerge returned
\begin{lstlisting}[basicstyle=\footnotesize\ttfamily,numbers=none]
private void decode() throws MalformedOSPFLSAException <@\texttt{\textcolor{red}{throws MalformedOSPFLSAException}}@>{
\end{lstlisting}
The doubled \<throws> statement is invalid.

Another example of this can be seen in 3880-48.
The exceptions span more than one line and the whole section is duplicated.
Git Merge returned 
\begin{lstlisting}[basicstyle=\footnotesize\ttfamily,numbers=none]
public FrameworkID getFrameworkID() throws InterruptedException, ExecutionException,
      InvalidProtocolBufferException {
\end{lstlisting}
while IntelliMerge returned
\begin{lstlisting}[basicstyle=\footnotesize\ttfamily,numbers=none]
public FrameworkID getFrameworkID() throws InterruptedException, ExecutionException,
  InvalidProtocolBufferException <@\texttt{\textcolor{red}{throws InterruptedException, ExecutionException,
  InvalidProtocolBufferException}}@> {
\end{lstlisting}

In general, IntelliMerge seems to struggle with types and named
elements (8323-1059). For example, in \<Main.java> one line uses
\<GameStates> (the old code) while another line uses \<GameStateEnum> (the
correctly merged code).

IntelliMerge also changed the name of a default constructor from \<UIImage> to \<U>
in \<UIIMage.java> of 8323-890.

\mysection{Compilable}

\subsubsection{IntelliMerge May Remove Docstrings (2967-105)}

\strut\\ While this does not break the code, it is not desirable behavior.
In \<RouterAddressTLV.java>, Git Merge returned
\begin{lstlisting}[basicstyle=\footnotesize\ttfamily,numbers=none]
/**
 * Router Address TLV from RFC 3630
 * 2.4.1. Router Address TLV

The Router Address TLV specifies a stable IP address of the
advertising router that is always reachable if there is any
connectivity to it; this is typically implemented as a "loopback
address".  The key attribute is that the address does not become
unusable if an interface is down.  In other protocols, this is known
as the "router ID," but for obvious reasons this nomenclature is
avoided here.  If a router advertises BGP routes with the BGP next
hop attribute set to the BGP router ID, then the Router Address
SHOULD be the same as the BGP router ID.
If IS-IS is also active in the domain, this address can also be used
to compute the mapping between the OSPF and IS-IS topologies.  For
example, suppose a router R is advertising both IS-IS and OSPF
Traffic Engineering LSAs, and suppose further that some router S is
building a single Traffic Engineering Database (TED) based on both
IS-IS and OSPF TE information.  R may then appear as two separate
nodes in S's TED.  However, if both the IS-IS and OSPF LSAs generated
by R contain the same Router Address, then S can determine that the
IS-IS TE LSA and the OSPF TE LSA from R are indeed from a single
router.

The router address TLV is type 1, has a length of 4, and a value that
is the four octet IP address.  It must appear in exactly one Traffic
Engineering LSA originated by a router.
 */

public class RouterAddressTLV extends OSPFTLV {
\end{lstlisting}
while IntelliMerge removed the docstring and only returned the class declaration:
\begin{lstlisting}[basicstyle=\footnotesize\ttfamily,numbers=none]
public class RouterAddressTLV extends OSPFTLV {
\end{lstlisting}

\subsubsection{IntelliMerge Expands Inline Declarations Into Separate Declarations (8323-890)}

In \<CatMovementSystem.java>, Ort returned
\begin{lstlisting}[basicstyle=\footnotesize\ttfamily,numbers=none]
private float maxVelocity = 0.0f, acceleration = 0.0f, foodBuffer = 0.0f;
\end{lstlisting}
while IntelliMerge returned
\begin{lstlisting}[basicstyle=\footnotesize\ttfamily,numbers=none]
private float maxVelocity = 0.0f;
    
private float acceleration = 0.0f;

private float foodBuffer = 0.0f;
\end{lstlisting}

\section{Hires Merge}

\subsection{Strengths}

There were 291 instances from our dataset where Hires Merge passed tests post-merge while Git Merge did not.

\subsubsection{Handling Refactorings With Multiple Inline Changes (3183-11)}

Overall, the strategy of Hires Merge is to break files into smaller "hunks" until they can be merged. This even includes breaking single lines into their characters and merging parts separately. One strength of this strategy is that it deals with refactorings quite effectively.

\begin{lstlisting}[basicstyle=\footnotesize\ttfamily,numbers=none]
HashSet<Range> ranges = new HashSet<@\texttt{\textcolor{red}{<>}}@>();
<@\basemarker@>
HashSet<Range> ranges = new HashSet<Range>();
<@\baserightsepmarker@>
<@\texttt{\textcolor{red}{Set}}@><Range> ranges = new HashSet<Range>();
<@\rightmarker@>
\end{lstlisting}
As seen in this example, a refactoring occurred where the left branch removes the second generic and the right branch changes the type to a Set. Looking at this example, we can see that the change in the left branch is inconsequential. However, Git Merge gets stuck. It observes all branches differing in one-line errors. However, Hires Merge breaks the line into smaller chunks that can be merged separately. It comes up with a correct merge:

\begin{lstlisting}[basicstyle=\footnotesize\ttfamily,numbers=none]
Set<Range> ranges = new HashSet<>();
\end{lstlisting}

Another example is in \<TriggerCallbackThread.java> of 44055-236 where Git Merge failed to merge and left behind conflict markers:
\begin{lstlisting}[basicstyle=\footnotesize\ttfamily,numbers=none]
<@\leftmarker@>
if(vaidateRetryCount(callbackParamList_bytes)){
    continue;
}
List<HandleCallbackParam> callbackParamList = (List<HandleCallbackParam>) XxlJobExecutor.getSerializer().deserialize(callbackParamList_bytes, HandleCallbackParam.class);
doCallback(callbackParamList, callbaclLogFile);
<@\basemarker@>
List<HandleCallbackParam> callbackParamList = (List<HandleCallbackParam>) XxlJobExecutor.getSerializer().deserialize(callbackParamList_bytes, HandleCallbackParam.class);

callbaclLogFile.delete();
doCallback(callbackParamList);
<@\baserightsepmarker@>
// avoid empty file
if(callbackParamList_bytes == null || callbackParamList_bytes.length < 1){
    callbaclLogFile.delete();
    continue;
}

List<HandleCallbackParam> callbackParamList = (List<HandleCallbackParam>) JdkSerializeTool.deserialize(callbackParamList_bytes, List.class);

callbaclLogFile.delete();
doCallback(callbackParamList);
<@\rightmarker@>
\end{lstlisting}
while Hires made a merge:
\begin{lstlisting}[basicstyle=\footnotesize\ttfamily,numbers=none]
// avoid empty file
if(callbackParamList_bytes == null || callbackParamList_bytes.length < 1){
    callbaclLogFile.delete();
    continue;
}

if(vaidateRetryCount(callbackParamList_bytes)){
    continue;
}
List<HandleCallbackParam> callbackParamList = (List<HandleCallbackParam>) JdkSerializeTool.deserialize(callbackParamList_bytes, List.class);
doCallback(callbackParamList, callbaclLogFile);
\end{lstlisting}
Hires's merge is the closest to what the programmer chose (they fixed the line \<callbacLogFile.delete()>, a typo). Hires was able to identify the two if statements from the left and right branches and merge them, then identified the signature change in \<doCallback> and merged it appropriately.

In \<HelloServer.java> of 1283-65, Git Merge left behind the following conflict markers:
\begin{lstlisting}[basicstyle=\footnotesize\ttfamily,numbers=none]
<@\leftmarker@>
configurePorts(builder -> builder.setPort(port)).
setContextPath(CTX_PATH).
withServices(builder -> builder.addService(new HelloServiceImpl(0), HELLO_SERVICE_PATH)).
configureExtraParams(builder -> builder.setRequestTimeout(50, TimeUnit.MILLISECONDS)).
<@\basemarker@>
configurePorts(builder -> builder.setPort(port)).
setContextPath(CTX_PATH).
withServices(builder -> builder.addService(new HelloServiceImpl(), HELLO_SERVICE_PATH)).
configureExtraParams(builder -> builder.setRequestTimeout(50, TimeUnit.MILLISECONDS)).
<@\baserightsepmarker@>
contextPath(CTX_PATH).
configure(builder -> builder.usePort(port).requestTimeout(50, TimeUnit.MILLISECONDS)).
service(builder -> builder.register(new HelloServiceImpl(), HELLO_SERVICE_PATH)).
<@\rightmarker@>
\end{lstlisting}
while Hires merged it:
\begin{lstlisting}[basicstyle=\footnotesize\ttfamily,numbers=none]
contextPath(CTX_PATH).
configure(builder -> builder.usePort(port).requestTimeout(50, TimeUnit.MILLISECONDS)).
<@\texttt{\textcolor{red}{service}}@>(builder -> builder.register(new HelloServiceImpl(<@\texttt{\textcolor{red}{0}}@>), HELLO_SERVICE_PATH)).
\end{lstlisting}
Here, Hires was able to take the right branch's change from \<withServices> to \<service> while also taking the correct usage of \<HelloServiceImpl> from the left branch, which has a parameter passed in the signature.

\subsubsection{Hires Merge Correctly Identifying Version Numbers (18228-77)}

Hires Merge can sometimes correctly address the version number conflict that frequently appears in \<pom.xml> files. Breaking up each line, Hires Merge checks that changes can be merged using Git Merge at the character level.

\begin{lstlisting}[basicstyle=\footnotesize\ttfamily,numbers=none]
<@\leftmarker@>
<version>2.4.1-SNAPSHOT</version>
<@\basemarker@>
<version>2.3.1-SNAPSHOT</version>
<@\baserightsepmarker@>
<version>2.4.3-SNAPSHOT</version>
<@\rightmarker@>
\end{lstlisting}

Git Merge fails to merge due to all three branches differing. However, Hires Merge determines that
the line can merge since at least 2 corresponding digits are identical
within numbers (2s, 4s, and 1s). However, after this process, it chooses the right branch. Hires Merge appears to choose the right branch over a character-level Git Merge. This, however, works since a character-level merge should output 2.3.3 (an invalid combination of the unique branches between the three). It just so happens that the right branch was correct in this case. This happens again in 11320-2 where Hires Merge deviates from its character level Git Merge to choose the correct left branch instead of an incorrect combination of the two branches.

\subsection{Weaknesses}

There were 12 instances from our dataset where Hires Merge did not pass tests when Git Merge passed.

\mysection{Non-Compilable}

\subsubsection{Hires Merge Can Fail When One of the Three
Branches is Empty in a Three-way Merge}

\strut\\ Even if two of the branches are
identical, when one is empty Hires merge sometimes leaves conflict markers
in the source code (causing an unresolved merge, or merge failure). This behavior can be observed
in 4807-44, 12369-43, 26421-309, 15682-296, 21770-73, 22315-39,
22266-41 with the base being empty and in 15259-251 and 10523-20 with the
right parent being empty. 

\mysection{Compilable}

\subsubsection{Hires Merge Incorrectly Identifying Version Numbers (25267-730)}

Hires merge often fails to get the version number correct when the version numbers differ by several digits.

\begin{lstlisting}[basicstyle=\footnotesize\ttfamily,numbers=none]
<@\leftmarker@>
<version>23.7.0</version>
<@\basemarker@>
<version>23.6.0</version>
<@\baserightsepmarker@>
<version>23.6.1</version>
<@\rightmarker@>
\end{lstlisting}
Hires Merge invented a nonexistent version number:
\begin{lstlisting}[basicstyle=\footnotesize\ttfamily,numbers=none]
<version>23.7.1</version>
\end{lstlisting}
while the programmer chose the more recent
\begin{lstlisting}[basicstyle=\footnotesize\ttfamily,numbers=none]
<version>23.7.0</version>
\end{lstlisting}
The Hires Merge strategy of using Git Merge at the character level chooses 23, 7, and then 1 (choosing the left's unique second option and the right's unique third option). Thus, it makes up a nonexistent number: 23.7.1. Moreover, choosing the right branch when the version numbers differ would also be an incorrect assumption in this case since the left is the most recent. This highlights an issue with the Hires Merge approach: context in the rest of the line matters. Ideally, a tool would address these conflicts by intelligently choosing the highest version number of the three. Until then, however, these types of errors might inhibit the adoption of Hires Merge.

Overall, merging inline is quite good at refactoring Java code that removes/adds generics that are unlikely to semantically change the code. However, this can risk running into problems like such cases where one programmer changes the type while another changes the variable name creating a cascade of issues. Git Merge might fail some inconsequential conflicts, but it avoids the headache of rolling back an incorrect merge.

\section{Git Merge Ignorespace}

\subsection{Strengths}

There were 158 instances from our dataset where Git Merge Ignorespace passed tests post-merge while Git Merge did not.

\subsubsection{Git Merge Can Produce Conflict Markers When Lines Have Mismatched Indentation (1690-69)}

\strut\\ A common trigger is when control flow branches are restructured. In
\<JdbcCreateTableBuilder.java>, Git Merge could not merge the
restructuring and left the following conflict markers:
\begin{lstlisting}[basicstyle=\footnotesize\ttfamily,numbers=none]
<@\leftmarker@>
        final String columnTypeString = queryRewriter.rewriteColumnType(columnType, columnSize);
        sb.append(columnTypeString);
    } else {
        sb.append(nativeType);
        if (columnSize != null) {
            sb.append('(');
            sb.append(columnSize.intValue());
            sb.append(')');
        }
    }
    if (column.isNullable() != null && !column.isNullable().booleanValue()) {
        sb.append(" NOT NULL");
    }
}
<@\basemarker@>
        final String columnTypeString = queryRewriter.rewriteColumnType(columnType);

        sb.append(columnTypeString);
    } else {
        sb.append(nativeType);
    }
    final Integer columnSize = column.getColumnSize();
    if (columnSize != null) {
        sb.append('(');
        sb.append(columnSize.intValue());
        sb.append(')');
    }
    if (column.isNullable() != null && !column.isNullable().booleanValue()) {
        sb.append(" NOT NULL");
    }
}
<@\baserightsepmarker@>
        final String columnTypeString = queryRewriter.rewriteColumnType(columnType);

        sb.append(columnTypeString);
    } else {
        sb.append(nativeType);
    }
    final Integer columnSize = column.getColumnSize();
    if (columnSize != null) {
        sb.append('(');
        sb.append(columnSize.intValue());
        sb.append(')');
    }
    if (column.isNullable() != null && !column.isNullable().booleanValue()) {
        sb.append(" NOT NULL");
    }
}
<@\rightmarker@>
\end{lstlisting}
while Git Merge Ignorespace was able to handle the changes
\begin{lstlisting}[basicstyle=\footnotesize\ttfamily,numbers=none]
	final String columnTypeString = queryRewriter.rewriteColumnType(columnType, columnSize);
        sb.append(columnTypeString);
    } else {
        sb.append(nativeType);
        if (columnSize != null) {
            sb.append('(');
            sb.append(columnSize.intValue());
            sb.append(')');
        }
    }
    if (column.isNullable() != null && !column.isNullable().booleanValue()) {
        sb.append(" NOT NULL");
    }
}
\end{lstlisting}
Notice how in the base and right branches, the middle \<if> block is outside the \<else> block, but in the left branch it is inside the \<else> block. This difference shifts all of those lines by some whitespace and this caused Git Merge to be confused when merging.

\subsubsection{Tab Inconsistency (21205-11)}

\strut\\ One issue Git Merge Ignorespace dealt with effectively was when developers used varying indentation. In code where the left and right indentation differed, most spacing-sensitive merge algorithms struggled to automatically merge functionally identical code.
\begin{lstlisting}[basicstyle=\footnotesize\ttfamily,numbers=none]
<@\leftmarker@>
  private String createSqlStatement(Table table) {
      final IQueryRewriter queryRewriter = getUpdateCallback().getDataContext().getQueryRewriter();
<@\basemarker@>
<@\baserightsepmarker@>
private String createSqlStatement(Table table) {
    final IQueryRewriter queryRewriter = getUpdateCallback().getDataContext().getQueryRewriter();
<@\rightmarker@>
\end{lstlisting}
This conflict represents a common tab width disagreement. One programmer chose to indent their inner createSqlStatement using a 2-space tab while the other did not. As such, Gitmerge fails to merge this conflict while Git Merge Ignorespace chooses to unambiguously take the right branch using the rule of taking the differing branch. This example was not unique (21205-11).

More generally, resolving tab spacing differences with a tool such as Git Merge Ignorespace should be advantageous for accepting semantically identical code without wasting the programmer's time. However, some caution is warranted as there is potential to introduce style inconsistencies without the programmer's knowledge.

\subsubsection{Extra Irrelevant Spaces (2955-73)}

\strut\\ Another issue Git Merge Ignorespace handled effectively was correctly disregarding unnecessary spaces left accidentally in the code. Spacing-sensitive merges, on the other hand, failed to merge these cases.
\begin{lstlisting}[basicstyle=\footnotesize\ttfamily,numbers=none]
<@\leftmarker@>
 * </p>
<@\basemarker@>
 *<@\textcolor{red}{\textvisiblespace}@>
<@\baserightsepmarker@>
 *
<@\rightmarker@>
\end{lstlisting}

The base had an extra space character at the end '* '. This created a problem, however, as the right had no space after the '*'. Therefore, when the left made a change, spacing-sensitive algorithms such as Git Merge incorrectly deduced all 3 branches as being different and failed to merge. Git Merge Ignorespace, however, ignored the irrelevant space and proceeded to merge correctly. The programmer also chose the left branch.

Overall, this type of difference highlights how Git Merge Ignorespace excels at passing merges with inconsequential space inconsistencies.

\subsubsection{Extra Lines (15035-451)}

\strut\\ Git Merge Ignorespace also successfully merged examples with extra lines in the codebase without conflict. This was fairly common when developers introduced extra lines at the EOF or as style changes after code or comment blocks. Also, extra lines can be deemed conflicts due to random spacing within the line.
\begin{lstlisting}[basicstyle=\footnotesize\ttfamily,numbers=none]
<@\leftmarker@>
<@\textcolor{red}{\textbackslash t}@>
<@\basemarker@>

<@\baserightsepmarker@>
<@\rightmarker@>
\end{lstlisting}
The left branch had an extra tab in the empty line, while the right branch deleted the blank line. Git Merge failed since decided all 3 branches differed. Git Merge Ignorespace, however, disregarded the tab and chose the right alongside the developer. From the programmer's perspective, an extra line with spacing within it should be considered identical to one without.

This example with extra lines was very common such as in indexes 21629-4, 44055-283. Git Merge Ignorespace works well with unimportant space inconsistencies. However, this approach could potentially lead to problems in whitespace-sensitive languages such as in Python. Developers might also prefer these to be reported as conflicts so they can manually ensure style consistency.

\subsubsection{Non-Spacing Sensitive Files (206658-5)}

\strut\\ Files such as \<pom.xml> files, \<changelogs>, \<README>s, \<JSON> files, and \<SQL> files are spacing-insensitive. Therefore, version control tools can afford more leniency when automatically merging these files when whitespace is in question.

\begin{lstlisting}[basicstyle=\footnotesize\ttfamily,numbers=none]
<@\leftmarker@>
  <jmock.version>2.8.3</jmock.version>
<@\basemarker@>
<@\baserightsepmarker@>
<jmock.version>2.8.3</jmock.version>
<@\rightmarker@>
\end{lstlisting}
 In this example, a \<jmock.version> property was added to a \<pom.xml> file that was not present in the base. However, the left indented the line 2 spaces while the right did not. This resulted in all three branches differing. This generated conflicts for spacing-sensitive algorithms. Git Merge Ignorespace, however, merged and passed tests by choosing the right version as did the programmer. This example was far from unique as non-spacing sensitive files made up a large number of Git Merge Ignorespace's successes (21559-418, 24859-106). Moreover, Git Merge Ignorespace also thrived when there were non-ambiguous spacing differences in comments (23370-83, 23581-136).

This single advantage Git Merge Ignorespace had in non-spacing-sensitive diffs likely contributed the most to its advantage over tools such as Git Merge. While trivial to manually resolve, many programmers might feel that the time saved by ignoring spaces outweighs the risks.

\subsection{Weaknesses}

There were 23 instances from our dataset where Hires Merge did not pass tests when Git Merge passed.

\mysection{Non-Compilable}

Unsurprisingly, most of Git Merge Ignorespace's differences came from indentation or
whitespace. These were usually harmless indentation errors that did not affect
the correctness of Java source code, however, there was \textit{one}
instance from the pool of non-test-passing repos where Git Merge Ignorespace retained an extra space in the middle of a
string which may impact the program. In \<QueryBuilderTest.java> of 17939-970, an extra space appeared in a
database query when it should have been removed. When Git Merge produced 
% Mike changed the actual text "query = " to "query=", to improve a line break.
\begin{lstlisting}[basicstyle=\footnotesize\ttfamily,numbers=none]
query="DELETE FROM foo USING TIMESTAMP 1240003134 WHERE k='value';";
\end{lstlisting}
Git Merge Ignorespace produced the following:
\begin{lstlisting}[basicstyle=\footnotesize\ttfamily,numbers=none]
query="DELETE <@\textcolor{red}{\textvisiblespace}@>FROM foo USING TIMESTAMP 1240003134 WHERE k='value';";
\end{lstlisting}
This instance happened to not change the code's intended behavior since database queries are whitespace friendly, however, it is generally not the case that extra spaces can be left in strings.

A more common example of when whitespace mattered was when merging YAML files (essentially files where indentation mattered). In \<swagger-spring.yaml> of 14378-60, Git Merge made the merge
\begin{lstlisting}[basicstyle=\footnotesize\ttfamily,numbers=none]
properties:
  id:
    type: "integer"
    format: "int64"
    xml:
      namespace: "http://com.wordnik/sample/model/category"
  name:
    type: "string"
    xml:
      namespace: "http://com.wordnik/sample/model/category"
xml:
  name: "Category"
  namespace: "http://com.wordnik/sample/model/category"
\end{lstlisting}
while Git Merge Ignorespace did
\begin{lstlisting}[basicstyle=\footnotesize\ttfamily,numbers=none]
properties:
  id:
    type: "integer"
    format: "int64"
    xml:
      namespace: "http://com.wordnik/sample/model/category"
  name:
    type: "string"
<@\texttt{\textcolor{red}{xml:}}@>
      namespace: "http://com.wordnik/sample/model/category"
xml:
  name: "Category"
  namespace: "http://com.wordnik/sample/model/category"
\end{lstlisting}
Since YAML is sensitive to indentation, the incorrect source file was
produced. Instead of adding \<xml> as a property of \<name> (that is, indented under \<name>),
Git Merge Ignorespace made \<xml> a sibling of \<properties>.

\mysection{Merge Failures}

Aside from whitespace differences, Git Merge Ignorespace sometimes
yields a conflict due to creating larger hunks when merging. For
instance, in \<Vimeo.java> of 3490-3, the function \<apiRequest> had a
change to its signature, and a couple of functions (\<getProxy> and
\<setProxy>) directly above it were moved. Git Merge Ignorespace incorrectly made
a single hunk containing \<getProxy>, \<setProxy>, and the changes in \<apiRequest>;
this caused a conflict since those functions had moved away as part of a
change independent from the \<apiRequest> changes.

A related situation was when Git Merge Ignorespace failed to merge two separate pieces of code that were not present in the base commit. For instance in \<CustomerRepository.java> of 26496-259, Git Merge merged
\begin{lstlisting}[basicstyle=\footnotesize\ttfamily,numbers=none]
public void removeOne(int id) {  
 this.dslContext.delete(Customer.CUSTOMER)  
    .where(Customer.CUSTOMER.ID.eq(id))  
    .execute();  
}  
  
public void removeGt(int id) {  
 this.dslContext.delete(Customer.CUSTOMER)  
    .where(Customer.CUSTOMER.ID.gt(id))  
    .execute();  
}  
  
public void modify(int id, String name, String email) {  
 this.dslContext.update(Customer.CUSTOMER)  
 .set(Customer.CUSTOMER.NAME, Customer.CUSTOMER.EMAIL)  
    .where(Customer.CUSTOMER.ID.eq(id))  
    .execute();  
}  
  
@Transactional  
public void save(Integer id, String name, String email) {  
  this.dslContext.insertInto(Customer.CUSTOMER)  
    .columns(Customer.CUSTOMER.ID, Customer.CUSTOMER.NAME, Customer.CUSTOMER.EMAIL)  
    .values(id, name, email).execute();  
}
\end{lstlisting}
while Git Merge Ignorespace had the same source code but within conflict markers
\begin{lstlisting}[basicstyle=\footnotesize\ttfamily,numbers=none]
<@\leftmarker@>
public void removeOne(int id) {  
  this.dslContext.delete(Customer.CUSTOMER)  
  .where(Customer.CUSTOMER.ID.eq(id))  
  .execute();  
}  public void removeGt(int id) {  
  this.dslContext.delete(Customer.CUSTOMER)  
  .where(Customer.CUSTOMER.ID.gt(id))  
  .execute();  
}  
public void modify(int id, String name, String email) {  
  this.dslContext.update(Customer.CUSTOMER)  
  .set(Customer.CUSTOMER.NAME, Customer.CUSTOMER.EMAIL)  
  .where(Customer.CUSTOMER.ID.eq(id))  
  .execute();  
<@\basemarker@>
<@\baserightsepmarker@>
@Transactional  
public void save(Integer id, String name, String email) {  
this.dslContext.insertInto(Customer.CUSTOMER)  
  .columns(Customer.CUSTOMER.ID, Customer.CUSTOMER.NAME, Customer.CUSTOMER.EMAIL)  
  .values(id, name, email).execute();  
<@\rightmarker@>
\end{lstlisting}
Git Merge was able to accept these independent additions while Git Merge Ignorespace
produced a conflict. This pattern occurred in 21770-251,
26522-672, and 36198-190.

\section{Adjacent}

\subsection{Strengths}

There were 264 instances from our dataset where Adjacent passed tests post-merge while Git Merge did not.

\subsubsection{Success on Unrelated Adjacent Lines (31280-110)}

\strut\\ Adjacent's strategy is to attempt to merge adjacent in-line conflicts using a Git-esque approach for each line. This met with success, particularly when changes were functionally unrelated.
\begin{lstlisting}[basicstyle=\footnotesize\ttfamily,numbers=none]
<@\leftmarker@>
public static File inputStreamToFile(InputStream ins, String name) throws Exception {
   File file = new File(System.getProperty("java.io.tmpdir") + name);
<@\basemarker@>
public static File inputStreamToFile(InputStream ins, String name) throws Exception{
   File file = new File(System.getProperty("java.io.tmpdir") + name);
<@\baserightsepmarker@>
public static File inputStreamToFile(InputStream ins, String name) throws Exception{
   File file = new File(System.getProperty("java.io.tmpdir") + File.separator + name);
<@\rightmarker@>
\end{lstlisting}
In this example, the first change was that an extra space was added between the method header and the exception in the left branch. However, the right branch changed the line inside the method.

Both the programmer and the adjacent algorithm decided on the final version including the extra space and the revised line. These unambiguous adjacent line changes were not unique.

\subsubsection{Refactoring on Adjacent Lines (1215-3280)}

\strut\\ Adjacent also successfully passed cases resolving refactoring particularly when variables were independent.
\begin{lstlisting}[basicstyle=\footnotesize\ttfamily,numbers=none]
<@\leftmarker@>
String comments = SourcesHelper.readerToString(reader);
CompilationUnit cu = new <@\texttt{\textcolor{red}{JavaParser().setSource}}@>(comments).parse();
<@\basemarker@>
String comments = SourcesHelper.readerToString(reader);
CompilationUnit cu = new InstanceJavaParser(comments).parse();
<@\baserightsepmarker@>
String comments = <@\texttt{\textcolor{red}{readerToString}}@>(reader);
CompilationUnit cu = new InstanceJavaParser(comments).parse();
<@\rightmarker@>
\end{lstlisting}

In this example, the right branch refactors the comments string, while the left branch refactors cu. Git Merge fails to merge due to its policy of not merging adjacent lines that conflict across multiple branches. However, adjacent merges successfully merge, and this is justified due to the lines being unrelated:
\begin{lstlisting}[basicstyle=\footnotesize\ttfamily,numbers=none]
String comments = readerToString(reader);
CompilationUnit cu = new JavaParser().setSource(comments).parse();
\end{lstlisting}
This is also the same output that the programmer chose.

\subsection{Weaknesses}

There were 5 instances from our dataset where Adjacent did not pass tests when Git Merge passed.

\mysection{Compilable}

\subsubsection{Adjacent Lines are Interdependent (5184-31)}

\strut\\ Perhaps the most critical flaw in the adjacent strategy as a whole is the idea that different branches would invalidate each other's changes. In this example, the left branch replaces the local \<dnsEntry> field with a shared \<dnsEntry> list. This also creates the need to synchronize the entire block as written in the left branch.
\begin{lstlisting}[basicstyle=\footnotesize\ttfamily,numbers=none]
<@\leftmarker@>
<@\texttt{\textcolor{red}{synchronized (cacheMap) \{}}@>
    List<DNSEntry> entryList = <@\texttt{\textcolor{red}{cacheMap}}@>.get(dnsEntry.getKey());
    if (entryList != null) {    
        entryList.remove(dnsEntry);
<@\basemarker@>
List<DNSEntry> entryList = this.get(dnsEntry.getKey());
if (entryList != null) {
    synchronized (entryList) {
        entryList.remove(dnsEntry);
<@\baserightsepmarker@>
List<DNSEntry> entryList = this.get(dnsEntry.getKey());
if (entryList != null) {
    synchronized (entryList) {
        <@\texttt{\textcolor{red}{result =}}@> entryList.remove(dnsEntry);
<@\rightmarker@>
    }
}
/* Remove from DNS cache when no records remain with this key */
if (result && entryList.isEmpty()) {
    this.remove(dnsEntry.getKey());
\end{lstlisting}
However, this also necessitates moving the outer if-statement inside the synchronized block.

Adjacent passes the following code readily:
\begin{lstlisting}[basicstyle=\footnotesize\ttfamily,numbers=none]
synchronized (cacheMap) {
    List<DNSEntry> entryList = cacheMap.get(dnsEntry.getKey());
    if (entryList != null) {
        result = entryList.remove(dnsEntry);
    }
}
/* Remove from DNS cache when no records remain with this key */
if (result && entryList.isEmpty()) {
    this.remove(dnsEntry.getKey());
}
\end{lstlisting}
while the programmer prevents the parallelism error with:
\begin{lstlisting}[basicstyle=\footnotesize\ttfamily,numbers=none]
synchronized (cacheMap) {
    List<DNSEntry> entryList = cacheMap.get(dnsEntry.getKey());
    if (entryList != null) {
        result = entryList.remove(dnsEntry);
    }
    /* Remove from DNS cache when no records remain with this key */
    if (result && entryList.isEmpty()) {
        cacheMap.remove(dnsEntry.getKey());
    }
\end{lstlisting}
Git Merge failing was an ideal scenario. This way, the programmer had a chance to address the issue before it made it into production.

While this example might be uncommon, such an incorrect merge would be disastrous as the programmer might be forced to search hours for the tiny if-statement causing dirty writes. Even worse, this Heisenbug might persist indefinitely, popping in and out of existence in rare multi-threaded circumstances.

More generally, a tool like adjacent might save time merging in the short run. However, bugs like this could end up negating the benefit. This illustrates a more general weakness in the adjacent strategy: the context of adjacent lines matters.

\subsubsection{Duplicating Adjacent Lines of Code (34607-547)}

\strut\\ One challenge with algorithms merging adjacent lines is that it can be difficult to tell whether a changed line was added or modified compared to the other branch.
\begin{lstlisting}[basicstyle=\footnotesize\ttfamily,numbers=none]
<@\leftmarker@>
private final int[][] outgoingEdges;
private final int[][] incomingEdges;
private final E[] edgeCache;
<@\basemarker@>
private int[][] outgoingEdges;
private int[][] incomingEdges;
<@\baserightsepmarker@>
private int[][] outgoingEdges;
private int[][] incomingEdges;
private E[] edgeCache;
<@\rightmarker@>
\end{lstlisting}
In this example, the final keyword was added to all three variables in the left branch. Further, the left and right branches initialized a variable \<edgeCache>. This introduced a challenge for the adjacent tool as it duplicated the new variable:
\begin{lstlisting}[basicstyle=\footnotesize\ttfamily,numbers=none]
private final int[][] outgoingEdges;
private final int[][] incomingEdges;
private final E[] edgeCache;
private E[] edgeCache;
\end{lstlisting}
However, the programmer only included \<edgeCache> once:
\begin{lstlisting}[basicstyle=\footnotesize\ttfamily,numbers=none]
private final int[][] outgoingEdges;
private final int[][] incomingEdges;
private final E[] edgeCache;
\end{lstlisting}

Perhaps one of the biggest challenges for an adjacent tool is identifying whether adjacent similar lines are both new or just modified versions of the same line. Even checking lines for similarity would necessitate some arbitrary cutoff. How different does a line have to be before it is separate? A thorough programmer would probably rather avoid the error and make the merge manually. This example was not unique (25892-14, 4807-43).

% ------------------------------
% \bibliographystyle{plain} % We choose the ACM-Reference-Format reference style, defined in ACM-Reference-Format.bst
% \bibliography{refs,plume-bib/bibstring-unabbrev,plume-bib/ernst,plume-bib/version-control,plume-bib/crossrefs}

\section{Version Numbers}

The Version Numbers tool starts with the output of Git Merge.  In only one
case, 9364-16, did it underperform Git Merge.  Git Merge yielded a
conflict, so was classified as ``unhandled''.  Version Numbers resolved the
conflict in exactly the same way the programmer did, but the tests failed.
(One branch added a failing test case to a disabled test suite, and the
other branch enabled the test suite.)  The programmer committed a fix 6
minutes after performing the merge.

\section{Imports}

\label{sec:last-qualitative}

The Imports tool starts with the output of Git Merge.  It never introduces mistakes and often corrects mistakes.
However, our Effort Reduction sometimes ranks it lower.
In one example, Git Merge made a mistake for one hunk (a clean but incorrect merge), because the left and right branches introduced the same function in different parts of the program.  However, Git Merge left, as a conflict, a hunk in the import statements of a different file.  So, Git Merge's merge was classified as ``unhandled'' (because some hunks were unhandled) rather than ``incorrect''.  The Imports tool was able to resolve the import conflict, and it did nothing with the file that had an incorrect merge within its code.  Since there were no merge conflicts remaining, the Imports tool was classified as a clean merge and the tests were run.  However, the tests failed due to the incorrect merge, so the Imports tool's merge was classified as ``incorrect''.  This is worse than the ``Unhandled'' classification of Git Merge, even though any programmer would prefer the Imports merge.  This illustrates a weakness of our proxy for correctness.

%% file: macros.tex
%% Comment or uncomment this line.
\def\notodocomments{}

\newcommand{\todo}[1]{{\color{red}\bfseries [[#1]]}}
% Don't show todo commands if the \notodocomments macro is defined.
\ifdefined\notodocomments
  \renewcommand{\todo}[1]{\relax}
\fi

% Use like: \ifanonymous{ANONYMOUS TEXT}\else{NON-ANONNYMOUS TEXT}\fi
% where the "\else{NON-ANONNYMOUS TEXT}" may be omitted.
\newif\ifanonymous
%% Comment or uncomment this line
\anonymoustrue

\newcommand{\anonurl}[1]{\ifanonymous URL removed for anonymity.\else\url{#1}\fi}
\newcommand{\footnoteanonurl}[1]{\footnote{\anonurl{#1}}}
\newcommand{\Plumelib}{\ifanonymous IVn\else Plume-lib\fi\xspace}

% \|name| or \mathid{name} denotes identifiers and slots in formulas
\def\|#1|{\mathid{#1}}
\newcommand{\mathid}[1]{\ensuremath{\mathit{#1}}}
% \<name> or \codeid{name} denotes computer code identifiers
\def\<#1>{\codeid{#1}}
\protected\def\codeid#1{\ifmmode{\mbox{\sf{#1}}}\else{\sf #1}\fi}
% Alternate definitions for \codeid, using typewriter font instead of sans-serif.
% \protected\def\codeid#1{\ifmmode{\mbox{\ttfamily{#1}}}\else{\ttfamily #1}\fi}
% \protected\def\codeid#1{\ifmmode{\mbox{\smaller\ttfamily{#1}}}\else{\smaller\ttfamily #1}\fi}

% Reduce indentation in lists.
\setlength{\leftmargini}{.5\leftmargini}
\setlength{\leftmarginii}{.5\leftmarginii}
\setlength{\leftmarginiii}{.5\leftmarginiii}

% TODO: why not just \paragraph?
\newcommand{\mysection}[1]{\vspace{5pt}\noindent{\normalsize\textbf{#1}}}

\newcommand{\precaptionspace}{\vspace{-5pt}}
\newcommand{\halfprecaptionspace}{\vspace{-2.5pt}}
% \newcommand{\precaptionspace}{\vspace{-10pt}}
% \newcommand{\halfprecaptionspace}{\vspace{-5pt}}

% Determined by visual inspection. :(
\newcommand{\ortSporkIntersection}{\ensuremath{\mathord{\sim 2}}}

\newcommand{\leftmarker}{\texttt{\textcolor{blue}{<\relax<\relax<\relax<\relax<\relax<\relax< LEFT}}}
\newcommand{\basemarker}{\texttt{\textcolor{blue}{||||||| BASE}}}
\newcommand{\baserightsepmarker}{\texttt{\textcolor{blue}{=======}}}
\newcommand{\rightmarker}{\texttt{\textcolor{blue}{>\relax>\relax>\relax>\relax>\relax>\relax> RIGHT}}}

\addtolength{\textfloatsep}{-.35\textfloatsep}
\addtolength{\dbltextfloatsep}{-.35\dbltextfloatsep}
\addtolength{\floatsep}{-.35\floatsep}
\addtolength{\dblfloatsep}{-.35\dblfloatsep}

%% file: findings-list.tex
\label{sec:findings-list}

Accounting for the \textbf{cost of incorrect merges} changes results,
compared to previous experiments.  The best merge tool depends on the
relative cost $k$ of incorrect merges to unhandled merges
(\cref{fig:costplot-tools}).  Spork is the best merge tool if incorrect
merges cost no more than unhandled merges ($k=1$).  But by $k =
\ortSporkIntersection$, Spork is the \emph{worst} tool other than IntelliMerge.
Future research should determine the value of $k$ and should develop
heuristics to reward tools for producing unhandled merges that are easy to
manually resolve, or producing incorrect merges that are easy to debug.

Some \textbf{previous proposals} underperformed their claims.
IntelliMerge~\cite{ShenZZLJW2019} is rarely applicable.  When it is
applicable, it produces more incorrect merges than Git Merge does.  Merging
edits on adjacent lines, as recommended by \cite{NguyenI2017}, improves
performance if $k< 6$ and degrades performance if $k> 6$.

\textbf{Non-main branches} have more challenging merges than main branches
do.  In order to reflect expected real-world performance, evaluations
should include both types of merges.

A diff algorithm's \textbf{readability} does not materially affect its
success for merging.  Handling whitespace is important, however:  about 1/8
of conflicts are due to whitespace alone (\cref{fig:git-heatmap}).

Current merge tools work conflict-by-conflict (except when calling
out to a refactoring-detection tool).
Accounting for the \textbf{context} of a conflict
(e.g., other code in the same file)
leads to better merges.

A merge algorithm that merely augments Git's handling of \textbf{import
  statements} usually outperforms Spork, IntelliMerge, and other tools.
This suggests that finding simple solutions to common problems is a more effective approach
to building a merge tool than complex algorithms that handle relatively
uncommon cases.  The research community should reward the former as well as
the latter.

% \vspace{-5pt}

% LocalWords:  Spork IntelliMerge Git's myers Merge's underperformed

%% file: git-merge-variants.tex
Git's resolution algorithms (in Git parlance, the strategies)
handle criss-cross merges differently.  A criss-cross merge is one
in which there is no unique base commit --- that is, the parents have no
unique nearest common ancestor.  Without loss of generality, assume there
are two nearest common ancestors, B1 and B2.

\begin{itemize}
\item
  resolve: arbitrarily chooses B1 or B2 as the base.  It produces a
  confusing conflict when one parent renames a file and the other changes
  the file.
\item
  % Sources: http://blog.plasticscm.com/2011/09/merge-recursive-strategy.html
  recursive:  recursively merges B1 and B2 to create a new commit B.  B
  is not added to the repository, but it is used as the base for the merge of
  parent 1 and parent 2.  In addition, the recursive strategy handles file
  renaming.
\item
  ort:  a reimplementation of the recursive strategy,
  with the same
  concepts and high-level algorithm.
  % ; ``ort'' stands for ``ostensibly recursive's twin''.
  %% Ort improves correctness by fixing some long-standing
  %% bugs that were hard to correct in the recursive strategy's
  %% implementation.
  %% Ort improves performance by reading less data, by
  %% caching and re-using computations, and via use of heuristics (such as to
  %% detect file renaming).
  Ort improves correctness, performance, and code structure, compared to
  the recursive strategy.
  Ort was added to Git in 2021 and became Git's
  default merge strategy in 2023.
\end{itemize}

%% TODO: I can save a line here.

An alignment algorithm (in Git parlance, a diff algorithm) tries to find
longest common subsequences between parent 1 and parent 2; the \emph{changes}
are what appear between those common subsequences.  Git's recursive resolution
algorithm permits customizing the diff algorithm.  The options are:
\begin{itemize}
\item myers: a basic greedy algorithm.  It is the default.
\item minimal: a slower algorithm that tries to produce diffs that are as
  small as possible.
\item patience: an algorithm designed to improve readability and avoid
  spurious matches.  It focuses on low-frequency high-content lines, and
  looks for longest matches that include them.  Then, the patience diff is
  recursively invoked on the unmatched text that is before and after the
  match.
\item histogram: similar to patience, but constructs a histogram of element
  occurrences to use with a heuristic when unique common elements cannot be used
  to match up a sequence.
\end{itemize}
% \vspace{-5pt}

\noindent
The ort strategy always uses the histogram diff algorithm.
% , and this cannot be overridden by the user.

% LocalWords:  Git's criss B1 B2 ort recursive's myers

%% file: results/combined/defs.tex
\def\combinedReposInitial{42092\xspace}
\def\combinedReposValid{4072\xspace}
\def\combinedMergesInitial{294714\xspace}
\def\combinedMergesPer{100\xspace}
\def\combinedMergesNonTrivial{85979\xspace}
\def\combinedReposNonTrivial{2712\xspace}
\def\combinedMergesJavaDiff{21860\xspace}
\def\combinedReposJavaDiff{1653\xspace}
\def\combinedMergesJavaDiffAndParentsPass{6045\xspace}
\def\combinedReposJavaDiffAndParentsPass{1120\xspace}
\def\combinedReposSampled{1120\xspace}
\def\combinedMergesSampled{6035\xspace}
\def\combinedReposYieldedFull{1\xspace}
\def\combinedReposTotal{1116\xspace}
\def\combinedMergesTotal{5983\xspace}

% Results
\def\combinedAverageTriesUntilPass{1.0138133934137565\xspace}
\def\combinedNumberofMergesWith1TriesUntilPass{90637\xspace}
\def\combinedNumberofMergesWith3TriesUntilPass{167\xspace}
\def\combinedNumberofMergesWith5TriesUntilPass{70\xspace}
\def\combinedNumberofMergesWith2TriesUntilPass{514\xspace}
\def\combinedNumberofMergesWith4TriesUntilPass{45\xspace}
\def\combinedSporkOverOrtCorrect{512\xspace}
\def\combinedSporkOverOrtIncorrect{486\xspace}
\def\combinedMainBranchMerges{3524\xspace}
\def\combinedMainBranchMergesPercent{59\xspace}
\def\combinedOtherBranchMerges{2459\xspace}
\def\combinedOtherBranchMergesPercent{41\xspace}
\def\combinedReposJava{42092\xspace}

% Timeout
\def\combinedParentTestTimeout{30\xspace}
\def\combinedMergeTestTimeout{45\xspace}

%% file: results/greatest_hits/defs.tex
\def\greatestHitsReposInitial{16816\xspace}
\def\greatestHitsReposValid{157\xspace}
\def\greatestHitsMergesInitial{111707\xspace}
\def\greatestHitsMergesPer{100\xspace}
\def\greatestHitsMergesNonTrivial{18257\xspace}
\def\greatestHitsReposNonTrivial{152\xspace}
\def\greatestHitsMergesJavaDiff{3706\xspace}
\def\greatestHitsReposJavaDiff{131\xspace}
\def\greatestHitsMergesJavaDiffAndParentsPass{1169\xspace}
\def\greatestHitsReposJavaDiffAndParentsPass{106\xspace}
\def\greatestHitsReposSampled{106\xspace}
\def\greatestHitsMergesSampled{1169\xspace}
\def\greatestHitsReposYieldedFull{0\xspace}
\def\greatestHitsReposTotal{104\xspace}
\def\greatestHitsMergesTotal{1149\xspace}

% Results
\def\greatestHitsAverageTriesUntilPass{1.0011848341232228\xspace}
\def\greatestHitsNumberofMergesWith1TriesUntilPass{16026\xspace}
\def\greatestHitsNumberofMergesWith2TriesUntilPass{5\xspace}
\def\greatestHitsNumberofMergesWith4TriesUntilPass{4\xspace}
\def\greatestHitsNumberofMergesWith3TriesUntilPass{1\xspace}
\def\greatestHitsSporkOverOrtCorrect{96\xspace}
\def\greatestHitsSporkOverOrtIncorrect{84\xspace}
\def\greatestHitsMainBranchMerges{523\xspace}
\def\greatestHitsMainBranchMergesPercent{46\xspace}
\def\greatestHitsOtherBranchMerges{626\xspace}
\def\greatestHitsOtherBranchMergesPercent{54\xspace}
\def\greatestHitsReposJava{16816\xspace}

% Timeout
\def\greatestHitsParentTestTimeout{30\xspace}
\def\greatestHitsMergeTestTimeout{45\xspace}

%% file: results/combined/tables/git/table_summary.tex
% Do not edit.  This file is automatically generated.
\begin{tabular}{l|c c|c c|c c|}
        Tool & \multicolumn{6}{c|}{Merges} \\ \cline{2-7}
        & \multicolumn{2}{c|}{Correct} &
        \multicolumn{2}{c|}{Unhandled} &
        \multicolumn{2}{c|}{Incorrect} \\
        & \# & \% & \# & \% & \# & \% \\
        \hline
Gitmerge-ort                     &  2748 &  46\% &  3078 &  51\% &   157 &   3\% \\
Gitmerge-ort-ignorespace         &  2889 &  48\% &  2905 &  49\% &   189 &   3\% \\
Gitmerge-recursive-histogram     &  2748 &  46\% &  3078 &  51\% &   157 &   3\% \\
Gitmerge-recursive-minimal       &  2748 &  46\% &  3078 &  51\% &   157 &   3\% \\
Gitmerge-recursive-myers         &  2748 &  46\% &  3078 &  51\% &   157 &   3\% \\
Gitmerge-recursive-patience      &  2751 &  46\% &  3074 &  51\% &   158 &   3\% \\
Gitmerge-resolve                 &  2703 &  45\% &  3124 &  52\% &   156 &   3\% \\
\end{tabular}

%% file: results/combined/plots/git/cost_without_manual.pgf
%% Creator: Matplotlib, PGF backend
%%
%% To include the figure in your LaTeX document, write
%%   \input{<filename>.pgf}
%%
%% Make sure the required packages are loaded in your preamble
%%   \usepackage{pgf}
%%
%% Also ensure that all the required font packages are loaded; for instance,
%% the lmodern package is sometimes necessary when using math font.
%%   \usepackage{lmodern}
%%
%% Figures using additional raster images can only be included by \input if
%% they are in the same directory as the main LaTeX file. For loading figures
%% from other directories you can use the `import` package
%% and then include the figures with
%%   \import{<path to file>}{<filename>.pgf}
%%
%% Matplotlib used the following preamble
%%   \def\mathdefault#1{#1}
%%   \everymath=\expandafter{\the\everymath\displaystyle}
%%
%%   \makeatletter\@ifpackageloaded{underscore}{}{\usepackage[strings]{underscore}}\makeatother
%%
\begingroup%
\makeatletter%
\begin{pgfpicture}%
\pgfpathrectangle{\pgfpointorigin}{\pgfqpoint{6.400000in}{4.800000in}}%
\pgfusepath{use as bounding box, clip}%
\begin{pgfscope}%
\pgfsetbuttcap%
\pgfsetmiterjoin%
\definecolor{currentfill}{rgb}{1.000000,1.000000,1.000000}%
\pgfsetfillcolor{currentfill}%
\pgfsetlinewidth{0.000000pt}%
\definecolor{currentstroke}{rgb}{1.000000,1.000000,1.000000}%
\pgfsetstrokecolor{currentstroke}%
\pgfsetdash{}{0pt}%
\pgfpathmoveto{\pgfqpoint{0.000000in}{0.000000in}}%
\pgfpathlineto{\pgfqpoint{6.400000in}{0.000000in}}%
\pgfpathlineto{\pgfqpoint{6.400000in}{4.800000in}}%
\pgfpathlineto{\pgfqpoint{0.000000in}{4.800000in}}%
\pgfpathlineto{\pgfqpoint{0.000000in}{0.000000in}}%
\pgfpathclose%
\pgfusepath{fill}%
\end{pgfscope}%
\begin{pgfscope}%
\pgfsetbuttcap%
\pgfsetmiterjoin%
\definecolor{currentfill}{rgb}{1.000000,1.000000,1.000000}%
\pgfsetfillcolor{currentfill}%
\pgfsetlinewidth{0.000000pt}%
\definecolor{currentstroke}{rgb}{0.000000,0.000000,0.000000}%
\pgfsetstrokecolor{currentstroke}%
\pgfsetstrokeopacity{0.000000}%
\pgfsetdash{}{0pt}%
\pgfpathmoveto{\pgfqpoint{0.673149in}{0.549691in}}%
\pgfpathlineto{\pgfqpoint{6.250000in}{0.549691in}}%
\pgfpathlineto{\pgfqpoint{6.250000in}{4.601775in}}%
\pgfpathlineto{\pgfqpoint{0.673149in}{4.601775in}}%
\pgfpathlineto{\pgfqpoint{0.673149in}{0.549691in}}%
\pgfpathclose%
\pgfusepath{fill}%
\end{pgfscope}%
\begin{pgfscope}%
\pgfsetbuttcap%
\pgfsetroundjoin%
\definecolor{currentfill}{rgb}{0.000000,0.000000,0.000000}%
\pgfsetfillcolor{currentfill}%
\pgfsetlinewidth{0.803000pt}%
\definecolor{currentstroke}{rgb}{0.000000,0.000000,0.000000}%
\pgfsetstrokecolor{currentstroke}%
\pgfsetdash{}{0pt}%
\pgfsys@defobject{currentmarker}{\pgfqpoint{0.000000in}{-0.048611in}}{\pgfqpoint{0.000000in}{0.000000in}}{%
\pgfpathmoveto{\pgfqpoint{0.000000in}{0.000000in}}%
\pgfpathlineto{\pgfqpoint{0.000000in}{-0.048611in}}%
\pgfusepath{stroke,fill}%
}%
\begin{pgfscope}%
\pgfsys@transformshift{0.673149in}{0.549691in}%
\pgfsys@useobject{currentmarker}{}%
\end{pgfscope}%
\end{pgfscope}%
\begin{pgfscope}%
\definecolor{textcolor}{rgb}{0.000000,0.000000,0.000000}%
\pgfsetstrokecolor{textcolor}%
\pgfsetfillcolor{textcolor}%
\pgftext[x=0.673149in,y=0.452469in,,top]{\color{textcolor}{\rmfamily\fontsize{10.000000}{12.000000}\selectfont\catcode`\^=\active\def^{\ifmmode\sp\else\^{}\fi}\catcode`\%=\active\def%{\%}$\mathdefault{0}$}}%
\end{pgfscope}%
\begin{pgfscope}%
\pgfsetbuttcap%
\pgfsetroundjoin%
\definecolor{currentfill}{rgb}{0.000000,0.000000,0.000000}%
\pgfsetfillcolor{currentfill}%
\pgfsetlinewidth{0.803000pt}%
\definecolor{currentstroke}{rgb}{0.000000,0.000000,0.000000}%
\pgfsetstrokecolor{currentstroke}%
\pgfsetdash{}{0pt}%
\pgfsys@defobject{currentmarker}{\pgfqpoint{0.000000in}{-0.048611in}}{\pgfqpoint{0.000000in}{0.000000in}}{%
\pgfpathmoveto{\pgfqpoint{0.000000in}{0.000000in}}%
\pgfpathlineto{\pgfqpoint{0.000000in}{-0.048611in}}%
\pgfusepath{stroke,fill}%
}%
\begin{pgfscope}%
\pgfsys@transformshift{1.565445in}{0.549691in}%
\pgfsys@useobject{currentmarker}{}%
\end{pgfscope}%
\end{pgfscope}%
\begin{pgfscope}%
\definecolor{textcolor}{rgb}{0.000000,0.000000,0.000000}%
\pgfsetstrokecolor{textcolor}%
\pgfsetfillcolor{textcolor}%
\pgftext[x=1.565445in,y=0.452469in,,top]{\color{textcolor}{\rmfamily\fontsize{10.000000}{12.000000}\selectfont\catcode`\^=\active\def^{\ifmmode\sp\else\^{}\fi}\catcode`\%=\active\def%{\%}$\mathdefault{2}$}}%
\end{pgfscope}%
\begin{pgfscope}%
\pgfsetbuttcap%
\pgfsetroundjoin%
\definecolor{currentfill}{rgb}{0.000000,0.000000,0.000000}%
\pgfsetfillcolor{currentfill}%
\pgfsetlinewidth{0.803000pt}%
\definecolor{currentstroke}{rgb}{0.000000,0.000000,0.000000}%
\pgfsetstrokecolor{currentstroke}%
\pgfsetdash{}{0pt}%
\pgfsys@defobject{currentmarker}{\pgfqpoint{0.000000in}{-0.048611in}}{\pgfqpoint{0.000000in}{0.000000in}}{%
\pgfpathmoveto{\pgfqpoint{0.000000in}{0.000000in}}%
\pgfpathlineto{\pgfqpoint{0.000000in}{-0.048611in}}%
\pgfusepath{stroke,fill}%
}%
\begin{pgfscope}%
\pgfsys@transformshift{2.457741in}{0.549691in}%
\pgfsys@useobject{currentmarker}{}%
\end{pgfscope}%
\end{pgfscope}%
\begin{pgfscope}%
\definecolor{textcolor}{rgb}{0.000000,0.000000,0.000000}%
\pgfsetstrokecolor{textcolor}%
\pgfsetfillcolor{textcolor}%
\pgftext[x=2.457741in,y=0.452469in,,top]{\color{textcolor}{\rmfamily\fontsize{10.000000}{12.000000}\selectfont\catcode`\^=\active\def^{\ifmmode\sp\else\^{}\fi}\catcode`\%=\active\def%{\%}$\mathdefault{4}$}}%
\end{pgfscope}%
\begin{pgfscope}%
\pgfsetbuttcap%
\pgfsetroundjoin%
\definecolor{currentfill}{rgb}{0.000000,0.000000,0.000000}%
\pgfsetfillcolor{currentfill}%
\pgfsetlinewidth{0.803000pt}%
\definecolor{currentstroke}{rgb}{0.000000,0.000000,0.000000}%
\pgfsetstrokecolor{currentstroke}%
\pgfsetdash{}{0pt}%
\pgfsys@defobject{currentmarker}{\pgfqpoint{0.000000in}{-0.048611in}}{\pgfqpoint{0.000000in}{0.000000in}}{%
\pgfpathmoveto{\pgfqpoint{0.000000in}{0.000000in}}%
\pgfpathlineto{\pgfqpoint{0.000000in}{-0.048611in}}%
\pgfusepath{stroke,fill}%
}%
\begin{pgfscope}%
\pgfsys@transformshift{3.350037in}{0.549691in}%
\pgfsys@useobject{currentmarker}{}%
\end{pgfscope}%
\end{pgfscope}%
\begin{pgfscope}%
\definecolor{textcolor}{rgb}{0.000000,0.000000,0.000000}%
\pgfsetstrokecolor{textcolor}%
\pgfsetfillcolor{textcolor}%
\pgftext[x=3.350037in,y=0.452469in,,top]{\color{textcolor}{\rmfamily\fontsize{10.000000}{12.000000}\selectfont\catcode`\^=\active\def^{\ifmmode\sp\else\^{}\fi}\catcode`\%=\active\def%{\%}$\mathdefault{6}$}}%
\end{pgfscope}%
\begin{pgfscope}%
\pgfsetbuttcap%
\pgfsetroundjoin%
\definecolor{currentfill}{rgb}{0.000000,0.000000,0.000000}%
\pgfsetfillcolor{currentfill}%
\pgfsetlinewidth{0.803000pt}%
\definecolor{currentstroke}{rgb}{0.000000,0.000000,0.000000}%
\pgfsetstrokecolor{currentstroke}%
\pgfsetdash{}{0pt}%
\pgfsys@defobject{currentmarker}{\pgfqpoint{0.000000in}{-0.048611in}}{\pgfqpoint{0.000000in}{0.000000in}}{%
\pgfpathmoveto{\pgfqpoint{0.000000in}{0.000000in}}%
\pgfpathlineto{\pgfqpoint{0.000000in}{-0.048611in}}%
\pgfusepath{stroke,fill}%
}%
\begin{pgfscope}%
\pgfsys@transformshift{4.242334in}{0.549691in}%
\pgfsys@useobject{currentmarker}{}%
\end{pgfscope}%
\end{pgfscope}%
\begin{pgfscope}%
\definecolor{textcolor}{rgb}{0.000000,0.000000,0.000000}%
\pgfsetstrokecolor{textcolor}%
\pgfsetfillcolor{textcolor}%
\pgftext[x=4.242334in,y=0.452469in,,top]{\color{textcolor}{\rmfamily\fontsize{10.000000}{12.000000}\selectfont\catcode`\^=\active\def^{\ifmmode\sp\else\^{}\fi}\catcode`\%=\active\def%{\%}$\mathdefault{8}$}}%
\end{pgfscope}%
\begin{pgfscope}%
\pgfsetbuttcap%
\pgfsetroundjoin%
\definecolor{currentfill}{rgb}{0.000000,0.000000,0.000000}%
\pgfsetfillcolor{currentfill}%
\pgfsetlinewidth{0.803000pt}%
\definecolor{currentstroke}{rgb}{0.000000,0.000000,0.000000}%
\pgfsetstrokecolor{currentstroke}%
\pgfsetdash{}{0pt}%
\pgfsys@defobject{currentmarker}{\pgfqpoint{0.000000in}{-0.048611in}}{\pgfqpoint{0.000000in}{0.000000in}}{%
\pgfpathmoveto{\pgfqpoint{0.000000in}{0.000000in}}%
\pgfpathlineto{\pgfqpoint{0.000000in}{-0.048611in}}%
\pgfusepath{stroke,fill}%
}%
\begin{pgfscope}%
\pgfsys@transformshift{5.134630in}{0.549691in}%
\pgfsys@useobject{currentmarker}{}%
\end{pgfscope}%
\end{pgfscope}%
\begin{pgfscope}%
\definecolor{textcolor}{rgb}{0.000000,0.000000,0.000000}%
\pgfsetstrokecolor{textcolor}%
\pgfsetfillcolor{textcolor}%
\pgftext[x=5.134630in,y=0.452469in,,top]{\color{textcolor}{\rmfamily\fontsize{10.000000}{12.000000}\selectfont\catcode`\^=\active\def^{\ifmmode\sp\else\^{}\fi}\catcode`\%=\active\def%{\%}$\mathdefault{10}$}}%
\end{pgfscope}%
\begin{pgfscope}%
\pgfsetbuttcap%
\pgfsetroundjoin%
\definecolor{currentfill}{rgb}{0.000000,0.000000,0.000000}%
\pgfsetfillcolor{currentfill}%
\pgfsetlinewidth{0.803000pt}%
\definecolor{currentstroke}{rgb}{0.000000,0.000000,0.000000}%
\pgfsetstrokecolor{currentstroke}%
\pgfsetdash{}{0pt}%
\pgfsys@defobject{currentmarker}{\pgfqpoint{0.000000in}{-0.048611in}}{\pgfqpoint{0.000000in}{0.000000in}}{%
\pgfpathmoveto{\pgfqpoint{0.000000in}{0.000000in}}%
\pgfpathlineto{\pgfqpoint{0.000000in}{-0.048611in}}%
\pgfusepath{stroke,fill}%
}%
\begin{pgfscope}%
\pgfsys@transformshift{6.026926in}{0.549691in}%
\pgfsys@useobject{currentmarker}{}%
\end{pgfscope}%
\end{pgfscope}%
\begin{pgfscope}%
\definecolor{textcolor}{rgb}{0.000000,0.000000,0.000000}%
\pgfsetstrokecolor{textcolor}%
\pgfsetfillcolor{textcolor}%
\pgftext[x=6.026926in,y=0.452469in,,top]{\color{textcolor}{\rmfamily\fontsize{10.000000}{12.000000}\selectfont\catcode`\^=\active\def^{\ifmmode\sp\else\^{}\fi}\catcode`\%=\active\def%{\%}$\mathdefault{12}$}}%
\end{pgfscope}%
\begin{pgfscope}%
\definecolor{textcolor}{rgb}{0.000000,0.000000,0.000000}%
\pgfsetstrokecolor{textcolor}%
\pgfsetfillcolor{textcolor}%
\pgftext[x=3.461574in,y=0.273457in,,top]{\color{textcolor}{\rmfamily\fontsize{10.000000}{12.000000}\selectfont\catcode`\^=\active\def^{\ifmmode\sp\else\^{}\fi}\catcode`\%=\active\def%{\%}Incorrect merges cost factor $k$}}%
\end{pgfscope}%
\begin{pgfscope}%
\pgfsetbuttcap%
\pgfsetroundjoin%
\definecolor{currentfill}{rgb}{0.000000,0.000000,0.000000}%
\pgfsetfillcolor{currentfill}%
\pgfsetlinewidth{0.803000pt}%
\definecolor{currentstroke}{rgb}{0.000000,0.000000,0.000000}%
\pgfsetstrokecolor{currentstroke}%
\pgfsetdash{}{0pt}%
\pgfsys@defobject{currentmarker}{\pgfqpoint{-0.048611in}{0.000000in}}{\pgfqpoint{-0.000000in}{0.000000in}}{%
\pgfpathmoveto{\pgfqpoint{-0.000000in}{0.000000in}}%
\pgfpathlineto{\pgfqpoint{-0.048611in}{0.000000in}}%
\pgfusepath{stroke,fill}%
}%
\begin{pgfscope}%
\pgfsys@transformshift{0.673149in}{0.549691in}%
\pgfsys@useobject{currentmarker}{}%
\end{pgfscope}%
\end{pgfscope}%
\begin{pgfscope}%
\definecolor{textcolor}{rgb}{0.000000,0.000000,0.000000}%
\pgfsetstrokecolor{textcolor}%
\pgfsetfillcolor{textcolor}%
\pgftext[x=0.329012in, y=0.501466in, left, base]{\color{textcolor}{\rmfamily\fontsize{10.000000}{12.000000}\selectfont\catcode`\^=\active\def^{\ifmmode\sp\else\^{}\fi}\catcode`\%=\active\def%{\%}$\mathdefault{0.20}$}}%
\end{pgfscope}%
\begin{pgfscope}%
\pgfsetbuttcap%
\pgfsetroundjoin%
\definecolor{currentfill}{rgb}{0.000000,0.000000,0.000000}%
\pgfsetfillcolor{currentfill}%
\pgfsetlinewidth{0.803000pt}%
\definecolor{currentstroke}{rgb}{0.000000,0.000000,0.000000}%
\pgfsetstrokecolor{currentstroke}%
\pgfsetdash{}{0pt}%
\pgfsys@defobject{currentmarker}{\pgfqpoint{-0.048611in}{0.000000in}}{\pgfqpoint{-0.000000in}{0.000000in}}{%
\pgfpathmoveto{\pgfqpoint{-0.000000in}{0.000000in}}%
\pgfpathlineto{\pgfqpoint{-0.048611in}{0.000000in}}%
\pgfusepath{stroke,fill}%
}%
\begin{pgfscope}%
\pgfsys@transformshift{0.673149in}{1.225038in}%
\pgfsys@useobject{currentmarker}{}%
\end{pgfscope}%
\end{pgfscope}%
\begin{pgfscope}%
\definecolor{textcolor}{rgb}{0.000000,0.000000,0.000000}%
\pgfsetstrokecolor{textcolor}%
\pgfsetfillcolor{textcolor}%
\pgftext[x=0.329012in, y=1.176813in, left, base]{\color{textcolor}{\rmfamily\fontsize{10.000000}{12.000000}\selectfont\catcode`\^=\active\def^{\ifmmode\sp\else\^{}\fi}\catcode`\%=\active\def%{\%}$\mathdefault{0.25}$}}%
\end{pgfscope}%
\begin{pgfscope}%
\pgfsetbuttcap%
\pgfsetroundjoin%
\definecolor{currentfill}{rgb}{0.000000,0.000000,0.000000}%
\pgfsetfillcolor{currentfill}%
\pgfsetlinewidth{0.803000pt}%
\definecolor{currentstroke}{rgb}{0.000000,0.000000,0.000000}%
\pgfsetstrokecolor{currentstroke}%
\pgfsetdash{}{0pt}%
\pgfsys@defobject{currentmarker}{\pgfqpoint{-0.048611in}{0.000000in}}{\pgfqpoint{-0.000000in}{0.000000in}}{%
\pgfpathmoveto{\pgfqpoint{-0.000000in}{0.000000in}}%
\pgfpathlineto{\pgfqpoint{-0.048611in}{0.000000in}}%
\pgfusepath{stroke,fill}%
}%
\begin{pgfscope}%
\pgfsys@transformshift{0.673149in}{1.900386in}%
\pgfsys@useobject{currentmarker}{}%
\end{pgfscope}%
\end{pgfscope}%
\begin{pgfscope}%
\definecolor{textcolor}{rgb}{0.000000,0.000000,0.000000}%
\pgfsetstrokecolor{textcolor}%
\pgfsetfillcolor{textcolor}%
\pgftext[x=0.329012in, y=1.852160in, left, base]{\color{textcolor}{\rmfamily\fontsize{10.000000}{12.000000}\selectfont\catcode`\^=\active\def^{\ifmmode\sp\else\^{}\fi}\catcode`\%=\active\def%{\%}$\mathdefault{0.30}$}}%
\end{pgfscope}%
\begin{pgfscope}%
\pgfsetbuttcap%
\pgfsetroundjoin%
\definecolor{currentfill}{rgb}{0.000000,0.000000,0.000000}%
\pgfsetfillcolor{currentfill}%
\pgfsetlinewidth{0.803000pt}%
\definecolor{currentstroke}{rgb}{0.000000,0.000000,0.000000}%
\pgfsetstrokecolor{currentstroke}%
\pgfsetdash{}{0pt}%
\pgfsys@defobject{currentmarker}{\pgfqpoint{-0.048611in}{0.000000in}}{\pgfqpoint{-0.000000in}{0.000000in}}{%
\pgfpathmoveto{\pgfqpoint{-0.000000in}{0.000000in}}%
\pgfpathlineto{\pgfqpoint{-0.048611in}{0.000000in}}%
\pgfusepath{stroke,fill}%
}%
\begin{pgfscope}%
\pgfsys@transformshift{0.673149in}{2.575733in}%
\pgfsys@useobject{currentmarker}{}%
\end{pgfscope}%
\end{pgfscope}%
\begin{pgfscope}%
\definecolor{textcolor}{rgb}{0.000000,0.000000,0.000000}%
\pgfsetstrokecolor{textcolor}%
\pgfsetfillcolor{textcolor}%
\pgftext[x=0.329012in, y=2.527508in, left, base]{\color{textcolor}{\rmfamily\fontsize{10.000000}{12.000000}\selectfont\catcode`\^=\active\def^{\ifmmode\sp\else\^{}\fi}\catcode`\%=\active\def%{\%}$\mathdefault{0.35}$}}%
\end{pgfscope}%
\begin{pgfscope}%
\pgfsetbuttcap%
\pgfsetroundjoin%
\definecolor{currentfill}{rgb}{0.000000,0.000000,0.000000}%
\pgfsetfillcolor{currentfill}%
\pgfsetlinewidth{0.803000pt}%
\definecolor{currentstroke}{rgb}{0.000000,0.000000,0.000000}%
\pgfsetstrokecolor{currentstroke}%
\pgfsetdash{}{0pt}%
\pgfsys@defobject{currentmarker}{\pgfqpoint{-0.048611in}{0.000000in}}{\pgfqpoint{-0.000000in}{0.000000in}}{%
\pgfpathmoveto{\pgfqpoint{-0.000000in}{0.000000in}}%
\pgfpathlineto{\pgfqpoint{-0.048611in}{0.000000in}}%
\pgfusepath{stroke,fill}%
}%
\begin{pgfscope}%
\pgfsys@transformshift{0.673149in}{3.251080in}%
\pgfsys@useobject{currentmarker}{}%
\end{pgfscope}%
\end{pgfscope}%
\begin{pgfscope}%
\definecolor{textcolor}{rgb}{0.000000,0.000000,0.000000}%
\pgfsetstrokecolor{textcolor}%
\pgfsetfillcolor{textcolor}%
\pgftext[x=0.329012in, y=3.202855in, left, base]{\color{textcolor}{\rmfamily\fontsize{10.000000}{12.000000}\selectfont\catcode`\^=\active\def^{\ifmmode\sp\else\^{}\fi}\catcode`\%=\active\def%{\%}$\mathdefault{0.40}$}}%
\end{pgfscope}%
\begin{pgfscope}%
\pgfsetbuttcap%
\pgfsetroundjoin%
\definecolor{currentfill}{rgb}{0.000000,0.000000,0.000000}%
\pgfsetfillcolor{currentfill}%
\pgfsetlinewidth{0.803000pt}%
\definecolor{currentstroke}{rgb}{0.000000,0.000000,0.000000}%
\pgfsetstrokecolor{currentstroke}%
\pgfsetdash{}{0pt}%
\pgfsys@defobject{currentmarker}{\pgfqpoint{-0.048611in}{0.000000in}}{\pgfqpoint{-0.000000in}{0.000000in}}{%
\pgfpathmoveto{\pgfqpoint{-0.000000in}{0.000000in}}%
\pgfpathlineto{\pgfqpoint{-0.048611in}{0.000000in}}%
\pgfusepath{stroke,fill}%
}%
\begin{pgfscope}%
\pgfsys@transformshift{0.673149in}{3.926427in}%
\pgfsys@useobject{currentmarker}{}%
\end{pgfscope}%
\end{pgfscope}%
\begin{pgfscope}%
\definecolor{textcolor}{rgb}{0.000000,0.000000,0.000000}%
\pgfsetstrokecolor{textcolor}%
\pgfsetfillcolor{textcolor}%
\pgftext[x=0.329012in, y=3.878202in, left, base]{\color{textcolor}{\rmfamily\fontsize{10.000000}{12.000000}\selectfont\catcode`\^=\active\def^{\ifmmode\sp\else\^{}\fi}\catcode`\%=\active\def%{\%}$\mathdefault{0.45}$}}%
\end{pgfscope}%
\begin{pgfscope}%
\pgfsetbuttcap%
\pgfsetroundjoin%
\definecolor{currentfill}{rgb}{0.000000,0.000000,0.000000}%
\pgfsetfillcolor{currentfill}%
\pgfsetlinewidth{0.803000pt}%
\definecolor{currentstroke}{rgb}{0.000000,0.000000,0.000000}%
\pgfsetstrokecolor{currentstroke}%
\pgfsetdash{}{0pt}%
\pgfsys@defobject{currentmarker}{\pgfqpoint{-0.048611in}{0.000000in}}{\pgfqpoint{-0.000000in}{0.000000in}}{%
\pgfpathmoveto{\pgfqpoint{-0.000000in}{0.000000in}}%
\pgfpathlineto{\pgfqpoint{-0.048611in}{0.000000in}}%
\pgfusepath{stroke,fill}%
}%
\begin{pgfscope}%
\pgfsys@transformshift{0.673149in}{4.601775in}%
\pgfsys@useobject{currentmarker}{}%
\end{pgfscope}%
\end{pgfscope}%
\begin{pgfscope}%
\definecolor{textcolor}{rgb}{0.000000,0.000000,0.000000}%
\pgfsetstrokecolor{textcolor}%
\pgfsetfillcolor{textcolor}%
\pgftext[x=0.329012in, y=4.553549in, left, base]{\color{textcolor}{\rmfamily\fontsize{10.000000}{12.000000}\selectfont\catcode`\^=\active\def^{\ifmmode\sp\else\^{}\fi}\catcode`\%=\active\def%{\%}$\mathdefault{0.50}$}}%
\end{pgfscope}%
\begin{pgfscope}%
\definecolor{textcolor}{rgb}{0.000000,0.000000,0.000000}%
\pgfsetstrokecolor{textcolor}%
\pgfsetfillcolor{textcolor}%
\pgftext[x=0.273457in,y=2.575733in,,bottom,rotate=90.000000]{\color{textcolor}{\rmfamily\fontsize{10.000000}{12.000000}\selectfont\catcode`\^=\active\def^{\ifmmode\sp\else\^{}\fi}\catcode`\%=\active\def%{\%}\mbox{Effort Reduction}}}%
\end{pgfscope}%
\begin{pgfscope}%
\pgfpathrectangle{\pgfqpoint{0.673149in}{0.549691in}}{\pgfqpoint{5.576851in}{4.052084in}}%
\pgfusepath{clip}%
\pgfsetrectcap%
\pgfsetroundjoin%
\pgfsetlinewidth{2.007500pt}%
\definecolor{currentstroke}{rgb}{0.121569,0.466667,0.705882}%
\pgfsetstrokecolor{currentstroke}%
\pgfsetstrokeopacity{0.800000}%
\pgfsetdash{}{0pt}%
\pgfpathmoveto{\pgfqpoint{1.119297in}{4.052060in}}%
\pgfpathlineto{\pgfqpoint{5.540510in}{0.539691in}}%
\pgfpathlineto{\pgfqpoint{5.540510in}{0.539691in}}%
\pgfusepath{stroke}%
\end{pgfscope}%
\begin{pgfscope}%
\pgfpathrectangle{\pgfqpoint{0.673149in}{0.549691in}}{\pgfqpoint{5.576851in}{4.052084in}}%
\pgfusepath{clip}%
\pgfsetbuttcap%
\pgfsetroundjoin%
\pgfsetlinewidth{2.007500pt}%
\definecolor{currentstroke}{rgb}{1.000000,0.498039,0.054902}%
\pgfsetstrokecolor{currentstroke}%
\pgfsetstrokeopacity{0.800000}%
\pgfsetdash{{2.000000pt}{3.300000pt}}{0.000000pt}%
\pgfpathmoveto{\pgfqpoint{1.119297in}{4.370375in}}%
\pgfpathlineto{\pgfqpoint{5.124785in}{0.539691in}}%
\pgfpathlineto{\pgfqpoint{5.124785in}{0.539691in}}%
\pgfusepath{stroke}%
\end{pgfscope}%
\begin{pgfscope}%
\pgfpathrectangle{\pgfqpoint{0.673149in}{0.549691in}}{\pgfqpoint{5.576851in}{4.052084in}}%
\pgfusepath{clip}%
\pgfsetbuttcap%
\pgfsetroundjoin%
\pgfsetlinewidth{2.007500pt}%
\definecolor{currentstroke}{rgb}{0.172549,0.627451,0.172549}%
\pgfsetstrokecolor{currentstroke}%
\pgfsetstrokeopacity{0.800000}%
\pgfsetdash{{7.400000pt}{3.200000pt}}{0.000000pt}%
\pgfpathmoveto{\pgfqpoint{1.119297in}{4.052060in}}%
\pgfpathlineto{\pgfqpoint{5.540510in}{0.539691in}}%
\pgfpathlineto{\pgfqpoint{5.540510in}{0.539691in}}%
\pgfusepath{stroke}%
\end{pgfscope}%
\begin{pgfscope}%
\pgfpathrectangle{\pgfqpoint{0.673149in}{0.549691in}}{\pgfqpoint{5.576851in}{4.052084in}}%
\pgfusepath{clip}%
\pgfsetbuttcap%
\pgfsetroundjoin%
\pgfsetlinewidth{2.007500pt}%
\definecolor{currentstroke}{rgb}{0.839216,0.152941,0.156863}%
\pgfsetstrokecolor{currentstroke}%
\pgfsetstrokeopacity{0.800000}%
\pgfsetdash{{12.800000pt}{3.200000pt}{2.000000pt}{3.200000pt}}{0.000000pt}%
\pgfpathmoveto{\pgfqpoint{1.119297in}{4.052060in}}%
\pgfpathlineto{\pgfqpoint{5.540510in}{0.539691in}}%
\pgfpathlineto{\pgfqpoint{5.540510in}{0.539691in}}%
\pgfusepath{stroke}%
\end{pgfscope}%
\begin{pgfscope}%
\pgfpathrectangle{\pgfqpoint{0.673149in}{0.549691in}}{\pgfqpoint{5.576851in}{4.052084in}}%
\pgfusepath{clip}%
\pgfsetbuttcap%
\pgfsetroundjoin%
\pgfsetlinewidth{2.007500pt}%
\definecolor{currentstroke}{rgb}{0.580392,0.403922,0.741176}%
\pgfsetstrokecolor{currentstroke}%
\pgfsetstrokeopacity{0.800000}%
\pgfsetdash{{2.000000pt}{2.000000pt}}{0.000000pt}%
\pgfpathmoveto{\pgfqpoint{1.119297in}{4.052060in}}%
\pgfpathlineto{\pgfqpoint{5.540510in}{0.539691in}}%
\pgfpathlineto{\pgfqpoint{5.540510in}{0.539691in}}%
\pgfusepath{stroke}%
\end{pgfscope}%
\begin{pgfscope}%
\pgfpathrectangle{\pgfqpoint{0.673149in}{0.549691in}}{\pgfqpoint{5.576851in}{4.052084in}}%
\pgfusepath{clip}%
\pgfsetbuttcap%
\pgfsetroundjoin%
\pgfsetlinewidth{2.007500pt}%
\definecolor{currentstroke}{rgb}{0.549020,0.337255,0.294118}%
\pgfsetstrokecolor{currentstroke}%
\pgfsetstrokeopacity{0.800000}%
\pgfsetdash{{10.000000pt}{20.000000pt}}{0.000000pt}%
\pgfpathmoveto{\pgfqpoint{1.119297in}{4.058833in}}%
\pgfpathlineto{\pgfqpoint{5.520998in}{0.539691in}}%
\pgfpathlineto{\pgfqpoint{5.520998in}{0.539691in}}%
\pgfusepath{stroke}%
\end{pgfscope}%
\begin{pgfscope}%
\pgfpathrectangle{\pgfqpoint{0.673149in}{0.549691in}}{\pgfqpoint{5.576851in}{4.052084in}}%
\pgfusepath{clip}%
\pgfsetbuttcap%
\pgfsetroundjoin%
\pgfsetlinewidth{2.007500pt}%
\definecolor{currentstroke}{rgb}{0.890196,0.466667,0.760784}%
\pgfsetstrokecolor{currentstroke}%
\pgfsetstrokeopacity{0.800000}%
\pgfsetdash{{10.000000pt}{10.000000pt}}{0.000000pt}%
\pgfpathmoveto{\pgfqpoint{1.119297in}{3.950470in}}%
\pgfpathlineto{\pgfqpoint{5.440154in}{0.539691in}}%
\pgfpathlineto{\pgfqpoint{5.440154in}{0.539691in}}%
\pgfusepath{stroke}%
\end{pgfscope}%
\begin{pgfscope}%
\pgfsetrectcap%
\pgfsetmiterjoin%
\pgfsetlinewidth{0.803000pt}%
\definecolor{currentstroke}{rgb}{0.000000,0.000000,0.000000}%
\pgfsetstrokecolor{currentstroke}%
\pgfsetdash{}{0pt}%
\pgfpathmoveto{\pgfqpoint{0.673149in}{0.549691in}}%
\pgfpathlineto{\pgfqpoint{0.673149in}{4.601775in}}%
\pgfusepath{stroke}%
\end{pgfscope}%
\begin{pgfscope}%
\pgfsetrectcap%
\pgfsetmiterjoin%
\pgfsetlinewidth{0.803000pt}%
\definecolor{currentstroke}{rgb}{0.000000,0.000000,0.000000}%
\pgfsetstrokecolor{currentstroke}%
\pgfsetdash{}{0pt}%
\pgfpathmoveto{\pgfqpoint{6.250000in}{0.549691in}}%
\pgfpathlineto{\pgfqpoint{6.250000in}{4.601775in}}%
\pgfusepath{stroke}%
\end{pgfscope}%
\begin{pgfscope}%
\pgfsetrectcap%
\pgfsetmiterjoin%
\pgfsetlinewidth{0.803000pt}%
\definecolor{currentstroke}{rgb}{0.000000,0.000000,0.000000}%
\pgfsetstrokecolor{currentstroke}%
\pgfsetdash{}{0pt}%
\pgfpathmoveto{\pgfqpoint{0.673149in}{0.549691in}}%
\pgfpathlineto{\pgfqpoint{6.250000in}{0.549691in}}%
\pgfusepath{stroke}%
\end{pgfscope}%
\begin{pgfscope}%
\pgfsetrectcap%
\pgfsetmiterjoin%
\pgfsetlinewidth{0.803000pt}%
\definecolor{currentstroke}{rgb}{0.000000,0.000000,0.000000}%
\pgfsetstrokecolor{currentstroke}%
\pgfsetdash{}{0pt}%
\pgfpathmoveto{\pgfqpoint{0.673149in}{4.601775in}}%
\pgfpathlineto{\pgfqpoint{6.250000in}{4.601775in}}%
\pgfusepath{stroke}%
\end{pgfscope}%
\begin{pgfscope}%
\pgfsetbuttcap%
\pgfsetmiterjoin%
\definecolor{currentfill}{rgb}{1.000000,1.000000,1.000000}%
\pgfsetfillcolor{currentfill}%
\pgfsetfillopacity{0.800000}%
\pgfsetlinewidth{1.003750pt}%
\definecolor{currentstroke}{rgb}{0.800000,0.800000,0.800000}%
\pgfsetstrokecolor{currentstroke}%
\pgfsetstrokeopacity{0.800000}%
\pgfsetdash{}{0pt}%
\pgfpathmoveto{\pgfqpoint{3.914156in}{3.134954in}}%
\pgfpathlineto{\pgfqpoint{6.152778in}{3.134954in}}%
\pgfpathquadraticcurveto{\pgfqpoint{6.180556in}{3.134954in}}{\pgfqpoint{6.180556in}{3.162732in}}%
\pgfpathlineto{\pgfqpoint{6.180556in}{4.504552in}}%
\pgfpathquadraticcurveto{\pgfqpoint{6.180556in}{4.532330in}}{\pgfqpoint{6.152778in}{4.532330in}}%
\pgfpathlineto{\pgfqpoint{3.914156in}{4.532330in}}%
\pgfpathquadraticcurveto{\pgfqpoint{3.886378in}{4.532330in}}{\pgfqpoint{3.886378in}{4.504552in}}%
\pgfpathlineto{\pgfqpoint{3.886378in}{3.162732in}}%
\pgfpathquadraticcurveto{\pgfqpoint{3.886378in}{3.134954in}}{\pgfqpoint{3.914156in}{3.134954in}}%
\pgfpathlineto{\pgfqpoint{3.914156in}{3.134954in}}%
\pgfpathclose%
\pgfusepath{stroke,fill}%
\end{pgfscope}%
\begin{pgfscope}%
\pgfsetrectcap%
\pgfsetroundjoin%
\pgfsetlinewidth{2.007500pt}%
\definecolor{currentstroke}{rgb}{0.121569,0.466667,0.705882}%
\pgfsetstrokecolor{currentstroke}%
\pgfsetstrokeopacity{0.800000}%
\pgfsetdash{}{0pt}%
\pgfpathmoveto{\pgfqpoint{3.941934in}{4.428164in}}%
\pgfpathlineto{\pgfqpoint{4.080823in}{4.428164in}}%
\pgfpathlineto{\pgfqpoint{4.219712in}{4.428164in}}%
\pgfusepath{stroke}%
\end{pgfscope}%
\begin{pgfscope}%
\definecolor{textcolor}{rgb}{0.000000,0.000000,0.000000}%
\pgfsetstrokecolor{textcolor}%
\pgfsetfillcolor{textcolor}%
\pgftext[x=4.330823in,y=4.379552in,left,base]{\color{textcolor}{\rmfamily\fontsize{10.000000}{12.000000}\selectfont\catcode`\^=\active\def^{\ifmmode\sp\else\^{}\fi}\catcode`\%=\active\def%{\%}Gitmerge-ort}}%
\end{pgfscope}%
\begin{pgfscope}%
\pgfsetbuttcap%
\pgfsetroundjoin%
\pgfsetlinewidth{2.007500pt}%
\definecolor{currentstroke}{rgb}{1.000000,0.498039,0.054902}%
\pgfsetstrokecolor{currentstroke}%
\pgfsetstrokeopacity{0.800000}%
\pgfsetdash{{2.000000pt}{3.300000pt}}{0.000000pt}%
\pgfpathmoveto{\pgfqpoint{3.941934in}{4.234491in}}%
\pgfpathlineto{\pgfqpoint{4.080823in}{4.234491in}}%
\pgfpathlineto{\pgfqpoint{4.219712in}{4.234491in}}%
\pgfusepath{stroke}%
\end{pgfscope}%
\begin{pgfscope}%
\definecolor{textcolor}{rgb}{0.000000,0.000000,0.000000}%
\pgfsetstrokecolor{textcolor}%
\pgfsetfillcolor{textcolor}%
\pgftext[x=4.330823in,y=4.185880in,left,base]{\color{textcolor}{\rmfamily\fontsize{10.000000}{12.000000}\selectfont\catcode`\^=\active\def^{\ifmmode\sp\else\^{}\fi}\catcode`\%=\active\def%{\%}Gitmerge-ort-ignorespace}}%
\end{pgfscope}%
\begin{pgfscope}%
\pgfsetbuttcap%
\pgfsetroundjoin%
\pgfsetlinewidth{2.007500pt}%
\definecolor{currentstroke}{rgb}{0.172549,0.627451,0.172549}%
\pgfsetstrokecolor{currentstroke}%
\pgfsetstrokeopacity{0.800000}%
\pgfsetdash{{7.400000pt}{3.200000pt}}{0.000000pt}%
\pgfpathmoveto{\pgfqpoint{3.941934in}{4.040818in}}%
\pgfpathlineto{\pgfqpoint{4.080823in}{4.040818in}}%
\pgfpathlineto{\pgfqpoint{4.219712in}{4.040818in}}%
\pgfusepath{stroke}%
\end{pgfscope}%
\begin{pgfscope}%
\definecolor{textcolor}{rgb}{0.000000,0.000000,0.000000}%
\pgfsetstrokecolor{textcolor}%
\pgfsetfillcolor{textcolor}%
\pgftext[x=4.330823in,y=3.992207in,left,base]{\color{textcolor}{\rmfamily\fontsize{10.000000}{12.000000}\selectfont\catcode`\^=\active\def^{\ifmmode\sp\else\^{}\fi}\catcode`\%=\active\def%{\%}Gitmerge-recursive-histogram}}%
\end{pgfscope}%
\begin{pgfscope}%
\pgfsetbuttcap%
\pgfsetroundjoin%
\pgfsetlinewidth{2.007500pt}%
\definecolor{currentstroke}{rgb}{0.839216,0.152941,0.156863}%
\pgfsetstrokecolor{currentstroke}%
\pgfsetstrokeopacity{0.800000}%
\pgfsetdash{{12.800000pt}{3.200000pt}{2.000000pt}{3.200000pt}}{0.000000pt}%
\pgfpathmoveto{\pgfqpoint{3.941934in}{3.847145in}}%
\pgfpathlineto{\pgfqpoint{4.080823in}{3.847145in}}%
\pgfpathlineto{\pgfqpoint{4.219712in}{3.847145in}}%
\pgfusepath{stroke}%
\end{pgfscope}%
\begin{pgfscope}%
\definecolor{textcolor}{rgb}{0.000000,0.000000,0.000000}%
\pgfsetstrokecolor{textcolor}%
\pgfsetfillcolor{textcolor}%
\pgftext[x=4.330823in,y=3.798534in,left,base]{\color{textcolor}{\rmfamily\fontsize{10.000000}{12.000000}\selectfont\catcode`\^=\active\def^{\ifmmode\sp\else\^{}\fi}\catcode`\%=\active\def%{\%}Gitmerge-recursive-minimal}}%
\end{pgfscope}%
\begin{pgfscope}%
\pgfsetbuttcap%
\pgfsetroundjoin%
\pgfsetlinewidth{2.007500pt}%
\definecolor{currentstroke}{rgb}{0.580392,0.403922,0.741176}%
\pgfsetstrokecolor{currentstroke}%
\pgfsetstrokeopacity{0.800000}%
\pgfsetdash{{2.000000pt}{2.000000pt}}{0.000000pt}%
\pgfpathmoveto{\pgfqpoint{3.941934in}{3.653472in}}%
\pgfpathlineto{\pgfqpoint{4.080823in}{3.653472in}}%
\pgfpathlineto{\pgfqpoint{4.219712in}{3.653472in}}%
\pgfusepath{stroke}%
\end{pgfscope}%
\begin{pgfscope}%
\definecolor{textcolor}{rgb}{0.000000,0.000000,0.000000}%
\pgfsetstrokecolor{textcolor}%
\pgfsetfillcolor{textcolor}%
\pgftext[x=4.330823in,y=3.604861in,left,base]{\color{textcolor}{\rmfamily\fontsize{10.000000}{12.000000}\selectfont\catcode`\^=\active\def^{\ifmmode\sp\else\^{}\fi}\catcode`\%=\active\def%{\%}Gitmerge-recursive-myers}}%
\end{pgfscope}%
\begin{pgfscope}%
\pgfsetbuttcap%
\pgfsetroundjoin%
\pgfsetlinewidth{2.007500pt}%
\definecolor{currentstroke}{rgb}{0.549020,0.337255,0.294118}%
\pgfsetstrokecolor{currentstroke}%
\pgfsetstrokeopacity{0.800000}%
\pgfsetdash{{10.000000pt}{20.000000pt}}{0.000000pt}%
\pgfpathmoveto{\pgfqpoint{3.941934in}{3.459800in}}%
\pgfpathlineto{\pgfqpoint{4.080823in}{3.459800in}}%
\pgfpathlineto{\pgfqpoint{4.219712in}{3.459800in}}%
\pgfusepath{stroke}%
\end{pgfscope}%
\begin{pgfscope}%
\definecolor{textcolor}{rgb}{0.000000,0.000000,0.000000}%
\pgfsetstrokecolor{textcolor}%
\pgfsetfillcolor{textcolor}%
\pgftext[x=4.330823in,y=3.411189in,left,base]{\color{textcolor}{\rmfamily\fontsize{10.000000}{12.000000}\selectfont\catcode`\^=\active\def^{\ifmmode\sp\else\^{}\fi}\catcode`\%=\active\def%{\%}Gitmerge-recursive-patience}}%
\end{pgfscope}%
\begin{pgfscope}%
\pgfsetbuttcap%
\pgfsetroundjoin%
\pgfsetlinewidth{2.007500pt}%
\definecolor{currentstroke}{rgb}{0.890196,0.466667,0.760784}%
\pgfsetstrokecolor{currentstroke}%
\pgfsetstrokeopacity{0.800000}%
\pgfsetdash{{10.000000pt}{10.000000pt}}{0.000000pt}%
\pgfpathmoveto{\pgfqpoint{3.941934in}{3.266127in}}%
\pgfpathlineto{\pgfqpoint{4.080823in}{3.266127in}}%
\pgfpathlineto{\pgfqpoint{4.219712in}{3.266127in}}%
\pgfusepath{stroke}%
\end{pgfscope}%
\begin{pgfscope}%
\definecolor{textcolor}{rgb}{0.000000,0.000000,0.000000}%
\pgfsetstrokecolor{textcolor}%
\pgfsetfillcolor{textcolor}%
\pgftext[x=4.330823in,y=3.217516in,left,base]{\color{textcolor}{\rmfamily\fontsize{10.000000}{12.000000}\selectfont\catcode`\^=\active\def^{\ifmmode\sp\else\^{}\fi}\catcode`\%=\active\def%{\%}Gitmerge-resolve}}%
\end{pgfscope}%
\end{pgfpicture}%
\makeatother%
\endgroup%

%% file: results/combined/tables/tools/table_summary.tex
% Do not edit.  This file is automatically generated.
\begin{tabular}{l|c c|c c|c c|}
        Tool & \multicolumn{6}{c|}{Merges} \\ \cline{2-7}
        & \multicolumn{2}{c|}{Correct} &
        \multicolumn{2}{c|}{Unhandled} &
        \multicolumn{2}{c|}{Incorrect} \\
        & \# & \% & \# & \% & \# & \% \\
        \hline
Gitmerge-ort                     &  2748 &  46\% &  3078 &  51\% &   157 &   3\% \\
Gitmerge-ort-ignorespace         &  2889 &  48\% &  2905 &  49\% &   189 &   3\% \\
Hires-Merge                      &  3040 &  51\% &  2721 &  45\% &   222 &   4\% \\
Spork                            &  3260 &  54\% &  2080 &  35\% &   643 &  11\% \\
IntelliMerge                     &  1434 &  24\% &  1582 &  26\% &  2967 &  50\% \\
Adjacent                         &  3073 &  51\% &  2692 &  45\% &   218 &   4\% \\
Imports                          &  2904 &  49\% &  2910 &  49\% &   169 &   3\% \\
Version Numbers                  &  2782 &  46\% &  3044 &  51\% &   157 &   3\% \\
IVn                              &  3011 &  50\% &  2803 &  47\% &   169 &   3\% \\
IVn-ignorespace                  &  3116 &  52\% &  2665 &  45\% &   202 &   3\% \\
\end{tabular}

%% file: results/combined/plots/tools/cost_with_manual.pgf
%% Creator: Matplotlib, PGF backend
%%
%% To include the figure in your LaTeX document, write
%%   \input{<filename>.pgf}
%%
%% Make sure the required packages are loaded in your preamble
%%   \usepackage{pgf}
%%
%% Also ensure that all the required font packages are loaded; for instance,
%% the lmodern package is sometimes necessary when using math font.
%%   \usepackage{lmodern}
%%
%% Figures using additional raster images can only be included by \input if
%% they are in the same directory as the main LaTeX file. For loading figures
%% from other directories you can use the `import` package
%% and then include the figures with
%%   \import{<path to file>}{<filename>.pgf}
%%
%% Matplotlib used the following preamble
%%   \def\mathdefault#1{#1}
%%   \everymath=\expandafter{\the\everymath\displaystyle}
%%
%%   \makeatletter\@ifpackageloaded{underscore}{}{\usepackage[strings]{underscore}}\makeatother
%%
\begingroup%
\makeatletter%
\begin{pgfpicture}%
\pgfpathrectangle{\pgfpointorigin}{\pgfqpoint{6.400000in}{4.800000in}}%
\pgfusepath{use as bounding box, clip}%
\begin{pgfscope}%
\pgfsetbuttcap%
\pgfsetmiterjoin%
\definecolor{currentfill}{rgb}{1.000000,1.000000,1.000000}%
\pgfsetfillcolor{currentfill}%
\pgfsetlinewidth{0.000000pt}%
\definecolor{currentstroke}{rgb}{1.000000,1.000000,1.000000}%
\pgfsetstrokecolor{currentstroke}%
\pgfsetdash{}{0pt}%
\pgfpathmoveto{\pgfqpoint{0.000000in}{0.000000in}}%
\pgfpathlineto{\pgfqpoint{6.400000in}{0.000000in}}%
\pgfpathlineto{\pgfqpoint{6.400000in}{4.800000in}}%
\pgfpathlineto{\pgfqpoint{0.000000in}{4.800000in}}%
\pgfpathlineto{\pgfqpoint{0.000000in}{0.000000in}}%
\pgfpathclose%
\pgfusepath{fill}%
\end{pgfscope}%
\begin{pgfscope}%
\pgfsetbuttcap%
\pgfsetmiterjoin%
\definecolor{currentfill}{rgb}{1.000000,1.000000,1.000000}%
\pgfsetfillcolor{currentfill}%
\pgfsetlinewidth{0.000000pt}%
\definecolor{currentstroke}{rgb}{0.000000,0.000000,0.000000}%
\pgfsetstrokecolor{currentstroke}%
\pgfsetstrokeopacity{0.000000}%
\pgfsetdash{}{0pt}%
\pgfpathmoveto{\pgfqpoint{0.603704in}{0.549691in}}%
\pgfpathlineto{\pgfqpoint{6.250000in}{0.549691in}}%
\pgfpathlineto{\pgfqpoint{6.250000in}{4.601775in}}%
\pgfpathlineto{\pgfqpoint{0.603704in}{4.601775in}}%
\pgfpathlineto{\pgfqpoint{0.603704in}{0.549691in}}%
\pgfpathclose%
\pgfusepath{fill}%
\end{pgfscope}%
\begin{pgfscope}%
\pgfsetbuttcap%
\pgfsetroundjoin%
\definecolor{currentfill}{rgb}{0.000000,0.000000,0.000000}%
\pgfsetfillcolor{currentfill}%
\pgfsetlinewidth{0.803000pt}%
\definecolor{currentstroke}{rgb}{0.000000,0.000000,0.000000}%
\pgfsetstrokecolor{currentstroke}%
\pgfsetdash{}{0pt}%
\pgfsys@defobject{currentmarker}{\pgfqpoint{0.000000in}{-0.048611in}}{\pgfqpoint{0.000000in}{0.000000in}}{%
\pgfpathmoveto{\pgfqpoint{0.000000in}{0.000000in}}%
\pgfpathlineto{\pgfqpoint{0.000000in}{-0.048611in}}%
\pgfusepath{stroke,fill}%
}%
\begin{pgfscope}%
\pgfsys@transformshift{0.603704in}{0.549691in}%
\pgfsys@useobject{currentmarker}{}%
\end{pgfscope}%
\end{pgfscope}%
\begin{pgfscope}%
\definecolor{textcolor}{rgb}{0.000000,0.000000,0.000000}%
\pgfsetstrokecolor{textcolor}%
\pgfsetfillcolor{textcolor}%
\pgftext[x=0.603704in,y=0.452469in,,top]{\color{textcolor}{\rmfamily\fontsize{10.000000}{12.000000}\selectfont\catcode`\^=\active\def^{\ifmmode\sp\else\^{}\fi}\catcode`\%=\active\def%{\%}$\mathdefault{0.0}$}}%
\end{pgfscope}%
\begin{pgfscope}%
\pgfsetbuttcap%
\pgfsetroundjoin%
\definecolor{currentfill}{rgb}{0.000000,0.000000,0.000000}%
\pgfsetfillcolor{currentfill}%
\pgfsetlinewidth{0.803000pt}%
\definecolor{currentstroke}{rgb}{0.000000,0.000000,0.000000}%
\pgfsetstrokecolor{currentstroke}%
\pgfsetdash{}{0pt}%
\pgfsys@defobject{currentmarker}{\pgfqpoint{0.000000in}{-0.048611in}}{\pgfqpoint{0.000000in}{0.000000in}}{%
\pgfpathmoveto{\pgfqpoint{0.000000in}{0.000000in}}%
\pgfpathlineto{\pgfqpoint{0.000000in}{-0.048611in}}%
\pgfusepath{stroke,fill}%
}%
\begin{pgfscope}%
\pgfsys@transformshift{1.353880in}{0.549691in}%
\pgfsys@useobject{currentmarker}{}%
\end{pgfscope}%
\end{pgfscope}%
\begin{pgfscope}%
\definecolor{textcolor}{rgb}{0.000000,0.000000,0.000000}%
\pgfsetstrokecolor{textcolor}%
\pgfsetfillcolor{textcolor}%
\pgftext[x=1.353880in,y=0.452469in,,top]{\color{textcolor}{\rmfamily\fontsize{10.000000}{12.000000}\selectfont\catcode`\^=\active\def^{\ifmmode\sp\else\^{}\fi}\catcode`\%=\active\def%{\%}$\mathdefault{2.5}$}}%
\end{pgfscope}%
\begin{pgfscope}%
\pgfsetbuttcap%
\pgfsetroundjoin%
\definecolor{currentfill}{rgb}{0.000000,0.000000,0.000000}%
\pgfsetfillcolor{currentfill}%
\pgfsetlinewidth{0.803000pt}%
\definecolor{currentstroke}{rgb}{0.000000,0.000000,0.000000}%
\pgfsetstrokecolor{currentstroke}%
\pgfsetdash{}{0pt}%
\pgfsys@defobject{currentmarker}{\pgfqpoint{0.000000in}{-0.048611in}}{\pgfqpoint{0.000000in}{0.000000in}}{%
\pgfpathmoveto{\pgfqpoint{0.000000in}{0.000000in}}%
\pgfpathlineto{\pgfqpoint{0.000000in}{-0.048611in}}%
\pgfusepath{stroke,fill}%
}%
\begin{pgfscope}%
\pgfsys@transformshift{2.104056in}{0.549691in}%
\pgfsys@useobject{currentmarker}{}%
\end{pgfscope}%
\end{pgfscope}%
\begin{pgfscope}%
\definecolor{textcolor}{rgb}{0.000000,0.000000,0.000000}%
\pgfsetstrokecolor{textcolor}%
\pgfsetfillcolor{textcolor}%
\pgftext[x=2.104056in,y=0.452469in,,top]{\color{textcolor}{\rmfamily\fontsize{10.000000}{12.000000}\selectfont\catcode`\^=\active\def^{\ifmmode\sp\else\^{}\fi}\catcode`\%=\active\def%{\%}$\mathdefault{5.0}$}}%
\end{pgfscope}%
\begin{pgfscope}%
\pgfsetbuttcap%
\pgfsetroundjoin%
\definecolor{currentfill}{rgb}{0.000000,0.000000,0.000000}%
\pgfsetfillcolor{currentfill}%
\pgfsetlinewidth{0.803000pt}%
\definecolor{currentstroke}{rgb}{0.000000,0.000000,0.000000}%
\pgfsetstrokecolor{currentstroke}%
\pgfsetdash{}{0pt}%
\pgfsys@defobject{currentmarker}{\pgfqpoint{0.000000in}{-0.048611in}}{\pgfqpoint{0.000000in}{0.000000in}}{%
\pgfpathmoveto{\pgfqpoint{0.000000in}{0.000000in}}%
\pgfpathlineto{\pgfqpoint{0.000000in}{-0.048611in}}%
\pgfusepath{stroke,fill}%
}%
\begin{pgfscope}%
\pgfsys@transformshift{2.854232in}{0.549691in}%
\pgfsys@useobject{currentmarker}{}%
\end{pgfscope}%
\end{pgfscope}%
\begin{pgfscope}%
\definecolor{textcolor}{rgb}{0.000000,0.000000,0.000000}%
\pgfsetstrokecolor{textcolor}%
\pgfsetfillcolor{textcolor}%
\pgftext[x=2.854232in,y=0.452469in,,top]{\color{textcolor}{\rmfamily\fontsize{10.000000}{12.000000}\selectfont\catcode`\^=\active\def^{\ifmmode\sp\else\^{}\fi}\catcode`\%=\active\def%{\%}$\mathdefault{7.5}$}}%
\end{pgfscope}%
\begin{pgfscope}%
\pgfsetbuttcap%
\pgfsetroundjoin%
\definecolor{currentfill}{rgb}{0.000000,0.000000,0.000000}%
\pgfsetfillcolor{currentfill}%
\pgfsetlinewidth{0.803000pt}%
\definecolor{currentstroke}{rgb}{0.000000,0.000000,0.000000}%
\pgfsetstrokecolor{currentstroke}%
\pgfsetdash{}{0pt}%
\pgfsys@defobject{currentmarker}{\pgfqpoint{0.000000in}{-0.048611in}}{\pgfqpoint{0.000000in}{0.000000in}}{%
\pgfpathmoveto{\pgfqpoint{0.000000in}{0.000000in}}%
\pgfpathlineto{\pgfqpoint{0.000000in}{-0.048611in}}%
\pgfusepath{stroke,fill}%
}%
\begin{pgfscope}%
\pgfsys@transformshift{3.604409in}{0.549691in}%
\pgfsys@useobject{currentmarker}{}%
\end{pgfscope}%
\end{pgfscope}%
\begin{pgfscope}%
\definecolor{textcolor}{rgb}{0.000000,0.000000,0.000000}%
\pgfsetstrokecolor{textcolor}%
\pgfsetfillcolor{textcolor}%
\pgftext[x=3.604409in,y=0.452469in,,top]{\color{textcolor}{\rmfamily\fontsize{10.000000}{12.000000}\selectfont\catcode`\^=\active\def^{\ifmmode\sp\else\^{}\fi}\catcode`\%=\active\def%{\%}$\mathdefault{10.0}$}}%
\end{pgfscope}%
\begin{pgfscope}%
\pgfsetbuttcap%
\pgfsetroundjoin%
\definecolor{currentfill}{rgb}{0.000000,0.000000,0.000000}%
\pgfsetfillcolor{currentfill}%
\pgfsetlinewidth{0.803000pt}%
\definecolor{currentstroke}{rgb}{0.000000,0.000000,0.000000}%
\pgfsetstrokecolor{currentstroke}%
\pgfsetdash{}{0pt}%
\pgfsys@defobject{currentmarker}{\pgfqpoint{0.000000in}{-0.048611in}}{\pgfqpoint{0.000000in}{0.000000in}}{%
\pgfpathmoveto{\pgfqpoint{0.000000in}{0.000000in}}%
\pgfpathlineto{\pgfqpoint{0.000000in}{-0.048611in}}%
\pgfusepath{stroke,fill}%
}%
\begin{pgfscope}%
\pgfsys@transformshift{4.354585in}{0.549691in}%
\pgfsys@useobject{currentmarker}{}%
\end{pgfscope}%
\end{pgfscope}%
\begin{pgfscope}%
\definecolor{textcolor}{rgb}{0.000000,0.000000,0.000000}%
\pgfsetstrokecolor{textcolor}%
\pgfsetfillcolor{textcolor}%
\pgftext[x=4.354585in,y=0.452469in,,top]{\color{textcolor}{\rmfamily\fontsize{10.000000}{12.000000}\selectfont\catcode`\^=\active\def^{\ifmmode\sp\else\^{}\fi}\catcode`\%=\active\def%{\%}$\mathdefault{12.5}$}}%
\end{pgfscope}%
\begin{pgfscope}%
\pgfsetbuttcap%
\pgfsetroundjoin%
\definecolor{currentfill}{rgb}{0.000000,0.000000,0.000000}%
\pgfsetfillcolor{currentfill}%
\pgfsetlinewidth{0.803000pt}%
\definecolor{currentstroke}{rgb}{0.000000,0.000000,0.000000}%
\pgfsetstrokecolor{currentstroke}%
\pgfsetdash{}{0pt}%
\pgfsys@defobject{currentmarker}{\pgfqpoint{0.000000in}{-0.048611in}}{\pgfqpoint{0.000000in}{0.000000in}}{%
\pgfpathmoveto{\pgfqpoint{0.000000in}{0.000000in}}%
\pgfpathlineto{\pgfqpoint{0.000000in}{-0.048611in}}%
\pgfusepath{stroke,fill}%
}%
\begin{pgfscope}%
\pgfsys@transformshift{5.104761in}{0.549691in}%
\pgfsys@useobject{currentmarker}{}%
\end{pgfscope}%
\end{pgfscope}%
\begin{pgfscope}%
\definecolor{textcolor}{rgb}{0.000000,0.000000,0.000000}%
\pgfsetstrokecolor{textcolor}%
\pgfsetfillcolor{textcolor}%
\pgftext[x=5.104761in,y=0.452469in,,top]{\color{textcolor}{\rmfamily\fontsize{10.000000}{12.000000}\selectfont\catcode`\^=\active\def^{\ifmmode\sp\else\^{}\fi}\catcode`\%=\active\def%{\%}$\mathdefault{15.0}$}}%
\end{pgfscope}%
\begin{pgfscope}%
\pgfsetbuttcap%
\pgfsetroundjoin%
\definecolor{currentfill}{rgb}{0.000000,0.000000,0.000000}%
\pgfsetfillcolor{currentfill}%
\pgfsetlinewidth{0.803000pt}%
\definecolor{currentstroke}{rgb}{0.000000,0.000000,0.000000}%
\pgfsetstrokecolor{currentstroke}%
\pgfsetdash{}{0pt}%
\pgfsys@defobject{currentmarker}{\pgfqpoint{0.000000in}{-0.048611in}}{\pgfqpoint{0.000000in}{0.000000in}}{%
\pgfpathmoveto{\pgfqpoint{0.000000in}{0.000000in}}%
\pgfpathlineto{\pgfqpoint{0.000000in}{-0.048611in}}%
\pgfusepath{stroke,fill}%
}%
\begin{pgfscope}%
\pgfsys@transformshift{5.854937in}{0.549691in}%
\pgfsys@useobject{currentmarker}{}%
\end{pgfscope}%
\end{pgfscope}%
\begin{pgfscope}%
\definecolor{textcolor}{rgb}{0.000000,0.000000,0.000000}%
\pgfsetstrokecolor{textcolor}%
\pgfsetfillcolor{textcolor}%
\pgftext[x=5.854937in,y=0.452469in,,top]{\color{textcolor}{\rmfamily\fontsize{10.000000}{12.000000}\selectfont\catcode`\^=\active\def^{\ifmmode\sp\else\^{}\fi}\catcode`\%=\active\def%{\%}$\mathdefault{17.5}$}}%
\end{pgfscope}%
\begin{pgfscope}%
\definecolor{textcolor}{rgb}{0.000000,0.000000,0.000000}%
\pgfsetstrokecolor{textcolor}%
\pgfsetfillcolor{textcolor}%
\pgftext[x=3.426852in,y=0.273457in,,top]{\color{textcolor}{\rmfamily\fontsize{10.000000}{12.000000}\selectfont\catcode`\^=\active\def^{\ifmmode\sp\else\^{}\fi}\catcode`\%=\active\def%{\%}Incorrect merges cost factor $k$}}%
\end{pgfscope}%
\begin{pgfscope}%
\pgfsetbuttcap%
\pgfsetroundjoin%
\definecolor{currentfill}{rgb}{0.000000,0.000000,0.000000}%
\pgfsetfillcolor{currentfill}%
\pgfsetlinewidth{0.803000pt}%
\definecolor{currentstroke}{rgb}{0.000000,0.000000,0.000000}%
\pgfsetstrokecolor{currentstroke}%
\pgfsetdash{}{0pt}%
\pgfsys@defobject{currentmarker}{\pgfqpoint{-0.048611in}{0.000000in}}{\pgfqpoint{-0.000000in}{0.000000in}}{%
\pgfpathmoveto{\pgfqpoint{-0.000000in}{0.000000in}}%
\pgfpathlineto{\pgfqpoint{-0.048611in}{0.000000in}}%
\pgfusepath{stroke,fill}%
}%
\begin{pgfscope}%
\pgfsys@transformshift{0.603704in}{0.680403in}%
\pgfsys@useobject{currentmarker}{}%
\end{pgfscope}%
\end{pgfscope}%
\begin{pgfscope}%
\definecolor{textcolor}{rgb}{0.000000,0.000000,0.000000}%
\pgfsetstrokecolor{textcolor}%
\pgfsetfillcolor{textcolor}%
\pgftext[x=0.329012in, y=0.632178in, left, base]{\color{textcolor}{\rmfamily\fontsize{10.000000}{12.000000}\selectfont\catcode`\^=\active\def^{\ifmmode\sp\else\^{}\fi}\catcode`\%=\active\def%{\%}$\mathdefault{0.0}$}}%
\end{pgfscope}%
\begin{pgfscope}%
\pgfsetbuttcap%
\pgfsetroundjoin%
\definecolor{currentfill}{rgb}{0.000000,0.000000,0.000000}%
\pgfsetfillcolor{currentfill}%
\pgfsetlinewidth{0.803000pt}%
\definecolor{currentstroke}{rgb}{0.000000,0.000000,0.000000}%
\pgfsetstrokecolor{currentstroke}%
\pgfsetdash{}{0pt}%
\pgfsys@defobject{currentmarker}{\pgfqpoint{-0.048611in}{0.000000in}}{\pgfqpoint{-0.000000in}{0.000000in}}{%
\pgfpathmoveto{\pgfqpoint{-0.000000in}{0.000000in}}%
\pgfpathlineto{\pgfqpoint{-0.048611in}{0.000000in}}%
\pgfusepath{stroke,fill}%
}%
\begin{pgfscope}%
\pgfsys@transformshift{0.603704in}{1.333965in}%
\pgfsys@useobject{currentmarker}{}%
\end{pgfscope}%
\end{pgfscope}%
\begin{pgfscope}%
\definecolor{textcolor}{rgb}{0.000000,0.000000,0.000000}%
\pgfsetstrokecolor{textcolor}%
\pgfsetfillcolor{textcolor}%
\pgftext[x=0.329012in, y=1.285740in, left, base]{\color{textcolor}{\rmfamily\fontsize{10.000000}{12.000000}\selectfont\catcode`\^=\active\def^{\ifmmode\sp\else\^{}\fi}\catcode`\%=\active\def%{\%}$\mathdefault{0.1}$}}%
\end{pgfscope}%
\begin{pgfscope}%
\pgfsetbuttcap%
\pgfsetroundjoin%
\definecolor{currentfill}{rgb}{0.000000,0.000000,0.000000}%
\pgfsetfillcolor{currentfill}%
\pgfsetlinewidth{0.803000pt}%
\definecolor{currentstroke}{rgb}{0.000000,0.000000,0.000000}%
\pgfsetstrokecolor{currentstroke}%
\pgfsetdash{}{0pt}%
\pgfsys@defobject{currentmarker}{\pgfqpoint{-0.048611in}{0.000000in}}{\pgfqpoint{-0.000000in}{0.000000in}}{%
\pgfpathmoveto{\pgfqpoint{-0.000000in}{0.000000in}}%
\pgfpathlineto{\pgfqpoint{-0.048611in}{0.000000in}}%
\pgfusepath{stroke,fill}%
}%
\begin{pgfscope}%
\pgfsys@transformshift{0.603704in}{1.987527in}%
\pgfsys@useobject{currentmarker}{}%
\end{pgfscope}%
\end{pgfscope}%
\begin{pgfscope}%
\definecolor{textcolor}{rgb}{0.000000,0.000000,0.000000}%
\pgfsetstrokecolor{textcolor}%
\pgfsetfillcolor{textcolor}%
\pgftext[x=0.329012in, y=1.939302in, left, base]{\color{textcolor}{\rmfamily\fontsize{10.000000}{12.000000}\selectfont\catcode`\^=\active\def^{\ifmmode\sp\else\^{}\fi}\catcode`\%=\active\def%{\%}$\mathdefault{0.2}$}}%
\end{pgfscope}%
\begin{pgfscope}%
\pgfsetbuttcap%
\pgfsetroundjoin%
\definecolor{currentfill}{rgb}{0.000000,0.000000,0.000000}%
\pgfsetfillcolor{currentfill}%
\pgfsetlinewidth{0.803000pt}%
\definecolor{currentstroke}{rgb}{0.000000,0.000000,0.000000}%
\pgfsetstrokecolor{currentstroke}%
\pgfsetdash{}{0pt}%
\pgfsys@defobject{currentmarker}{\pgfqpoint{-0.048611in}{0.000000in}}{\pgfqpoint{-0.000000in}{0.000000in}}{%
\pgfpathmoveto{\pgfqpoint{-0.000000in}{0.000000in}}%
\pgfpathlineto{\pgfqpoint{-0.048611in}{0.000000in}}%
\pgfusepath{stroke,fill}%
}%
\begin{pgfscope}%
\pgfsys@transformshift{0.603704in}{2.641089in}%
\pgfsys@useobject{currentmarker}{}%
\end{pgfscope}%
\end{pgfscope}%
\begin{pgfscope}%
\definecolor{textcolor}{rgb}{0.000000,0.000000,0.000000}%
\pgfsetstrokecolor{textcolor}%
\pgfsetfillcolor{textcolor}%
\pgftext[x=0.329012in, y=2.592864in, left, base]{\color{textcolor}{\rmfamily\fontsize{10.000000}{12.000000}\selectfont\catcode`\^=\active\def^{\ifmmode\sp\else\^{}\fi}\catcode`\%=\active\def%{\%}$\mathdefault{0.3}$}}%
\end{pgfscope}%
\begin{pgfscope}%
\pgfsetbuttcap%
\pgfsetroundjoin%
\definecolor{currentfill}{rgb}{0.000000,0.000000,0.000000}%
\pgfsetfillcolor{currentfill}%
\pgfsetlinewidth{0.803000pt}%
\definecolor{currentstroke}{rgb}{0.000000,0.000000,0.000000}%
\pgfsetstrokecolor{currentstroke}%
\pgfsetdash{}{0pt}%
\pgfsys@defobject{currentmarker}{\pgfqpoint{-0.048611in}{0.000000in}}{\pgfqpoint{-0.000000in}{0.000000in}}{%
\pgfpathmoveto{\pgfqpoint{-0.000000in}{0.000000in}}%
\pgfpathlineto{\pgfqpoint{-0.048611in}{0.000000in}}%
\pgfusepath{stroke,fill}%
}%
\begin{pgfscope}%
\pgfsys@transformshift{0.603704in}{3.294651in}%
\pgfsys@useobject{currentmarker}{}%
\end{pgfscope}%
\end{pgfscope}%
\begin{pgfscope}%
\definecolor{textcolor}{rgb}{0.000000,0.000000,0.000000}%
\pgfsetstrokecolor{textcolor}%
\pgfsetfillcolor{textcolor}%
\pgftext[x=0.329012in, y=3.246426in, left, base]{\color{textcolor}{\rmfamily\fontsize{10.000000}{12.000000}\selectfont\catcode`\^=\active\def^{\ifmmode\sp\else\^{}\fi}\catcode`\%=\active\def%{\%}$\mathdefault{0.4}$}}%
\end{pgfscope}%
\begin{pgfscope}%
\pgfsetbuttcap%
\pgfsetroundjoin%
\definecolor{currentfill}{rgb}{0.000000,0.000000,0.000000}%
\pgfsetfillcolor{currentfill}%
\pgfsetlinewidth{0.803000pt}%
\definecolor{currentstroke}{rgb}{0.000000,0.000000,0.000000}%
\pgfsetstrokecolor{currentstroke}%
\pgfsetdash{}{0pt}%
\pgfsys@defobject{currentmarker}{\pgfqpoint{-0.048611in}{0.000000in}}{\pgfqpoint{-0.000000in}{0.000000in}}{%
\pgfpathmoveto{\pgfqpoint{-0.000000in}{0.000000in}}%
\pgfpathlineto{\pgfqpoint{-0.048611in}{0.000000in}}%
\pgfusepath{stroke,fill}%
}%
\begin{pgfscope}%
\pgfsys@transformshift{0.603704in}{3.948213in}%
\pgfsys@useobject{currentmarker}{}%
\end{pgfscope}%
\end{pgfscope}%
\begin{pgfscope}%
\definecolor{textcolor}{rgb}{0.000000,0.000000,0.000000}%
\pgfsetstrokecolor{textcolor}%
\pgfsetfillcolor{textcolor}%
\pgftext[x=0.329012in, y=3.899988in, left, base]{\color{textcolor}{\rmfamily\fontsize{10.000000}{12.000000}\selectfont\catcode`\^=\active\def^{\ifmmode\sp\else\^{}\fi}\catcode`\%=\active\def%{\%}$\mathdefault{0.5}$}}%
\end{pgfscope}%
\begin{pgfscope}%
\pgfsetbuttcap%
\pgfsetroundjoin%
\definecolor{currentfill}{rgb}{0.000000,0.000000,0.000000}%
\pgfsetfillcolor{currentfill}%
\pgfsetlinewidth{0.803000pt}%
\definecolor{currentstroke}{rgb}{0.000000,0.000000,0.000000}%
\pgfsetstrokecolor{currentstroke}%
\pgfsetdash{}{0pt}%
\pgfsys@defobject{currentmarker}{\pgfqpoint{-0.048611in}{0.000000in}}{\pgfqpoint{-0.000000in}{0.000000in}}{%
\pgfpathmoveto{\pgfqpoint{-0.000000in}{0.000000in}}%
\pgfpathlineto{\pgfqpoint{-0.048611in}{0.000000in}}%
\pgfusepath{stroke,fill}%
}%
\begin{pgfscope}%
\pgfsys@transformshift{0.603704in}{4.601775in}%
\pgfsys@useobject{currentmarker}{}%
\end{pgfscope}%
\end{pgfscope}%
\begin{pgfscope}%
\definecolor{textcolor}{rgb}{0.000000,0.000000,0.000000}%
\pgfsetstrokecolor{textcolor}%
\pgfsetfillcolor{textcolor}%
\pgftext[x=0.329012in, y=4.553549in, left, base]{\color{textcolor}{\rmfamily\fontsize{10.000000}{12.000000}\selectfont\catcode`\^=\active\def^{\ifmmode\sp\else\^{}\fi}\catcode`\%=\active\def%{\%}$\mathdefault{0.6}$}}%
\end{pgfscope}%
\begin{pgfscope}%
\definecolor{textcolor}{rgb}{0.000000,0.000000,0.000000}%
\pgfsetstrokecolor{textcolor}%
\pgfsetfillcolor{textcolor}%
\pgftext[x=0.273457in,y=2.575733in,,bottom,rotate=90.000000]{\color{textcolor}{\rmfamily\fontsize{10.000000}{12.000000}\selectfont\catcode`\^=\active\def^{\ifmmode\sp\else\^{}\fi}\catcode`\%=\active\def%{\%}\mbox{Effort Reduction}}}%
\end{pgfscope}%
\begin{pgfscope}%
\pgfpathrectangle{\pgfqpoint{0.603704in}{0.549691in}}{\pgfqpoint{5.646296in}{4.052084in}}%
\pgfusepath{clip}%
\pgfsetrectcap%
\pgfsetroundjoin%
\pgfsetlinewidth{2.007500pt}%
\definecolor{currentstroke}{rgb}{0.121569,0.466667,0.705882}%
\pgfsetstrokecolor{currentstroke}%
\pgfsetstrokeopacity{0.800000}%
\pgfsetdash{}{0pt}%
\pgfpathmoveto{\pgfqpoint{0.903775in}{3.682222in}}%
\pgfpathlineto{\pgfqpoint{6.250000in}{0.626658in}}%
\pgfpathlineto{\pgfqpoint{6.250000in}{0.626658in}}%
\pgfusepath{stroke}%
\end{pgfscope}%
\begin{pgfscope}%
\pgfpathrectangle{\pgfqpoint{0.603704in}{0.549691in}}{\pgfqpoint{5.646296in}{4.052084in}}%
\pgfusepath{clip}%
\pgfsetbuttcap%
\pgfsetroundjoin%
\pgfsetlinewidth{2.007500pt}%
\definecolor{currentstroke}{rgb}{1.000000,0.498039,0.054902}%
\pgfsetstrokecolor{currentstroke}%
\pgfsetstrokeopacity{0.800000}%
\pgfsetdash{{2.000000pt}{3.300000pt}}{0.000000pt}%
\pgfpathmoveto{\pgfqpoint{0.903775in}{3.836245in}}%
\pgfpathlineto{\pgfqpoint{5.695081in}{0.539691in}}%
\pgfpathlineto{\pgfqpoint{5.695081in}{0.539691in}}%
\pgfusepath{stroke}%
\end{pgfscope}%
\begin{pgfscope}%
\pgfpathrectangle{\pgfqpoint{0.603704in}{0.549691in}}{\pgfqpoint{5.646296in}{4.052084in}}%
\pgfusepath{clip}%
\pgfsetbuttcap%
\pgfsetroundjoin%
\pgfsetlinewidth{2.007500pt}%
\definecolor{currentstroke}{rgb}{0.172549,0.627451,0.172549}%
\pgfsetstrokecolor{currentstroke}%
\pgfsetstrokeopacity{0.800000}%
\pgfsetdash{{7.400000pt}{3.200000pt}}{0.000000pt}%
\pgfpathmoveto{\pgfqpoint{0.903775in}{4.001193in}}%
\pgfpathlineto{\pgfqpoint{5.186962in}{0.539691in}}%
\pgfpathlineto{\pgfqpoint{5.186962in}{0.539691in}}%
\pgfusepath{stroke}%
\end{pgfscope}%
\begin{pgfscope}%
\pgfpathrectangle{\pgfqpoint{0.603704in}{0.549691in}}{\pgfqpoint{5.646296in}{4.052084in}}%
\pgfusepath{clip}%
\pgfsetbuttcap%
\pgfsetroundjoin%
\pgfsetlinewidth{2.007500pt}%
\definecolor{currentstroke}{rgb}{0.839216,0.152941,0.156863}%
\pgfsetstrokecolor{currentstroke}%
\pgfsetstrokeopacity{0.800000}%
\pgfsetdash{{12.800000pt}{3.200000pt}{2.000000pt}{3.200000pt}}{0.000000pt}%
\pgfpathmoveto{\pgfqpoint{0.903775in}{4.241513in}}%
\pgfpathlineto{\pgfqpoint{2.485241in}{0.539691in}}%
\pgfpathlineto{\pgfqpoint{2.485241in}{0.539691in}}%
\pgfusepath{stroke}%
\end{pgfscope}%
\begin{pgfscope}%
\pgfpathrectangle{\pgfqpoint{0.603704in}{0.549691in}}{\pgfqpoint{5.646296in}{4.052084in}}%
\pgfusepath{clip}%
\pgfsetbuttcap%
\pgfsetroundjoin%
\pgfsetlinewidth{2.007500pt}%
\definecolor{currentstroke}{rgb}{0.580392,0.403922,0.741176}%
\pgfsetstrokecolor{currentstroke}%
\pgfsetstrokeopacity{0.800000}%
\pgfsetdash{{2.000000pt}{2.000000pt}}{0.000000pt}%
\pgfpathmoveto{\pgfqpoint{0.903775in}{2.246855in}}%
\pgfpathlineto{\pgfqpoint{1.061831in}{0.539691in}}%
\pgfpathlineto{\pgfqpoint{1.061831in}{0.539691in}}%
\pgfusepath{stroke}%
\end{pgfscope}%
\begin{pgfscope}%
\pgfpathrectangle{\pgfqpoint{0.603704in}{0.549691in}}{\pgfqpoint{5.646296in}{4.052084in}}%
\pgfusepath{clip}%
\pgfsetbuttcap%
\pgfsetroundjoin%
\pgfsetlinewidth{2.007500pt}%
\definecolor{currentstroke}{rgb}{0.549020,0.337255,0.294118}%
\pgfsetstrokecolor{currentstroke}%
\pgfsetstrokeopacity{0.800000}%
\pgfsetdash{{10.000000pt}{20.000000pt}}{0.000000pt}%
\pgfpathmoveto{\pgfqpoint{0.903775in}{4.037241in}}%
\pgfpathlineto{\pgfqpoint{5.310976in}{0.539691in}}%
\pgfpathlineto{\pgfqpoint{5.310976in}{0.539691in}}%
\pgfusepath{stroke}%
\end{pgfscope}%
\begin{pgfscope}%
\pgfpathrectangle{\pgfqpoint{0.603704in}{0.549691in}}{\pgfqpoint{5.646296in}{4.052084in}}%
\pgfusepath{clip}%
\pgfsetbuttcap%
\pgfsetroundjoin%
\pgfsetlinewidth{2.007500pt}%
\definecolor{currentstroke}{rgb}{0.890196,0.466667,0.760784}%
\pgfsetstrokecolor{currentstroke}%
\pgfsetstrokeopacity{0.800000}%
\pgfsetdash{{10.000000pt}{10.000000pt}}{0.000000pt}%
\pgfpathmoveto{\pgfqpoint{0.903775in}{3.852631in}}%
\pgfpathlineto{\pgfqpoint{6.250000in}{0.563520in}}%
\pgfpathlineto{\pgfqpoint{6.250000in}{0.563520in}}%
\pgfusepath{stroke}%
\end{pgfscope}%
\begin{pgfscope}%
\pgfpathrectangle{\pgfqpoint{0.603704in}{0.549691in}}{\pgfqpoint{5.646296in}{4.052084in}}%
\pgfusepath{clip}%
\pgfsetbuttcap%
\pgfsetroundjoin%
\pgfsetlinewidth{2.007500pt}%
\definecolor{currentstroke}{rgb}{0.498039,0.498039,0.498039}%
\pgfsetstrokecolor{currentstroke}%
\pgfsetstrokeopacity{0.800000}%
\pgfsetdash{{6.000000pt}{10.000000pt}{2.000000pt}{10.000000pt}}{0.000000pt}%
\pgfpathmoveto{\pgfqpoint{0.903775in}{3.719362in}}%
\pgfpathlineto{\pgfqpoint{6.250000in}{0.663798in}}%
\pgfpathlineto{\pgfqpoint{6.250000in}{0.663798in}}%
\pgfusepath{stroke}%
\end{pgfscope}%
\begin{pgfscope}%
\pgfpathrectangle{\pgfqpoint{0.603704in}{0.549691in}}{\pgfqpoint{5.646296in}{4.052084in}}%
\pgfusepath{clip}%
\pgfsetrectcap%
\pgfsetroundjoin%
\pgfsetlinewidth{2.007500pt}%
\definecolor{currentstroke}{rgb}{0.737255,0.741176,0.133333}%
\pgfsetstrokecolor{currentstroke}%
\pgfsetstrokeopacity{0.800000}%
\pgfsetdash{}{0pt}%
\pgfpathmoveto{\pgfqpoint{0.903775in}{3.969514in}}%
\pgfpathlineto{\pgfqpoint{6.250000in}{0.680403in}}%
\pgfpathlineto{\pgfqpoint{6.250000in}{0.680403in}}%
\pgfusepath{stroke}%
\end{pgfscope}%
\begin{pgfscope}%
\pgfpathrectangle{\pgfqpoint{0.603704in}{0.549691in}}{\pgfqpoint{5.646296in}{4.052084in}}%
\pgfusepath{clip}%
\pgfsetbuttcap%
\pgfsetroundjoin%
\pgfsetlinewidth{2.007500pt}%
\definecolor{currentstroke}{rgb}{0.090196,0.745098,0.811765}%
\pgfsetstrokecolor{currentstroke}%
\pgfsetstrokeopacity{0.800000}%
\pgfsetdash{{2.000000pt}{3.300000pt}}{0.000000pt}%
\pgfpathmoveto{\pgfqpoint{0.903775in}{4.084212in}}%
\pgfpathlineto{\pgfqpoint{5.723937in}{0.539691in}}%
\pgfpathlineto{\pgfqpoint{5.723937in}{0.539691in}}%
\pgfusepath{stroke}%
\end{pgfscope}%
\begin{pgfscope}%
\pgfpathrectangle{\pgfqpoint{0.603704in}{0.549691in}}{\pgfqpoint{5.646296in}{4.052084in}}%
\pgfusepath{clip}%
\pgfsetrectcap%
\pgfsetroundjoin%
\pgfsetlinewidth{1.505625pt}%
\definecolor{currentstroke}{rgb}{1.000000,0.000000,0.000000}%
\pgfsetstrokecolor{currentstroke}%
\pgfsetdash{}{0pt}%
\pgfpathmoveto{\pgfqpoint{0.903775in}{0.680403in}}%
\pgfpathlineto{\pgfqpoint{6.250000in}{0.680403in}}%
\pgfpathlineto{\pgfqpoint{6.250000in}{0.680403in}}%
\pgfusepath{stroke}%
\end{pgfscope}%
\begin{pgfscope}%
\pgfsetrectcap%
\pgfsetmiterjoin%
\pgfsetlinewidth{0.803000pt}%
\definecolor{currentstroke}{rgb}{0.000000,0.000000,0.000000}%
\pgfsetstrokecolor{currentstroke}%
\pgfsetdash{}{0pt}%
\pgfpathmoveto{\pgfqpoint{0.603704in}{0.549691in}}%
\pgfpathlineto{\pgfqpoint{0.603704in}{4.601775in}}%
\pgfusepath{stroke}%
\end{pgfscope}%
\begin{pgfscope}%
\pgfsetrectcap%
\pgfsetmiterjoin%
\pgfsetlinewidth{0.803000pt}%
\definecolor{currentstroke}{rgb}{0.000000,0.000000,0.000000}%
\pgfsetstrokecolor{currentstroke}%
\pgfsetdash{}{0pt}%
\pgfpathmoveto{\pgfqpoint{6.250000in}{0.549691in}}%
\pgfpathlineto{\pgfqpoint{6.250000in}{4.601775in}}%
\pgfusepath{stroke}%
\end{pgfscope}%
\begin{pgfscope}%
\pgfsetrectcap%
\pgfsetmiterjoin%
\pgfsetlinewidth{0.803000pt}%
\definecolor{currentstroke}{rgb}{0.000000,0.000000,0.000000}%
\pgfsetstrokecolor{currentstroke}%
\pgfsetdash{}{0pt}%
\pgfpathmoveto{\pgfqpoint{0.603704in}{0.549691in}}%
\pgfpathlineto{\pgfqpoint{6.250000in}{0.549691in}}%
\pgfusepath{stroke}%
\end{pgfscope}%
\begin{pgfscope}%
\pgfsetrectcap%
\pgfsetmiterjoin%
\pgfsetlinewidth{0.803000pt}%
\definecolor{currentstroke}{rgb}{0.000000,0.000000,0.000000}%
\pgfsetstrokecolor{currentstroke}%
\pgfsetdash{}{0pt}%
\pgfpathmoveto{\pgfqpoint{0.603704in}{4.601775in}}%
\pgfpathlineto{\pgfqpoint{6.250000in}{4.601775in}}%
\pgfusepath{stroke}%
\end{pgfscope}%
\begin{pgfscope}%
\pgfsetbuttcap%
\pgfsetmiterjoin%
\definecolor{currentfill}{rgb}{1.000000,1.000000,1.000000}%
\pgfsetfillcolor{currentfill}%
\pgfsetfillopacity{0.800000}%
\pgfsetlinewidth{1.003750pt}%
\definecolor{currentstroke}{rgb}{0.800000,0.800000,0.800000}%
\pgfsetstrokecolor{currentstroke}%
\pgfsetstrokeopacity{0.800000}%
\pgfsetdash{}{0pt}%
\pgfpathmoveto{\pgfqpoint{4.177660in}{2.360263in}}%
\pgfpathlineto{\pgfqpoint{6.152778in}{2.360263in}}%
\pgfpathquadraticcurveto{\pgfqpoint{6.180556in}{2.360263in}}{\pgfqpoint{6.180556in}{2.388041in}}%
\pgfpathlineto{\pgfqpoint{6.180556in}{4.504552in}}%
\pgfpathquadraticcurveto{\pgfqpoint{6.180556in}{4.532330in}}{\pgfqpoint{6.152778in}{4.532330in}}%
\pgfpathlineto{\pgfqpoint{4.177660in}{4.532330in}}%
\pgfpathquadraticcurveto{\pgfqpoint{4.149882in}{4.532330in}}{\pgfqpoint{4.149882in}{4.504552in}}%
\pgfpathlineto{\pgfqpoint{4.149882in}{2.388041in}}%
\pgfpathquadraticcurveto{\pgfqpoint{4.149882in}{2.360263in}}{\pgfqpoint{4.177660in}{2.360263in}}%
\pgfpathlineto{\pgfqpoint{4.177660in}{2.360263in}}%
\pgfpathclose%
\pgfusepath{stroke,fill}%
\end{pgfscope}%
\begin{pgfscope}%
\pgfsetrectcap%
\pgfsetroundjoin%
\pgfsetlinewidth{2.007500pt}%
\definecolor{currentstroke}{rgb}{0.121569,0.466667,0.705882}%
\pgfsetstrokecolor{currentstroke}%
\pgfsetstrokeopacity{0.800000}%
\pgfsetdash{}{0pt}%
\pgfpathmoveto{\pgfqpoint{4.205437in}{4.428164in}}%
\pgfpathlineto{\pgfqpoint{4.344326in}{4.428164in}}%
\pgfpathlineto{\pgfqpoint{4.483215in}{4.428164in}}%
\pgfusepath{stroke}%
\end{pgfscope}%
\begin{pgfscope}%
\definecolor{textcolor}{rgb}{0.000000,0.000000,0.000000}%
\pgfsetstrokecolor{textcolor}%
\pgfsetfillcolor{textcolor}%
\pgftext[x=4.594326in,y=4.379552in,left,base]{\color{textcolor}{\rmfamily\fontsize{10.000000}{12.000000}\selectfont\catcode`\^=\active\def^{\ifmmode\sp\else\^{}\fi}\catcode`\%=\active\def%{\%}Gitmerge-ort}}%
\end{pgfscope}%
\begin{pgfscope}%
\pgfsetbuttcap%
\pgfsetroundjoin%
\pgfsetlinewidth{2.007500pt}%
\definecolor{currentstroke}{rgb}{1.000000,0.498039,0.054902}%
\pgfsetstrokecolor{currentstroke}%
\pgfsetstrokeopacity{0.800000}%
\pgfsetdash{{2.000000pt}{3.300000pt}}{0.000000pt}%
\pgfpathmoveto{\pgfqpoint{4.205437in}{4.234491in}}%
\pgfpathlineto{\pgfqpoint{4.344326in}{4.234491in}}%
\pgfpathlineto{\pgfqpoint{4.483215in}{4.234491in}}%
\pgfusepath{stroke}%
\end{pgfscope}%
\begin{pgfscope}%
\definecolor{textcolor}{rgb}{0.000000,0.000000,0.000000}%
\pgfsetstrokecolor{textcolor}%
\pgfsetfillcolor{textcolor}%
\pgftext[x=4.594326in,y=4.185880in,left,base]{\color{textcolor}{\rmfamily\fontsize{10.000000}{12.000000}\selectfont\catcode`\^=\active\def^{\ifmmode\sp\else\^{}\fi}\catcode`\%=\active\def%{\%}Gitmerge-ort-ignorespace}}%
\end{pgfscope}%
\begin{pgfscope}%
\pgfsetbuttcap%
\pgfsetroundjoin%
\pgfsetlinewidth{2.007500pt}%
\definecolor{currentstroke}{rgb}{0.172549,0.627451,0.172549}%
\pgfsetstrokecolor{currentstroke}%
\pgfsetstrokeopacity{0.800000}%
\pgfsetdash{{7.400000pt}{3.200000pt}}{0.000000pt}%
\pgfpathmoveto{\pgfqpoint{4.205437in}{4.040818in}}%
\pgfpathlineto{\pgfqpoint{4.344326in}{4.040818in}}%
\pgfpathlineto{\pgfqpoint{4.483215in}{4.040818in}}%
\pgfusepath{stroke}%
\end{pgfscope}%
\begin{pgfscope}%
\definecolor{textcolor}{rgb}{0.000000,0.000000,0.000000}%
\pgfsetstrokecolor{textcolor}%
\pgfsetfillcolor{textcolor}%
\pgftext[x=4.594326in,y=3.992207in,left,base]{\color{textcolor}{\rmfamily\fontsize{10.000000}{12.000000}\selectfont\catcode`\^=\active\def^{\ifmmode\sp\else\^{}\fi}\catcode`\%=\active\def%{\%}Hires-Merge}}%
\end{pgfscope}%
\begin{pgfscope}%
\pgfsetbuttcap%
\pgfsetroundjoin%
\pgfsetlinewidth{2.007500pt}%
\definecolor{currentstroke}{rgb}{0.839216,0.152941,0.156863}%
\pgfsetstrokecolor{currentstroke}%
\pgfsetstrokeopacity{0.800000}%
\pgfsetdash{{12.800000pt}{3.200000pt}{2.000000pt}{3.200000pt}}{0.000000pt}%
\pgfpathmoveto{\pgfqpoint{4.205437in}{3.847145in}}%
\pgfpathlineto{\pgfqpoint{4.344326in}{3.847145in}}%
\pgfpathlineto{\pgfqpoint{4.483215in}{3.847145in}}%
\pgfusepath{stroke}%
\end{pgfscope}%
\begin{pgfscope}%
\definecolor{textcolor}{rgb}{0.000000,0.000000,0.000000}%
\pgfsetstrokecolor{textcolor}%
\pgfsetfillcolor{textcolor}%
\pgftext[x=4.594326in,y=3.798534in,left,base]{\color{textcolor}{\rmfamily\fontsize{10.000000}{12.000000}\selectfont\catcode`\^=\active\def^{\ifmmode\sp\else\^{}\fi}\catcode`\%=\active\def%{\%}Spork}}%
\end{pgfscope}%
\begin{pgfscope}%
\pgfsetbuttcap%
\pgfsetroundjoin%
\pgfsetlinewidth{2.007500pt}%
\definecolor{currentstroke}{rgb}{0.580392,0.403922,0.741176}%
\pgfsetstrokecolor{currentstroke}%
\pgfsetstrokeopacity{0.800000}%
\pgfsetdash{{2.000000pt}{2.000000pt}}{0.000000pt}%
\pgfpathmoveto{\pgfqpoint{4.205437in}{3.653472in}}%
\pgfpathlineto{\pgfqpoint{4.344326in}{3.653472in}}%
\pgfpathlineto{\pgfqpoint{4.483215in}{3.653472in}}%
\pgfusepath{stroke}%
\end{pgfscope}%
\begin{pgfscope}%
\definecolor{textcolor}{rgb}{0.000000,0.000000,0.000000}%
\pgfsetstrokecolor{textcolor}%
\pgfsetfillcolor{textcolor}%
\pgftext[x=4.594326in,y=3.604861in,left,base]{\color{textcolor}{\rmfamily\fontsize{10.000000}{12.000000}\selectfont\catcode`\^=\active\def^{\ifmmode\sp\else\^{}\fi}\catcode`\%=\active\def%{\%}IntelliMerge}}%
\end{pgfscope}%
\begin{pgfscope}%
\pgfsetbuttcap%
\pgfsetroundjoin%
\pgfsetlinewidth{2.007500pt}%
\definecolor{currentstroke}{rgb}{0.549020,0.337255,0.294118}%
\pgfsetstrokecolor{currentstroke}%
\pgfsetstrokeopacity{0.800000}%
\pgfsetdash{{10.000000pt}{20.000000pt}}{0.000000pt}%
\pgfpathmoveto{\pgfqpoint{4.205437in}{3.459800in}}%
\pgfpathlineto{\pgfqpoint{4.344326in}{3.459800in}}%
\pgfpathlineto{\pgfqpoint{4.483215in}{3.459800in}}%
\pgfusepath{stroke}%
\end{pgfscope}%
\begin{pgfscope}%
\definecolor{textcolor}{rgb}{0.000000,0.000000,0.000000}%
\pgfsetstrokecolor{textcolor}%
\pgfsetfillcolor{textcolor}%
\pgftext[x=4.594326in,y=3.411189in,left,base]{\color{textcolor}{\rmfamily\fontsize{10.000000}{12.000000}\selectfont\catcode`\^=\active\def^{\ifmmode\sp\else\^{}\fi}\catcode`\%=\active\def%{\%}Adjacent}}%
\end{pgfscope}%
\begin{pgfscope}%
\pgfsetbuttcap%
\pgfsetroundjoin%
\pgfsetlinewidth{2.007500pt}%
\definecolor{currentstroke}{rgb}{0.890196,0.466667,0.760784}%
\pgfsetstrokecolor{currentstroke}%
\pgfsetstrokeopacity{0.800000}%
\pgfsetdash{{10.000000pt}{10.000000pt}}{0.000000pt}%
\pgfpathmoveto{\pgfqpoint{4.205437in}{3.266127in}}%
\pgfpathlineto{\pgfqpoint{4.344326in}{3.266127in}}%
\pgfpathlineto{\pgfqpoint{4.483215in}{3.266127in}}%
\pgfusepath{stroke}%
\end{pgfscope}%
\begin{pgfscope}%
\definecolor{textcolor}{rgb}{0.000000,0.000000,0.000000}%
\pgfsetstrokecolor{textcolor}%
\pgfsetfillcolor{textcolor}%
\pgftext[x=4.594326in,y=3.217516in,left,base]{\color{textcolor}{\rmfamily\fontsize{10.000000}{12.000000}\selectfont\catcode`\^=\active\def^{\ifmmode\sp\else\^{}\fi}\catcode`\%=\active\def%{\%}Imports}}%
\end{pgfscope}%
\begin{pgfscope}%
\pgfsetbuttcap%
\pgfsetroundjoin%
\pgfsetlinewidth{2.007500pt}%
\definecolor{currentstroke}{rgb}{0.498039,0.498039,0.498039}%
\pgfsetstrokecolor{currentstroke}%
\pgfsetstrokeopacity{0.800000}%
\pgfsetdash{{6.000000pt}{10.000000pt}{2.000000pt}{10.000000pt}}{0.000000pt}%
\pgfpathmoveto{\pgfqpoint{4.205437in}{3.072454in}}%
\pgfpathlineto{\pgfqpoint{4.344326in}{3.072454in}}%
\pgfpathlineto{\pgfqpoint{4.483215in}{3.072454in}}%
\pgfusepath{stroke}%
\end{pgfscope}%
\begin{pgfscope}%
\definecolor{textcolor}{rgb}{0.000000,0.000000,0.000000}%
\pgfsetstrokecolor{textcolor}%
\pgfsetfillcolor{textcolor}%
\pgftext[x=4.594326in,y=3.023843in,left,base]{\color{textcolor}{\rmfamily\fontsize{10.000000}{12.000000}\selectfont\catcode`\^=\active\def^{\ifmmode\sp\else\^{}\fi}\catcode`\%=\active\def%{\%}Version Numbers}}%
\end{pgfscope}%
\begin{pgfscope}%
\pgfsetrectcap%
\pgfsetroundjoin%
\pgfsetlinewidth{2.007500pt}%
\definecolor{currentstroke}{rgb}{0.737255,0.741176,0.133333}%
\pgfsetstrokecolor{currentstroke}%
\pgfsetstrokeopacity{0.800000}%
\pgfsetdash{}{0pt}%
\pgfpathmoveto{\pgfqpoint{4.205437in}{2.878781in}}%
\pgfpathlineto{\pgfqpoint{4.344326in}{2.878781in}}%
\pgfpathlineto{\pgfqpoint{4.483215in}{2.878781in}}%
\pgfusepath{stroke}%
\end{pgfscope}%
\begin{pgfscope}%
\definecolor{textcolor}{rgb}{0.000000,0.000000,0.000000}%
\pgfsetstrokecolor{textcolor}%
\pgfsetfillcolor{textcolor}%
\pgftext[x=4.594326in,y=2.830170in,left,base]{\color{textcolor}{\rmfamily\fontsize{10.000000}{12.000000}\selectfont\catcode`\^=\active\def^{\ifmmode\sp\else\^{}\fi}\catcode`\%=\active\def%{\%}IVn}}%
\end{pgfscope}%
\begin{pgfscope}%
\pgfsetbuttcap%
\pgfsetroundjoin%
\pgfsetlinewidth{2.007500pt}%
\definecolor{currentstroke}{rgb}{0.090196,0.745098,0.811765}%
\pgfsetstrokecolor{currentstroke}%
\pgfsetstrokeopacity{0.800000}%
\pgfsetdash{{2.000000pt}{3.300000pt}}{0.000000pt}%
\pgfpathmoveto{\pgfqpoint{4.205437in}{2.685109in}}%
\pgfpathlineto{\pgfqpoint{4.344326in}{2.685109in}}%
\pgfpathlineto{\pgfqpoint{4.483215in}{2.685109in}}%
\pgfusepath{stroke}%
\end{pgfscope}%
\begin{pgfscope}%
\definecolor{textcolor}{rgb}{0.000000,0.000000,0.000000}%
\pgfsetstrokecolor{textcolor}%
\pgfsetfillcolor{textcolor}%
\pgftext[x=4.594326in,y=2.636497in,left,base]{\color{textcolor}{\rmfamily\fontsize{10.000000}{12.000000}\selectfont\catcode`\^=\active\def^{\ifmmode\sp\else\^{}\fi}\catcode`\%=\active\def%{\%}IVn-ignorespace}}%
\end{pgfscope}%
\begin{pgfscope}%
\pgfsetrectcap%
\pgfsetroundjoin%
\pgfsetlinewidth{1.505625pt}%
\definecolor{currentstroke}{rgb}{1.000000,0.000000,0.000000}%
\pgfsetstrokecolor{currentstroke}%
\pgfsetdash{}{0pt}%
\pgfpathmoveto{\pgfqpoint{4.205437in}{2.491436in}}%
\pgfpathlineto{\pgfqpoint{4.344326in}{2.491436in}}%
\pgfpathlineto{\pgfqpoint{4.483215in}{2.491436in}}%
\pgfusepath{stroke}%
\end{pgfscope}%
\begin{pgfscope}%
\definecolor{textcolor}{rgb}{0.000000,0.000000,0.000000}%
\pgfsetstrokecolor{textcolor}%
\pgfsetfillcolor{textcolor}%
\pgftext[x=4.594326in,y=2.442825in,left,base]{\color{textcolor}{\rmfamily\fontsize{10.000000}{12.000000}\selectfont\catcode`\^=\active\def^{\ifmmode\sp\else\^{}\fi}\catcode`\%=\active\def%{\%}Manual Merging}}%
\end{pgfscope}%
\end{pgfpicture}%
\makeatother%
\endgroup%

%% file: results/combined/tables/tools/table_run_time.tex
% Do not edit.  This file is automatically generated.
\begin{tabular}{c|c|c|c}
    & \multicolumn{3}{c}{Run time (seconds)} \\
    Tool & Mean & Median & Max \\
    \hline
    Gitmerge-ort                     & 0.04 & 0.04 & 1.19 \\
    Gitmerge-ort-ignorespace         & 0.04 & 0.04 & 1.22 \\
    Hires-Merge                      & 0.23 & 0.13 & 17.3 \\
    Spork                            & 2.55 & 1.06 & 653 \\
    IntelliMerge                     & 1.04 & 0.50 & 89.3 \\
    Adjacent                         & 0.17 & 0.05 & 76.9 \\
    Imports                          & 0.11 & 0.07 & 7.86 \\
    Version Numbers                  & 0.07 & 0.05 & 2.75 \\
    IVn                              & 0.12 & 0.08 & 7.97 \\
    IVn-ignorespace                  & 0.12 & 0.08 & 8.08 \\
\end{tabular}

%% file: results/combined/tables/tools/table_feature_main_summary.tex
% Do not edit.  This file is automatically generated.
\setlength{\tabcolsep}{.285\tabcolsep}
\begin{tabular}{c|cc|cc|cc}
            Tool &
            \multicolumn{6}{c}{Merges} \\ \cline{2-7}
            &
            \multicolumn{2}{c|}{Correct} &
            \multicolumn{2}{c|}{Unhandled} &
            \multicolumn{2}{c}{Incorrect} \\
            &
            \multicolumn{1}{c}{Main} &
            \multicolumn{1}{c|}{Other} &
            \multicolumn{1}{c}{Main} &
            \multicolumn{1}{c|}{Other} &
            \multicolumn{1}{c}{Main} &
            \multicolumn{1}{c}{Other} \\
            \hline
            Gitmerge-ort                     &  53\% &  35\% &  44\% &  62\% &   3\% &   2\% \\
            Gitmerge-ort-ignorespace         &  56\% &  38\% &  41\% &  59\% &   3\% &   3\% \\
            Hires-Merge                      &  57\% &  41\% &  39\% &  55\% &   4\% &   4\% \\
            Spork                            &  62\% &  44\% &  28\% &  45\% &  11\% &  11\% \\
            IntelliMerge                     &  27\% &  19\% &  21\% &  34\% &  52\% &  47\% \\
            Adjacent                         &  58\% &  42\% &  39\% &  54\% &   4\% &   4\% \\
            Imports                          &  56\% &  38\% &  41\% &  59\% &   3\% &   3\% \\
            Version Numbers                  &  54\% &  36\% &  44\% &  61\% &   3\% &   2\% \\
            IVn                              &  57\% &  41\% &  40\% &  57\% &   3\% &   3\% \\
            IVn-ignorespace                  &  59\% &  42\% &  38\% &  54\% &   3\% &   4\% \\
\end{tabular}

%% file: paper-qualitative.tex
We manually examined hundreds of merges in which two tools produced
different results.

For each tool $X$, we created two pools of merges:  a pool where $X$ failed
and all others succeeded, and a pool where $X$ succeeded and all others
failed.  We randomly chose merges from each pool, so we saw
examples of each tool doing well and doing poorly.

For each selected merge, we compared the base, left, and right
versions, the programmer merge, and the results of merge tools.  We
primarily used the \<diff> and \<diff3> tools for these comparisons.  Every
evaluation was performed by one author and reviewed by at least two other
authors.  Disagreements were resolved by discussion.\looseness=-1

The appendix~\cite{ScheschFYRE2024:TR} shows our analysis of 75 merges.
Here we present a subset of them.
Each merge in our dataset has an index such as ``123-45''.  We show edits
in diff3 format, which gives the left parent, then the base, then the right
parent.

\subsection{Hires Merge}

\subsubsection{Handling Refactorings With Multiple Inline Changes (3183-11)}
Hires Merge works character-wise.  This strategy deals with refactorings quite effectively.

\begin{lstlisting}[basicstyle=\footnotesize\ttfamily,numbers=none]
<@\leftmarker@>
HashSet<Range> ranges = new HashSet<@\texttt{\textcolor{red}{<>}}@>();
<@\basemarker@>
HashSet<Range> ranges = new HashSet<Range>();
<@\baserightsepmarker@>
<@\texttt{\textcolor{red}{Set}}@><Range> ranges = new HashSet<Range>();
<@\rightmarker@>
\end{lstlisting}

Git Merge gets stuck because the left and right edited the same line.
Hires Merge comes up with a correct merge:
\begin{lstlisting}[basicstyle=\footnotesize\ttfamily,numbers=none]
<@\texttt{\textcolor{red}{Set}}@><Range> ranges = new HashSet<@\texttt{\textcolor{red}{<>}}@>();
\end{lstlisting}

% The programmer also gets a semantically identical
% \begin{lstlisting}[basicstyle=\footnotesize\ttfamily,numbers=none]
% Set<Range> ranges = new HashSet<>();
% \end{lstlisting}

\subsubsection{Hires Merge Incorrectly Identifying Version Numbers (25267-730)}

Merging character-by-character loses context.  In this merge:

\begin{lstlisting}[basicstyle=\footnotesize\ttfamily,numbers=none]
<@\leftmarker@>
<version>23.7.0</version>
<@\basemarker@>
<version>23.6.0</version>
<@\baserightsepmarker@>
<version>23.6.1</version>
<@\rightmarker@>
\end{lstlisting}
Hires Merge invented a nonexistent version number:
\begin{lstlisting}[basicstyle=\footnotesize\ttfamily,numbers=none]
<version>23.7.1</version>
\end{lstlisting}

% while the programmer chose:
% \begin{lstlisting}[basicstyle=\footnotesize\ttfamily,numbers=none]
% <version>23.7.0</version>
% \end{lstlisting}
%
% The Hires Merge strategy of using Git Merge at the character level chooses
% 23, 7, and then 1 (choosing the left's unique second option the right's
% unique third option). Thus, it makes up a nonexistent number:
% 23.7.1. Moreover, choosing the right branch when the version numbers differ
% would also be an incorrect assumption in this case since the left is the
% most recent. This highlights an issue with the Hires Merge approach:
% context in the rest of the line matters. Ideally, a tool would address
% these conflicts by intelligently choosing the highest version number of the
% three. Until then, however, Git Merge would probably cause less headaches.

% \vspace{-5pt}

\subsection{Adjacent}

\subsubsection{Refactoring on Adjacent Lines (1215-3280)}

Adjacent successfully merged scenarios involving refactoring, particularly when variables were independent.
\begin{lstlisting}[basicstyle=\footnotesize\ttfamily,numbers=none]
<@\leftmarker@>
String comments = SourcesHelper.readerToString(reader);
CompilationUnit cu = new <@\texttt{\textcolor{red}{JavaParser().setSource}}@>(comments).parse();
<@\basemarker@>
String comments = SourcesHelper.readerToString(reader);
CompilationUnit cu = new InstanceJavaParser(comments).parse();
<@\baserightsepmarker@>
String comments = <@\texttt{\textcolor{red}{readerToString}}@>(reader);
CompilationUnit cu = new InstanceJavaParser(comments).parse();
<@\rightmarker@>
\end{lstlisting}

% In this example, the right branch refactors the comments string, while the left branch refactors cu. Git Merge fails to merge due to its policy of not merging adjacent lines that conflict across multiple branches. However, adjacent merges successfully merges, and this is justified due to the lines being unrelated:
% \begin{lstlisting}[basicstyle=\footnotesize\ttfamily,numbers=none]
% String comments = readerToString(reader);
% CompilationUnit cu = new JavaParser().setSource(comments).parse();
% \end{lstlisting}
% This is also the same output that the programmer chose.

\subsubsection{Adjacent Lines are Interdependent (5184-31)}

The key weakness in the adjacent strategy is its local view, disregarding context.
Consider this merge.  The left parent changed the variable being
synchronized upon.

\begin{lstlisting}[basicstyle=\footnotesize\ttfamily,numbers=none]
<@\leftmarker@>
<@\texttt{\textcolor{red}{synchronized (cacheMap) \{}}@>
    List<DNSEntry> entryList = <@\texttt{\textcolor{red}{cacheMap}}@>.get(dnsEntry.getKey());
    if (entryList != null) {    
        entryList.remove(dnsEntry);
<@\basemarker@>
List<DNSEntry> entryList = this.get(dnsEntry.getKey());
if (entryList != null) {
    synchronized (entryList) {
        entryList.remove(dnsEntry);
<@\baserightsepmarker@>
List<DNSEntry> entryList = this.get(dnsEntry.getKey());
if (entryList != null) {
    synchronized (entryList) {
        <@\texttt{\textcolor{red}{result =}}@> entryList.remove(dnsEntry);
<@\rightmarker@>
    }
}
/* Remove from DNS cache when no records remain with this key */
if (result && entryList.isEmpty()) {
    this.remove(dnsEntry.getKey());
\end{lstlisting}

Adjacent readily merges the code, but without moving the outer if-statement
inside the \<synchronized> block, leading to code that compiles but
contains a race condition.

\label{sec:race-condition-merge}

\begin{lstlisting}[basicstyle=\footnotesize\ttfamily,numbers=none]
synchronized (cacheMap) {
    List<DNSEntry> entryList = cacheMap.get(dnsEntry.getKey());
    if (entryList != null) {
        result = entryList.remove(dnsEntry);
    }
}
/* Remove from DNS cache when no records remain with this key */
if (result && entryList.isEmpty()) {
    this.remove(dnsEntry.getKey());
}
\end{lstlisting}

Git Merge left this conflict unhandled, forcing the programmer to do the
merge, which is a better outcome.

% \vspace{-5pt}

\subsection{Git Merge Ignorespace}

\subsubsection{Extra Irrelevant Spaces (2955-73)}
\label{sec:yaml-bad-merge}

Git Merge Ignorespace is not confused by inconsequential spaces.  It merges
code like this:

\pagebreak[3]

\begin{lstlisting}[basicstyle=\footnotesize\ttfamily,numbers=none]
<@\leftmarker@>
 * </p>
<@\basemarker@>
 *<@\textcolor{red}{\textvisiblespace}@>
<@\baserightsepmarker@>
 *
<@\rightmarker@>
\end{lstlisting}

Git Merge Ignorespace caused havoc when merging YAML files (e.g., 14378-60),
where indentation matters and there may be multiple occurrences of a key.

\iffalse
\todo{Is this example too long?}

 In \<swagger-spring.yaml> of 14378-60, Git Merge made the merge
\begin{lstlisting}[basicstyle=\footnotesize\ttfamily,numbers=none]
properties:
  id:
    type: "integer"
    format: "int64"
    xml:
      namespace: "http://com.wordnik/sample/model/category"
  name:
    type: "string"
    xml:
      namespace: "http://com.wordnik/sample/model/category"
xml:
  name: "Category"
  namespace: "http://com.wordnik/sample/model/category"
\end{lstlisting}
while Git Merge Ignorespace did
\begin{lstlisting}[basicstyle=\footnotesize\ttfamily,numbers=none]
properties:
  id:
    type: "integer"
    format: "int64"
    xml:
      namespace: "http://com.wordnik/sample/model/category"
  name:
    type: "string"
<@\texttt{\textcolor{red}{xml:}}@>
      namespace: "http://com.wordnik/sample/model/category"
xml:
  name: "Category"
  namespace: "http://com.wordnik/sample/model/category"
\end{lstlisting}
Since YAML is sensitive to indentation, the incorrect source file was
produced. Instead of adding \<xml> as a property of \<name> (that is, indented under \<name>),
Git Merge Ignorespace made \<xml> a sibling of \<properties>.
\fi

\subsection{Spork}

As explained in the appendix~\cite{ScheschFYRE2024:TR},
Spork sometimes produced uncompilable code, made
gratuitous formatting changes, or omitted method bodies.
Spork's maintainers acknowledged our bug reports but have not fixed them.
We spent well over a person-month trying to fix the bugs ourselves, but
were not able to address them all.  Then, we tried to refactor Spork to
eliminate its dependence on Spoon (which the Spork maintainers blamed for
some of Spork's bugs), but they were so entangled that we were unable to do
so.  We speculate that a better implementation of the Spork algorithm could
be a very effective merge tool.

\todo{Link to bug reports?}

\subsubsection{Overlapping Unique Additions (35091-165)}

Spork's strategy of parsing code into an AST tree and matching methods by
name was quite successful when different branches added
different methods at the same location.  Examples like this were
the bread and butter of Spork's successes.

\iffalse
\begin{lstlisting}[basicstyle=\footnotesize\ttfamily,numbers=none]
<@\leftmarker@>
<@\leftmarker@>
public void testAsStringColumn() {
  column1.appendCell("1923-10-20T10:15:30");
  column1.appendMissing();
  StringColumn sc = column1.asStringColumn();
  assertEquals("Game date strings", sc.name());
  assertEquals(2, sc.size());
  assertEquals("1923-10-20T10:15:30.000", sc.get(0));
  assertEquals(StringColumnType.missingValueIndicator(), sc.get(1));
<@\basemarker@>
<@\baserightsepmarker@>
public void testFormatter() {
  column1.setPrintFormatter(DateTimeFormatter.ISO_LOCAL_DATE_TIME, "NaT");
  column1.append(LocalDateTime.of(2000, 1, 1, 0, 0));
  column1.appendMissing();
  assertEquals("2000-01-01T00:00:00", column1.getString(0));
  assertEquals("NaT", column1.getString(1));
<@\rightmarker@>
\end{lstlisting}

Git Merge failed this conflict since the lines that the methods were added
overlapped. Spork, however, successfully merged both methods alongside the
programmer. The biggest weakness of line-based merging is that it ignores
the context of the conflicting file. A programmer's vein of thinking
alongside Spork, however, use this context to realize that both branches
were adding non-semantically overlapping changes. The Spork algorithm
thrives on clear cut examples where both branches added non-semantically
overlapping changes.
\fi

\subsection{Version Numbers}
\label{sec:qualitative-version-numbers}

The Version Numbers tool starts with the output of Git Merge.
%% This will be true with the new experiments.
It never underperformed Git Merge.
% In only one
% case, 9364-16, did it underperform Git Merge.  Git Merge yielded a
% conflict, so its output was classified as ``unhandled''.  Version Numbers
% resolved the conflict in exactly the same way the programmer did, but the
% tests failed, so its output was classified as ``incorrect'', which is worse
% than ``unhandled''.  In this particular case, one branch added a failing test case to a disabled test
% suite, and the other branch enabled the test suite.  The programmer
% committed a fix 6 minutes after performing the merge.

\subsection{Imports}
\label{sec:qualitative-imports}

\label{sec:last-qualitative}

The Imports tool starts with the output of Git Merge and only fixes merges
in \<import> statements.  It never introduces mistakes, because it parses
the entire file looking for uses of imports.  It can correct mistakes by
re-introducing \<import> statements that Git Merge removed by a clean but
incorrect merge.

% LocalWords:  diff3 basicstyle SourcesHelper readerToString setSource DNS
% LocalWords:  CompilationUnit InstanceJavaParser cacheMap DNSEntry getKey
% LocalWords:  entryList dnsEntry Ignorespace YAML yaml Ort int64 Spork sc
% LocalWords:  namespace ignorespace Spork's testAsStringColumn column1
% LocalWords:  appendCell appendMissing StringColumn asStringColumn NaT
% LocalWords:  assertEquals StringColumnType missingValueIndicator Merge's
% LocalWords:  testFormatter setPrintFormatter DateTimeFormatter getString
% LocalWords:  LocalDateTime underperform